\documentclass[journal]{IEEEtran}
\ifCLASSINFOpdf
\usepackage[pdftex]{graphicx}
\graphicspath{{../pdf/}{../jpeg/}}
\DeclareGraphicsExtensions{.pdf,.jpeg,.png}
\else
\usepackage[dvips]{graphicx}
\graphicspath{{../eps/}}
\DeclareGraphicsExtensions{.eps}
\fi

\usepackage{subfigure}
\usepackage{makecell,multirow,diagbox}
\usepackage{pifont}
\usepackage{cite}
\usepackage{color}

\usepackage{algorithmic}
\usepackage{algorithm}
\usepackage{epstopdf}
\usepackage{textcomp}
\usepackage{amsfonts}
\usepackage{booktabs}
\usepackage{pifont}
\usepackage{enumerate}
\usepackage[cmex10]{amsmath}

\usepackage{mathabx}
\usepackage{eufrak}
\usepackage{yfonts}
\usepackage{soul}
\usepackage{mathrsfs}
\soulregister\cite7 
\soulregister\citep7 
\soulregister\citet7 
\soulregister\ref7 
\soulregister\eqref7 
\soulregister\pageref7 

\begin{document}

\title{Joint Data Compression, Secure Multi-Part Collaborative Task Offloading and Resource Assignment in Ultra-Dense Networks}

\author{Tianqing~Zhou,~
        Kangle~Liu,~
        Dong~Qin,~
        Xuan~Li,~
        Nan~Jiang,~
        and Chunguo~Li~
\thanks{This work was supported by National Natural Science Foundation of China under Grant Nos. 62261020, 62171119, 62361026 and 62461036, Jiangxi Provincial Natural Science Foundation under Grant Nos. 20232ACB212005, 20224BAB202001 and 20232BAB202019, Key research and development plan of Jiangsu Province under Grant No. BE2021013-3. } 
\thanks{T. Zhou, K. Liu, X. Li and N. Jiang are with the School of Information and Software Engineering, East China Jiaotong University, Nanchang 330013, China (email: zhoutian930@163.com; liukl3181@163.com; lixuan@ecjtu.edu.cn; jiangnan1018@gmail.com).}
\thanks{D. Qin is with School of Information Engineering, Nanchang University, Nanchang 330031, China (e-mail: qindong@seu.edu.cn).}
\thanks{C. Li is with School of Information Science and Engineering, Southeast University, Nanjing 210096, China (email: chunguoli@seu.edu.cn).}}

\maketitle

\begin{abstract}
To enhance resource utilization and address interference issues in ultra-dense networks with mobile edge computing (MEC), a resource utilization approach is first introduced, which integrates orthogonal frequency division multiple access (OFDMA) and non-orthogonal multiple access (NOMA). Then, to minimize the energy consumed by ultra-densely deployed small base stations (SBSs) while ensuring proportional assignment of computational resources and the constraints related to processing delay and security breach cost, the joint optimization of channel selection, the number of subchannels, secure service assignment, multi-step computation offloading, device association, data compression (DC) control, power control, and frequency band partitioning is done for minimizing network-wide energy consumption (EC). Given that the current problem is nonlinear and involves integral optimization parameters, we have devised an adaptive genetic water wave optimization (AGWWO) algorithm by improving the traditional water wave optimization (WWO) algorithm using genetic operations. After that, the computational complexity, convergence, and parallel implementation of AGWWO algorithm are analyzed. Simulation results reveal that this algorithm effectively reduces network-wide EC while guaranteeing the constraints of processing delay and security breach cost.
\end{abstract}


\IEEEpeerreviewmaketitle

\section{Introduction}\label{sec1}

\IEEEPARstart{T}{he} rapid advancement of communication technologies has given rise to an array of computing-intensive and delay-sensitive applications, including virtual reality, augmented reality, smart cities, and smart vehicles\cite{J.Zhao2024,Q.Zhang2024}. To deal with these challenges, mobile edge computing (MEC) has emerged as a promising solution by offering abundant computational resources to users at the edge of the network \cite{J.Zhao2023Apr,J.Zhao2024Mar}. To improve spectrum utilization, enhance network coverage and support the connections of massive mobile devices (MDs), ultra-dense small base stations (SBSs) are deployed into MEC-enabled networks. Such ultra-dense MEC-enabled networks have attracted more and more attention because of shortened communicational distances and more utilized computational resources \cite{T.Zhou2021Dec}.
\par
However, the proliferation of base stations (BSs) will result in the underutilization of computational resources, severe network interferences and huge energy consumption (EC). Moreover, during wireless offloading, a large amount of offloaded tasks will result in a heavy communicational burden. Meanwhile, such offloaded data is vulnerable to attack. Generally, in ultra-dense MEC-enabled networks, how to effectively mitigate network interferences and support massive connections is an important issue, how to effectively select multiple BSs to serve each MD is a key issue, how to effectively protect offloaded tasks is a vital issue, and how to effectively reduce the size of offloaded data is another important issue. Furthermore, how to effectively assign communicational and computational resources to reduce network-wide EC is a significant and open topic, where such an EC refers to the energy consumed by MDs and BSs.

\subsection{Related Work}
In recent years, to tackle the aforementioned challenges, significant efforts have been dedicated to developing computation offloading mechanisms.
\par
Spectrum sharing can greatly improve the spectrum utilization of wireless networks, but it can cause severe network interference. Therefore, it is necessary to introduce reasonable spectrum resource management strategies. In \cite{L.Tan2022Mar}, Tan \textit{et al.} jointly optimized multi-user cooperative decision-making, computation offloading, and the assignment of communicational and computational resources to minimize EC of all mobile devices (MDs) for an orthogonal frequency division multiple access (OFDMA) system with equal bandwidth assignment. After introducing non-orthogonal multiple access (NOMA) into the multi-cell MEC networks, Lai \textit{et al.}\cite{P.Lai2022Dec} jointly optimized user and power assignment to minimize the system cost denoted as the weighted sum of computational resources and user power. In \cite{SMao2022Mar}, Mao \textit{et al.} jointly optimized cooperative computation offloading, wireless energy transfer, and resource assignment to minimize system EC for a time division multiple access (TDMA) system. In \cite{HLim2022}, Lim \textit{et al.} jointly optimized downloading and offloading time, user power, local computational resource and time, and downlink beamforming to maximize the system energy efficiency for a system based on both TDMA and space division multiple access (SDMA).
\par
Clearly, the previously mentioned single-step approach to computation offloading does not sufficiently leverage the available computational resources. Given this, there has been increasing attention on multi-step offloading. In \cite{Y.Dai2018Dec}, the optimization process integrated user association, a multi-step decision for offloading, user power assignment, and the management of computational resources, reducing the overall EC across the networks for a multi-task MEC system with OFDMA and sharing frequency bands. In \cite{T.Zhou2022Oct}, the joint optimization of device association, multi-step offloading decision, user power, and frequency band partitioning factor was done, minimizing the network-wide EC for ultra-dense multi-task IoT networks with OFDMA. In \cite{H.Zhang2022Aug}, the joint optimization of device association, channel assignment, and multi-part collaborative offloading decisions was done to reduce the average task processing delay for an MEC system with OFDMA and sharing frequency bands under the constraint of the network operator's affordable cost. In \cite{T.Zhou2023Aug}, the joint optimization of device association, multi-step offloading decision, user power, security service assignment, and channel selection was performed, minimizing the energy consumed by all IMDs (IoT MDs) for ultra-dense multi-task IoT networks with both OFDMA and NOMA.
\par
When the computation tasks are offloaded to edge servers or further partially offloaded to other edge servers for computing, these offloaded tasks will result in extra wireless transmission time. In resource-strained ultra-dense MEC networks, such offloading manners will give rise to great pressure on wireless communication systems. To overcome this challenge, data compression (DC) technology has been advocated to reduce the transmission workload, especially in ultra-dense networks. In \cite{T.N.2020Jan}, the joint optimization of compression ratio, computation offloading decision, and resource assignment was done, minimizing the maximum weighted energy and service delay cost of all users in heterogeneous fog computing networks with OFDMA. In \cite{W.Bai2021Mar}, the joint optimization of computation offloading decision, compression ratio, energy harvesting, and application scenarios was considered, minimizing the overall cost of the users in multi-layer fog computing networks with OFDMA. In \cite{I.A.E.2020Dec}, to minimize the weighted sum energy for multi-task IoT networks with OFDMA under latency constraints, radio and computational resources were jointly addressed, compression algorithms of joint photographic experts group (JPEG) and moving picture experts group 4 (MPEG4) were used to reduce the transfer overhead, and a security layer was introduced to protect the transmitted data from cyber-attacks.
\par
If the computation tasks are offloaded to edge servers or further partially offloaded to other edge servers for computing, these offloaded tasks will be very vulnerable to malicious attacks and eavesdropping. To tackle this issue, some security measures have been introduced into the computation offloading. In \cite{S.S.Y.2022Dec}, helpers were used to assist cell-edge users in partially offloading tasks to edge servers in a single-cell MEC system with massive multiple-input multiple-output (MIMO) and NOMA under security rate constraints. In \cite{J.B.Wang2020Aug}, the joint optimization of offloading decision, local computing capacity (CC), offloading power, and offloading timeslots was performed, minimizing the total system EC for a single-cell MEC system with OFDMA under security rate constraints. In \cite{Y.Zhou2020Apr}, the joint optimization of user power, user association, and edge CC assignment was done, minimizing task processing delay for a single-cell MEC system with full frequency reuse under security rate constraints. Besides physical layer security techniques, other work concentrated on the encryption of offloaded data. In \cite{W.Z.Zhang2021May}, a joint load balancing and computation offloading approach was proposed to minimize the weighted sum of time and energy for a multi-task, multi-layer edge cloud system with OFDMA, and new security layers were introduced to circumvent potential security issues and safeguard the vulnerability of offloaded data. In \cite{IElgendy2019}, the joint optimization of security decision, resource assignment, and offloading decision was mentioned, minimizing the sum of weighted EC and delay for an MEC system with OFDMA. After introducing a new security layer in the cloud, the computation offloading was dynamically performed according to the EC, execution time, and memory and CPU usage in \cite{IAElgendy2021Jan}. In \cite{MZahed2020}, the joint optimization of cooperative task offloading and caching, and security service assignment was proposed, minimizing the overall cost for networks with OFDMA.
\par
Among the above-mentioned efforts, besides the work \cite{T.Zhou2023Aug}, other secure, multi-task, and/or multi-step offloading mechanisms nearly always concentrated on the OFDMA or/and full frequency reuse. However, such resource utilization often results in low frequency-spectrum efficiency, and it is unfavorable to massive connections. In addition, a great many of them focused on a single-cell framework, which doesn't suit practical applications. Besides the work \cite{I.A.E.2020Dec}, other secure, multi-task, and/or multi-step offloading mechanisms rarely concentrated on reducing offloading time through data compress technologies. In this paper, unlike the work \cite{I.A.E.2020Dec}, we will investigate the secure and multi-task offloading after introducing both OFDMA and NOMA to improve the frequency-spectrum efficiency and support massive connections, and consider the multi-step offloading to fully utilize computational resources in ultra-dense networks. In addition, unlike the work \cite{T.Zhou2023Aug}, we will further consider the optimization of the frequency band partitioning factor and the number of subchannels under the given number of subchannels, minimize the network-wide EC but not local EC to reduce the huge EC caused by the deployment of ultra-dense SBSs, and finally design a new optimization algorithm to the formulated problem.

\begin{table*}[htbp]
  \centering
  \caption{NOTATIONS}
  \begin{tabular}{llll}
    \toprule[1pt]
    \textbf{Notations} & \textbf{Definitions} & \textbf{Notations} & \textbf{Definitions}\\ \midrule[0.5pt]
    \rule{0pt}{8pt}  $\cal J$, $J$  & Index-set and number of BSs, respectively                &$\mathcal{Q}$, $Q$& Index-set and number of cryptographic algorithms, respectively\\
    \rule{0pt}{8pt}  $\cal I$, $I$  & Index-set and number of MDs, respectively               &$\mathcal{F}$, $\varpi $ & System frequency band and its bandwidth, respectively\\
    \rule{0pt}{8pt}  $\bar {\cal J}$, $\bar{J}$ & Index-set and number of SBSs, respectively   &$\mathcal{F}_1$, $\mathcal{F}_2$& Frequency bands used by MBS and SBSs, respectively\\
    \rule{0pt}{8pt}  $\cal K$, $K$ & Index-set and number of tasks, respectively               &$\mathcal{A}_{q}$, $v_q$ & Cryptographic algorithm $q$ and its security level, respectively\\
    \rule{0pt}{8pt}  ${\bar \gamma  _q}$& The CC of encrypting one-bit task using $\mathcal{A}_{q}$      &$\mathcal{T}_{i,k}$, $\mathcal{E}_{i,j}$& Task $k$ of MD $i$, and MD $i$ associated with BS $j$, respectively\\
    \rule{0pt}{8pt}  ${\hat \gamma  _q}$& The CC of decrypting one-bit task using $\mathcal{A}_{q}$      &${{d}_{i,k}}$, $\mathcal{B}_{i,j}$ &  Size of $\mathcal{T}_{i,k}$, and BS $j$ selected by MD $i$, respectively\\
    \rule{0pt}{8pt}  ${\hat d _{i,j,k}}$ & Size of offloaded part of task $k$ from $\mathcal{E}_{i,j}$ to MBS   &$\mu$, ${\sigma ^2}$ & Frequency band partitioning factor and noise power, respectively\\
    \rule{0pt}{8pt}  $L$ & The number of clusters consisting of SBSs                           &${p_i}$, ${p_i^{\max}}$& Transmission power and maximal power of MD $i$, respectively\\
    \rule{0pt}{8pt}  $T$, $V$& Numbers of iterations and solitary waves, respectively          &$R_{i,j,n}$&Uplink data rate from MD $i$ to BS $j$ on subchannel $n$\\
    \rule{0pt}{8pt}  ${{\tau}_{i,0,k}^{\rm{MBS}}}$& Processing time of task $k$ of $\mathcal{E}_{i,0}$   &$\cal M$, $M$ &Index-set and the number of individuals, respectively\\
    \rule{0pt}{8pt}  $\hbar _{i,j}$ &Channel gain between MD $i$ and BS $j$            &${\tau}_i$, ${\tau}_i^{\max }$& Task processing time and deadline of MD $i$, respectively \\
    \rule{0pt}{8pt}  $\cal N$, $N$ & Index-set and number of subchannels, respectively         &${\tilde \gamma _q}$& The EC of encrypting or decrypting one-bit task using $\mathcal{A}_{q}$\\
    \rule{0pt}{8pt}  ${x_{i,j}}$ & Association index between MD $i$ and BS ${j}$              &$\hat \lambda ^{\rm{BS}}$,$\bar \lambda ^{\rm{BS}}$& Decompression and compression coefficients of BSs, respectively\\
    \rule{0pt}{8pt}  $\lambda ^{\rm{LOC}}$& Decompression coefficient of MDs                       &$\mathscr{C}_{i,j,k}^{\rm{LOC}}$ &The CC for compressing task $k$ at $\mathcal{E}_{i,j}$\\

        \rule{0pt}{8pt}\multirow{2}{*}{${\tilde{\mathcal{I}}_{i,j,n}}$}
                & {Index-set of MDs interfering with communications} &
\multirow{2}{*}{$\mathscr{C}_{i,j,k}^{\rm{MBS}}$} & {The CC for decompressing task $k$ offloaded from MD $i$ or SBS $j$}\\
                & { between SBS $j$ and MD $i$ on the subchannel $n$} & &  { at MBS}\\

    \rule{0pt}{8pt}  ${\bar{R}}$ & Wired backhauling rate from SBS to MBS                      &$\hat{\mathscr{C}}_{i,j,k}^{\rm{SBS}}$ &The CC for decompressing task $k$ at $\mathcal{B}_{i,j}$\\
    \rule{0pt}{8pt}  ${\hat{v} _{i,k}}$ & Security risk coefficient of $\mathcal{T}_{i,k}$         &$\bar{\mathscr{C}}_{i,j,k}^{\rm{SBS}}$&The CC for compressing task $k$ at $\mathcal{B}_{i,j}$\\
    \rule{0pt}{8pt}  ${\bar{v} _{i,k}}$& Expected security level of $\mathcal{T}_{i,k}$           &${\hat f _{i,0,k}}$& The CC assigned to task $k$ of $\mathcal{E}_{i,0}$ by MBS\\
    \rule{0pt}{8pt}  ${{\varphi}_{i,k}}$& Security breach cost of $\mathcal{T}_{i,k}$          &${\hat f _{i,j,k}}$& The CC assigned to task ${k}$ of $\mathcal{E}_{i,j}$ by MBS\\
    \rule{0pt}{8pt}  ${\hat \xi _j}$& The EC of each CPU cycle BS $j$                          &${\bar z _{i,j,k}},$ & The DC ratio of task ${k}$ of $\mathcal{E}_{i,j}$\\
    \rule{0pt}{8pt}  ${b_{i,n}}$ & Association index between MD $i$ and subchannel ${n}$      &${\hat z _{i,j,k}}$& The DC ratio of $\mathcal{T}_{i,k}$ transmitted from SBS $j$ to MBS\\
    \rule{0pt}{8pt}  $\tilde \xi $ &Power consumption per second on wired line                 &${\mathscr{P}_{i,k,q}}$& Probability that $\mathcal{A}_{q}$ fails in protecting $\mathcal{T}_{i,k}$\\
    \rule{0pt}{8pt}  ${\mathscr{F}^{{\rm{ave}}}}$  & Average fitness value of the population       &${\eta _{i,k}}$& Financial loss incurred if $\mathcal{T}_{i,k}$ is unprotected\\
    \rule{0pt}{8pt}  ${\mathscr F}_{m,{\bar m}}$ & Minimal fitness value of waves $m$ and $\bar m$   &${{c}_{i,k}}$& The CC of calculating one bit of $\mathcal{T}_{i,k}$\\
    \rule{0pt}{8pt}  $u $, $\mathscr{D}$ & Breaking coefficient and diversity, respectively  &${\bar d _{i,j,k}}$& Size of offloaded part of task $k$ from MD $i$ to BS $j$\\
    \rule{0pt}{8pt}  $\mathscr{D}_1 $, $\mathscr{D}_2 $ & Diversity thresholds                 &$\zeta$ &Random number obeying standard normal distribution\\
    \rule{0pt}{8pt}  ${{\mathcal{Q}}_{j}}$& The cluster that SBS $j$ belongs to                 &${\mathscr F}_m$, $\tilde{\mathscr F}_m$ & Fitness values of wave $m$ and temporary wave $m$, respectively\\
    \rule{0pt}{8pt}  ${\bar{\mathscr{P}}_{m,\bar m }}$ & Crossover probability of waves $m$ and $\bar m$   &${\psi _i}$, ${\psi _i^{\max}}$& Security breach cost of MD $i$ and its maximum, respectively\\
    \rule{0pt}{8pt}  ${\hat{\mathscr{P}}_m}$& Mutation probability of wave $m$                 &${y_{i,k,q}}$ & Security index between $\mathcal{T}_{i,k}$ and $\mathcal{A}_{q}$ \\
    \rule{0pt}{8pt}  $\tilde{\mathscr{P}}$ & Diversity-guided mutation probability of population &${\tau}_{i,j,k}^{\rm{LOC}}$ & Local execution time of task ${k}$ of $\mathcal{E}_{i,j}$\\
    \rule{0pt}{8pt}  ${\bar f _{i,j,k}}$& The CC assigned to $\mathcal{T}_{i,k}$ by BS $j$    &${{\tau}_{i,j,k}^{\rm{SBS}}}$ & Processing time of task ${k}$ of $\mathcal{E}_{i,j}$\\
    \rule{0pt}{8pt}  {${\xi}$}& Linearly modeling coefficient of DC algorithms        &${h_m}$, $h^{\max}$& Height and maximal height of wave $m$, respectively\\
    \rule{0pt}{8pt}  ${\mathscr{F}^{{\rm{max}}}}$& Maximum fitness values of the population    &${\varepsilon}_{i,j,k}^{\rm{LOC}}$&Local EC of task ${k}$ of $\mathcal{E}_{i,j}$\\
    \rule{0pt}{8pt}  ${\mathscr{F}^{{\rm{min}}}}$& Minimum fitness values of the population    &${\varepsilon}_{i,j,k}^{\rm{SBS}}$ &Processing EC of task ${k}$ of $\mathcal{E}_{i,j}$\\
    \rule{0pt}{8pt}  $\varsigma $ & Effective switched capacitance of chip architecture   &${\varepsilon}_{i,0,k}^{\rm{MBS}}$ &Processing EC of task $k$ of $\mathcal{E}_{i,0}$\\
    \rule{0pt}{8pt}  $ \dot{x}_{m,i}$& Index of BS associated with MD $i$ in wave $m$     &$\dot{y}_{m,i}$& Index of cryptographic algorithm used by virtual MD $i$ in wave $m$\\
    \rule{0pt}{8pt}  $\dot{p}_{m,i}$& Transmission power of MD $i$ in wave $m$             &$\dot{d}_{m,i}$& Size of data offloaded from virtual MD $i$ to BS in wave $m$\\
    \rule{0pt}{8pt}  ${\omega}$& The bandwidth of a subchannel                             &$\dot{z}_{m,i}$ & The DC ratio of the task  generated from virtual MD $i$ in wave $m$\\
    \rule{0pt}{8pt}  $\dot{\mu}_{m}$&Frequency band partitioning factor in wave $m$              &$\ddot{z}_{m,i}$ & The DC ratio of the task generated from SBS in wave $m$\\
    \rule{0pt}{8pt}  $\dot{N}_{m}$&Number of subchannels in wave $m$                              &${{o}_{i}}$, ${{\bar{o}}_{i}}$& Indices of subchannel and SBS selected by MD $i$, respectively \\
    \rule{0pt}{8pt}  $\ddot{d}_{m,i}$ & Size of data offloaded from SBS to MBS in wave $m$   &${f_i}$, $f_j^{\rm{BS}}$ & The CCs of MD $i$ and BS $j$, respectively \\
   \rule{0pt}{8pt}  $\dot{b}_{m,i}$& Index of subchannel selected by MD $i$ in wave $m$     &${\bar z ^{\min }}$, ${\bar z ^{\max }}$& Lower and upper bounds of any ${{{\bar z }_{i,j,k}}}$, respectively\\
    \bottomrule[0.5pt]
  \end{tabular}
  \label{tab1}
\end{table*}

\subsection{Contributions and Organization}
In ultra-dense networks, we will design a secure, green, and highly effective offloading mechanism after introducing both OFDMA and NOMA. Specifically, the contributions and work of this paper can be summarized as follows.
\begin{enumerate}[1)]
\item \textit{Secure data-compressed multi-step computation offloading model is established for multi-task ultra-dense networks with both OFDMA and NOMA.} In multi-task ultra-dense networks, an MD can execute a part of each task locally, and offload the remaining part to an associated BS for computing after compressing and encrypting. After decrypting and decompressing the received part, an SBS can perform its certain portion locally, and it can also offload the remaining portion to a nearby macro base station (MBS) for computing. Before further offloading, the SBS needs to compresses and encrypts it. After decrypting and decompressing the part received from MD or SBS, an MBS can execute directly. For all we know, such a computation offloading model should be newly investigated. Moreover, unlike the fixed bandwidth used by MDs associated with SBSs in \cite{T.Zhou2023Aug}, the bandwidth assigned to MDs by SBSs is tightly dependent on the optimized number of subchannels and frequency band partitioning factor in multi-task ultra-dense networks with both OFDMA and NOMA.
\item \textit{Joint optimization of DC, secure multi-step computation offloading, and resource assignment, minimizing network-wide EC in multi-task ultra-dense networks with both OFDMA and NOMA.} After introducing OFDMA, NOMA and SBS clustering, in order to minimize the energy consumed by ultra-densely deployed SBSs while ensuring proportional assignment of computational resources and the constraints of processing delay and security breach cost, the joint optimization of channel selection, the number of subchannels, secure service assignment, multi-step computation offloading, device association, DC control, power control, and frequency band partitioning is done for minimizing network-wide energy consumption (EC). To the best of our knowledge, such an optimization should be a new work.
\item \textit{A new effective algorithm is designed to solve the formulated problem.} Given that the formulated problem is nonlinear and involves integral optimization parameters, a closed-form solution cannot be obtained through convex optimization theorems. Therefore, we propose an adaptive genetic water wave optimization (AGWWO) algorithm to solve it. Specifically, to improve the global search capability and the stability of computation in large-scale data, the propagation behavior of the traditional water wave optimization (WWO) algorithm is replaced with genetic operations including selection, diversity-guided mutation, adaptive crossover and mutation. As far as we know, such an improved algorithm should be a new one.
\item \textit{Analyses of convergence, computational complexity, parallel implementation and simulation.} To demonstrate the performance of the developed algorithm, we make some detailed analyses focusing on convergence and computational complexity. In addition, some skillful operations are advocated for reducing the computational complexity. Furthermore, some other analyses are made towards the parallel implementation to guide practical applications of the designed algorithm. Finally, we compare our algorithm with existing methods in simulations to assess its effectiveness, providing deeper insights accordingly.
\end{enumerate}
\par
The remainder of this paper is structured as follows. Section \ref{sec2} presents the system model, covering aspects such as the network, communication, compression, security, and computation models. Section \ref{sec3} formulates a problem of minimizing network-wide EC while ensuring the constraints of latency and security costs. Section \ref{sec4} designs AGWWO to solve the formulated problem. Section \ref{sec5} offers comprehensive analyses of the convergence, complexity, and potential for parallel execution of the developed algorithm. Section \ref{sec6} presents the simulation results and their corresponding analyses. Finally, Section \ref{sec7} concludes the study and offers additional insights for future exploration.
\par
A summary of key notations used throughout this paper is provided in TABLE I.
\section{System Model}\label{sec2}
This section provides a detailed overview of the network, communication, DC, security, and computation models.
\subsection{Network model}
\begin{figure}[t]
	\centering
	\centerline{\includegraphics[width=3.7in]{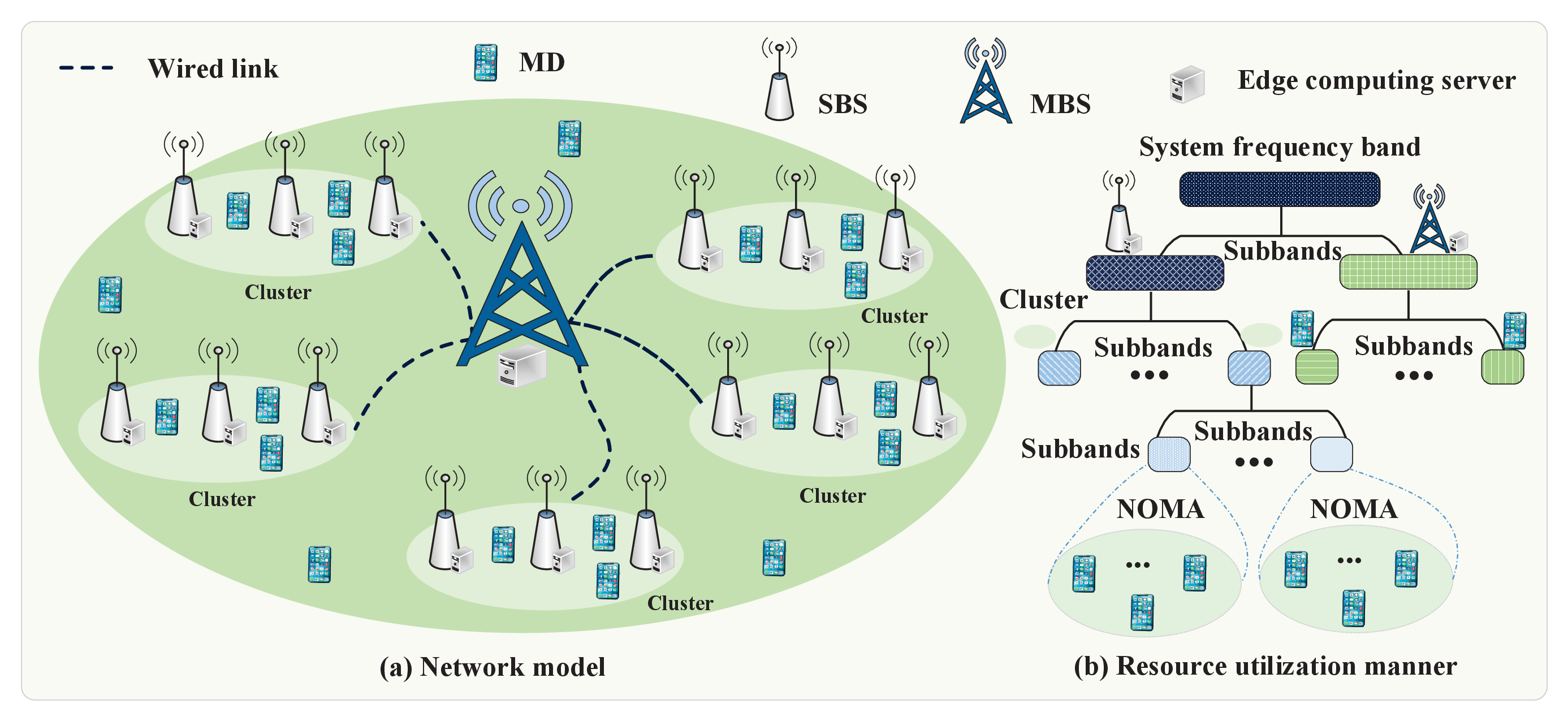}}
	\caption{Multi-task ultra-dense networks with both OFDMA and NOMA.}
	\label{fig1}
\end{figure}

The focus of this paper is on multi-user, multi-task ultra-dense networks, as depicted in Fig.\ref{fig1}(a). The networks consist of an MBS, densely deployed SBSs, and multiple MDs, where each BS is furnished with an MEC server, with all SBSs linked to the nearest MBS through a physical connection. Additionally, each MD must fulfill $K$ distinct computation-heavy and time-critical tasks within specified constraints on security breach costs and time. The MBS is indexed by 0, $\bar J$ SBSs are indexed in $\bar {\cal J}  = \left\{ {1,2, \cdots ,\bar J} \right\}$, $I$ MDs are indexed in ${\cal I} = \left\{ {1,2, \cdots ,I} \right\}$, and $K$ tasks are indexed in ${\cal K} = \left\{ {1,2, \cdots ,K} \right\}$. Now, the index-set of all BSs can be given by $ {\cal J} = \bar {\cal J}  \cup \left\{ 0 \right\}$.
\par
To mitigate interferences and improve spectrum efficiency, a resource utilization manner with both OFDMA and NOMA is utilized \cite{T.Zhou2023Aug}, which is illustrated in Fig.\ref{fig1}(b). After introducing the frequency band partitioning factor $\mu $ satisfying $0\leq\mu\leq1$, the system frequency band $\mathcal{F}$ is divided into ${\mathcal{F}_1}$ and ${\mathcal{F}_2}$ used by MBS and SBS separately. The widths of frequency bands $\mathcal{F}$, ${\mathcal{F}_1}$, and ${\mathcal{F}_2}$ are $\varpi $, $\mu \varpi $, and $\left( {1 - \mu } \right)\varpi $, respectively. According to the physical locations of SBSs, they are divided into $L$ clusters using the K-means approach. Then, the subband ${\mathcal{F}_2}$ is further divided into $L$ subbands used by different clusters separately. The subband occupied by each cluster is divided into $N$ subchannels (subbands) indexed in $\mathcal{N}=\{1,2,\cdots ,N\}$, where the bandwidth of each subchannel is $\omega={(1-\mu ){\varpi}}/{\left( LN \right)}$. The same subchannel can be shared by MDs associated with SBSs through a NOMA manner, and the frequency band ${\mathcal{F}_1}$ of MBS is equally assigned to its associated MDs.
\par
\begin{figure}[t]
	\centering
	\centerline{\includegraphics[width=3.7in]{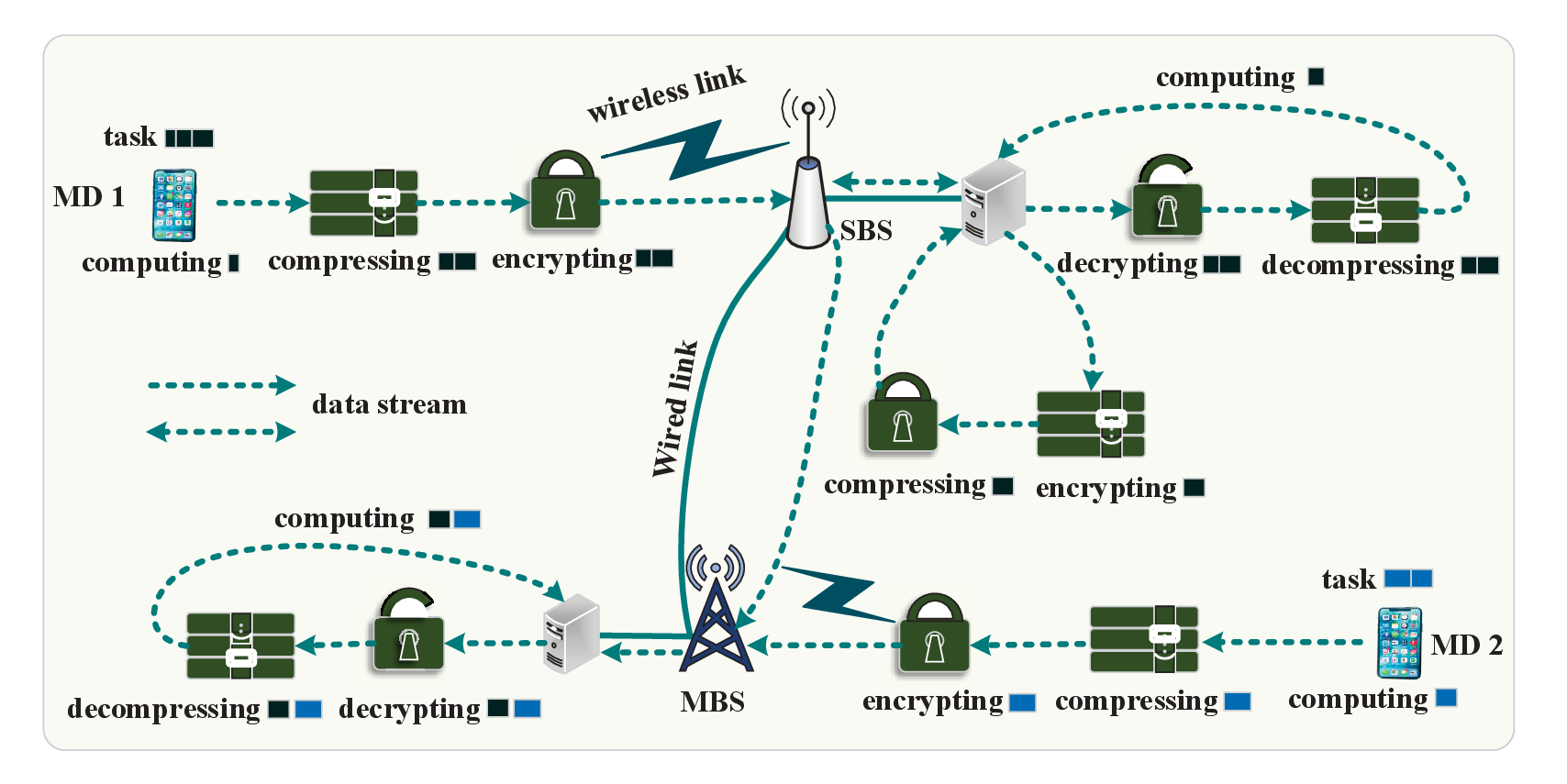}}
	\caption{The multi-step offloading procedure.}
	\label{fig2}
\end{figure}

To fully utilize the computational resources, alleviate network congestion, and protect offloaded data, a secure data-compressed multi-step computation offloading model is established for multi-task ultra-dense networks, which can be found in Fig.\ref{fig2}. As shown in Fig.\ref{fig2}, an MD associated with some SBS (e.g., MD 1) considers a two-step offloading manner. Specifically, a task of MD 1 is cut into three parts. One part is executed by MD 1 locally, other two parts are offloaded to its associated SBS after compressing and encrypting via a wireless link. After achieving these two parts through decrypting and decompressing, this SBS executes one part, and further offloads another part to its nearby MBS via a wired link after compressing and encrypting. After receiving this part, MBS decrypts, decompresses and executes it sequentially. In addition, an MD associated with some MBS (e.g., MD 2) considers a one-step offloading manner. Specifically, a task of MD 2 is cut into two parts. One part is executed by MD 2 locally, another part is offloaded to its associated MBS after compressing and encrypting via a wireless link. After decrypting and decompressing the part received from MD 2, this MBS executes it.


\subsection{Communication Model}
In the above-mentioned resource utilization manner, only the intra-cluster interferences exist for any MD associated with some SBS. According to the rule of uplink NOMA, such an MD will receive interferences from other MDs that have worse channel gains than it on the same subchannel. After receiving signals, SBSs decode them in the decreasing order of channel gains. Consequently, when MD $i$ is served by (associated with) SBS $s\in \bar{\mathcal{J}}$ on the subchannel $n$, its uplink data rate can be given by
\begin{equation}\label{eq1}
\left\{ \begin{aligned}
  & {{R}_{i,j,n}}={\omega}{{\log }_{2}}\Big( 1+\frac{{{p}_{i}}{{\hbar}_{i,j}}}{\sum\nolimits_{u\in {\tilde{\mathcal{I}}_{i,j,n}}}{{{p}_{u}}{{\hbar}_{u,s}}}+{{\sigma }^{2}}} \Big), \\
 & {\tilde{\mathcal{I}}_{i,j,n}}=\left\{ u\in \mathcal{I}\right\}\backslash \left\{ u=i \right\}: \\
 &\ \ \ \ \ \  \ \ {{\hbar}_{u,s}}\le {{\hbar}_{i,j}},\ \ \ {{o}_{u}}={{o}_{i}}=n,\ \ \ {{\bar{o}}_{u}},{{\bar{o}}_{i}}\in {{\mathcal{Q}}_{j}}, \\
\end{aligned} \right.
\end{equation}
where ${{o}_{i}}$ and ${{\bar{o}}_{i}}$ denote the indices of channel and SBS selected by MD $i$, respectively; $p_i$ is the transmission power of MD $i$; ${\bf{p}}=\left\{ {{p_i},\forall i \in {{\cal I}}} \right\}$; ${{\hbar}_{i,j}}$ denotes the channel gain between MD $i$ and BS $j$; $\sigma^2$ represents the noise power; ${{\mathcal{Q}}_{j}}$ is the cluster that SBS $j$ belongs to; {${\tilde{\mathcal{I}}_{i,j,n}}$ is the index-set of MDs interfering with communications between SBS $j$ and MD $i$ on the subchannel $n$. In \eqref{eq1}, when ${{o}_{u}}={{o}_{i}}=n$, ${{\bar{o}}_{u}}\ne {{\bar{o}}_{i}}$ and ${{\bar{o}}_{u}},{{\bar{o}}_{i}}\in {{\mathcal{Q}}_{j}}$, MD $i$ receives the interference from MD $u$ associated with a different BS in the same cluster on an identical channel. When ${{o}_{u}}={{o}_{i}}=n$, ${{\bar{o}}_{u}}={{\bar{o}}_{i}}$ and ${{\bar{o}}_{u}},{{\bar{o}}_{i}}\in {{\mathcal{Q}}_{j}}$, MD $i$ receives the interference from MD $u$ associated with the same BS on an identical channel.
\par
Since the frequency band of MBS is equally assigned to its associated MDs, there don't exist cross-tier and intra-tier interferences. To achieve the uniform expression and be favourable for the design of algorithms, it's assumed that there exist $N$ virtual subchannels indexed in $\mathcal{N}$ at each MBS, but only one channel in reality. Then, any MD associated with this MBS can transmit a task on any one of these subchannels, but does it on a real channel. Consequently, when MD $i$ is served by (associated with) MBS on subchannel $n$, its uplink data rate can be given by
\begin{equation}\label{eq2}
    {{R}_{i,0,n}}=\mu {{\varpi}}{{\big( \sum\nolimits_{u\in \mathcal{I}}{{{x}_{u,0}}} \big)}^{-1}}{{\log}_{2}}\left( 1+{{{p}_{i}}{{\hbar}_{i,0}}}/{{{\sigma }^{2}}} \right),
\end{equation}
where $\sum\nolimits_{u\in \mathcal{I}}{{{x}_{u,0}}}$ is the number of MDs associated with MBS; ${{x}_{i,j}}$ is an association index between MD $i$ and BS $j$; ${{x}_{i,j}}=1$ if MD $i$ is served by (associated with) the BS $j$; otherwise, ${{x}_{i,j}}=0$; ${\bf{X}}=\left\{ {{x_{i,j}},\forall i \in {{\cal I}},\forall j \in {{\cal J}}} \right\}$.
\subsection{DC model}
We assume that $\mathcal{T}_{i,k}$ is denoted as ${{{\cal D}}_{i,k}} \buildrel \Delta \over = \left( {{d_{i,k}},{c_{i,k}},{\tau}_i^{\max },{\bar{v} _{i,k}}} \right)$, where ${d_{i,k}}$ is the size of $\mathcal{T}_{i,k}$; ${c_{i,k}}$ represents the number of CPU cycles used for calculating one bit of ${{{\cal D}}_{i,k}}$; ${\tau}_i^{\max }$ is the execution deadline of MD $i$; ${\bar{v} _{i,k}}$ is the expected security level of $\mathcal{T}_{i,k}$.
\par
As revealed in the previous section, a secure data-compressed multi-step computation offloading model is established for multi-task ultra-dense networks. For any task $k$, when MD $i$ is associated with MBS, the part with size ${d_{i,k}} - {\bar d _{i,0,k}}$ is calculated locally, and the one with size ${\bar d _{i,0,k}}$ is transmitted from MD $i$ to MBS after successively compressing and encrypting. After receiving the compressed and encrypted data, MBS calculates the part with size ${\bar d _{i,0,k}}$ after successively decrypting and decompressing. When MD $i$ is associated with SBS $j$, the part with size ${d_{i,k}} - {\bar d _{i,j,k}}$ is calculated locally, and the one with size ${\bar d _{i,j,k}}$ is transmitted from MD $i$ to MBS after successively compressing and encrypting. After receiving the compressed and encrypted data, SBS calculates the part with size ${\bar d _{i,j,k}} - {\hat d_{i,j,k}}$ after successively decrypting and decompressing. Then, the part with size ${\hat d_{i,j,k}}$ is transmitted to MBS after successively compressing and encrypting, and MBS calculates this part after successively decrypting and decompressing. Significantly, $\bar {\bf{D}}=\{ {{{\bar d }_{i,j,k}},}{\forall i \in {\cal I},\forall j \in {{\cal J}},\forall k \in {\cal K}} \}$, and $\hat {\bf{D}}=\{ {{{\hat d}_{i,j,k}},\forall i \in {\cal I},\forall j \in {{\cal J}},}{\forall k \in }   {\cal K} \}$.
\par
According to the rules in \cite{T.N.2020Jan}, when the part with size ${\bar d _{i,j,k}}$ (a part of $\mathcal{T}_{i,k}$) is offloaded to BS $j$, the CC used for compressing this part at MD $i$ can be given by
\begin{equation}\label{eq3}
	\mathscr{C}_{i,j,k}^{\rm{LOC}} = {\xi} {\bar d _{i,j,k}}\left[ {\lambda _1^{\rm{LOC}}{{\left( {{{\bar z }_{i,j,k}}} \right)}^{\lambda _2^{\rm{LOC}}}} + \lambda _3^{\rm{LOC}}} \right],
\end{equation}
where ${\xi}$ is a constant coefficient; ${\bar z _{i,j,k}}$ is the compression ratio of the offloaded part of task $k$ from MD $i$ to BS $j$, which is defined as the ratio of raw data to compressed data; $\bar {\bf{Z}}=\left\{ {{\bar z }_{i,j,k}},\forall i \in {{\cal I}}, \forall j \in {\cal J}, \forall k \in {{\cal K}} \right\}$; $\lambda _1^{\rm{LOC}}$, $\lambda _2^{\rm{LOC}}$ and $\lambda _3^{\rm{LOC}}$ are constant compression coefficients respectively corresponding to the compression algorithms GNU zip (GZIP), BWT (Burrows-Wheeler transform) zip (BZ2), and JPEG at MDs.
\par
When the part with size ${\bar d _{i,j,k}}$ (a part of $\mathcal{T}_{i,k}$) is offloaded to MBS $j=0$ after compressing, the CC used for decompressing this part at MBS $j$ can be given by
\begin{equation}\label{eq4}
\mathscr{C}_{i,j,k}^{\rm{MBS}} = {\xi} {\bar d _{i,j,k}}\left[ {\hat \lambda _1^{\rm{BS}}{{\left( {{{\bar z}_{i,j,k}}} \right)}^{\hat \lambda _2^{\rm{BS}}}} + \hat \lambda _3^{\rm{BS}}} \right], \quad j = 0,
\end{equation}
where $\hat \lambda _1^{\rm{BS}}$, $\hat \lambda _2^{\rm{BS}}$ and $\hat \lambda _3^{\rm{BS}}$ are constant decompression coefficients respectively corresponding to the compression algorithms GZIP, BZ2, and JPEG at MBSs.
\par
When the part with size ${\hat d_{i,j,k}}$ (a part of ${\bar d _{i,j,k}}$) is offloaded from SBS $j$ to MBS after compressing, the CC used for decompressing this part at MBS $j$ can be given by
\begin{equation}\label{eq5}
\mathscr{C}_{i,j,k}^{\rm{MBS}} ={\xi} {\hat d_{i,j,k}}\left[ {\hat \lambda _1^{\rm{BS}}{{\left( {{{\hat z}_{i,j,k}}} \right)}^{\hat \lambda _2^{\rm{BS}}}} + \hat \lambda _3^{\rm{BS}}} \right], \quad j \in {\bar {\cal J}}.
\end{equation}
where ${\hat z_{i,j,k}}$ is the compression ratio of the offloaded part of task $k$ from SBS $j$ to MBS; $\hat {\bf{Z}}=\left\{ {{{\hat z}_{i,j,k}},\forall i \in {{\cal I}},\forall j \in {{\cal J}},\forall k \in {{\cal K}}} \right\}$.
\par
When the part with size ${\bar d _{i,j,k}}$ (a part of $\mathcal{T}_{i,k}$) is offloaded to SBS $j$ after compressing, the CC used for decompressing this part at SBS $j$ can be given by
\begin{equation}\label{eq6}
\hat {\mathscr{C}}_{i,j,k}^{\rm{SBS}} = {\xi} {\bar d _{i,j,k}}\left[ {\hat \lambda _1^{\rm{BS}}{{\left( {{{\bar z }_{i,j,k}}} \right)}^{\hat \lambda _2^{\rm{BS}}}} + \hat \lambda _3^{\rm{BS}}} \right],
\end{equation}
\par
When the part with size ${\bar d _{i,j,k}}$ (a part of $\mathcal{T}_{i,k}$) is offloaded to SBS $j$ after compressing, this SBS first decompresses it. Then, the part with size ${\hat d_{i,j,k}}$ (a part of the decompressed data) needs to be further offloaded to MBSs for computing. At this time, the CC used for compressing this part at SBS $j$ can be given by
\begin{equation}\label{eq7}
	\bar{\mathscr{C}} _{i,j,k}^{\rm{SBS}} = {\xi} {\hat d_{i,j,k}}\left[ {\bar \lambda _1^{\rm{BS}}{{\left( {{{\hat z}_{i,j,k}}} \right)}^{\bar \lambda _2^{\rm{BS}}}} + \bar \lambda _3^{\rm{BS}}} \right].
\end{equation}
where $\bar \lambda _1^{\rm{BS}}$, $\bar \lambda _2^{\rm{BS}}$ and $\bar \lambda _3^{\rm{BS}}$ are constant compression coefficients respectively corresponding to the compression algorithms GZIP, BZ2, and JPEG at SBSs.

\subsection{Security Model}
Similar to \cite{T.Zhou2023Aug}, $Q$ types of cryptographic algorithms indexed from 1 to $Q$ in the set ${\cal Q} = \left\{ {{\rm{1,2,}} \cdots ,Q} \right\}$ are introduced for protecting the offloaded data. These algorithms have different security levels, and the protection level (robustness) of the $\mathcal{A}_{q}$ is ${v_q}$. Although encrypting offloaded data can achieve secure communications in a certain probability, it will cause extra processing time and EC. The computation capacities that $\mathcal{A}_{q}$ encrypts and decrypts one-bit data of are ${\bar \gamma _q}$ (in CPU cycles/bit) and ${\hat \gamma _q}$ (in CPU cycles/bit), respectively. Here, the sets of them are $\boldsymbol{\bar \gamma}   = \{ {{{\bar \gamma  }_q},\forall q \in {{\cal Q}}} \}$ and $\boldsymbol{\hat \gamma}  =  \{{{\hat \gamma }_q},\forall q \in {{\cal Q}} \} $, respectively. In addition, we assume that the energy consumed by encrypting and decrypting one-bit data is the same in $\mathcal{A}_{q}$, denoted as ${\tilde \gamma _q}$ (in mJ/bit) in $\boldsymbol{\tilde \gamma}  = \{ {{{\tilde \gamma }_q},\forall q \in {{\cal Q}}} \}$.
\par
As defined in \cite{W.Jiang2015Aug}, the failure probability of $\mathcal{T}_{i,k}$, when employing $\mathcal{A}_{q}$, can be determined using
\begin{equation}\label{eq8}
	{{ \mathscr{P}}_{i,k,q}} = \left\{
	\begin{split}
		&1 - {e^{ - {\hat{v} _{i,k}}\left( {{\bar{v} _{i,k}} - {v_q}} \right)}} ,&{\rm{ if }}\,{v_q} < {\bar{v} _{i,k}}&,\\
		&0 ,&{\,\, \rm{otherwise}}&,  \\
	\end{split} \right.
\end{equation}
where ${\hat{v} _{i,k}}$ represents the security risk coefficient associated with $\mathcal{T}_{i,k}$. As shown in \eqref{eq8}, the $\mathcal{A}_{q}$ is considered successful in safeguarding $\mathcal{T}_{i,k}$ if its security level is equal to or greater than the expected level. However, if the security level falls below the expected level, the algorithm is deemed to have failed with a certain probability.
\par
According to the definitions in \cite{MZahed2020,T.Zhou2023Aug}, the security breach cost of $\mathcal{T}_{i,k}$ can be given by
\begin{equation}\label{eq9}
{\varphi _{i,k}} = \sum\nolimits_{j \in {{\cal J}}} {\sum\nolimits_{q \in {{\cal Q}}} {{\eta _{i,k}}{x_{i,j}}{y_{i,k,q}}{{ \mathscr{P}}_{i,k,q}}} } ,
\end{equation}
where ${\eta _{i,k}}$ represents the financial loss incurred if $\mathcal{T}_{i,k}$ is unprotected; ${y_{i,k,q}}$ signifies the security decision index for $\mathcal{T}_{i,k}$. In the event that $\mathcal{A}_{q}$ is applied to $\mathcal{T}_{i,k}$, ${y_{i,k,q}} = 1$; otherwise, it is 0; ${\bf{Y}}=\left\{ {{y_{i,k,q}},} \forall i  {\in {{\cal I}},\forall k \in {{\cal K}},\forall k \in {{\cal Q}}} \right\}$. The security breach cost of MD $i$ can be given by
\begin{equation}\label{eq10}
	\begin{split}
		{\psi _i} = \sum\limits_{k \in {{\cal K}}} {{\varphi _{i,k}}}
		= \sum\limits_{k \in {{\cal K}}} {\sum\limits_{j \in {{\cal J}}} {\sum\limits_{q \in {{\cal Q}}} {{\eta _{i,k}}{x_{i,j}}{y_{i,k,q}}{{ \mathscr{P}}_{i,k,q}}} } } .\\
	\end{split}
\end{equation}

\subsection{Computation Model}
A multi-step computation offloading model involves local computation, offloading to SBSs, directly and indirectly offloading to MBSs.

\subsubsection{Local computation}
For any task $k$, if MD $i$ is associated with BS $j$, the size of the data offloaded from MD $i$ to BS $j$ is ${\bar d _{i,j,k}}$. This MD needs to successively compress and encrypt the offloaded part, and calculate the remaining part with size ${d_{i,k}} - {\bar d _{i,j,k}}$. The local execution time ${\tau}_{i,j,k}^{\rm{LOC}}$ used for processing $\mathcal{T}_{i,k}$ can be given by
\begin{equation}\label{eq11}
\begin{split}
{\tau}_{i,j,k}^{\rm{LOC}} = &{{\left( {{d_{i,k}} - {{\bar d }_{i,j,k}}} \right){c_{i,k}}}}/{{{f_i}}} + {{\mathscr{C}_{i,j,k}^{\rm{LOC}}}}/{{{f_i}}}\\
&+\sum\nolimits_{q \in {{\cal Q}}} {{{{y_{i,k,q}}{{\bar \gamma  }_q}{{\bar d }_{i,j,k}}}}/({{{{\bar z }_{i,j,k}}{f_i}}})}.\\
\end{split}
\end{equation}
where ${f_i}$ is the CC of MD $i$; three items on the right of the equal sign are the computing time, encrypting time, and compressing time, respectively. Then, the local EC ${\varepsilon}_{i,j,k}^{\rm{LOC}}$ used for processing $\mathcal{T}_{i,k}$ can be given by
\begin{equation}\label{eq12}
\begin{split}
{\varepsilon}_{i,j,k}^{\rm{LOC}} =& \varsigma \left( {{d_{i,k}} - {{\bar d }_{i,j,k}}} \right){c_{i,k}}f_i^2 + \varsigma \mathscr{C}_{i,j,k}^{\rm{LOC}}f_i^2\\
	 &+ \sum\limits_{n \in {{\cal N}}}\frac{{{{b}_{i,n}}}{{p_i}{{\bar d }_{i,j,k}}}}{{{{\bar z }_{i,j,k}}{R_{i,j,n}}}} + \sum\limits_{q \in {{\cal Q}}} \frac{{y_{i,k,q}}{{\tilde \gamma }_q}{{\bar d }_{i,j,k}}}{{{\bar z }_{i,j,k}}} ,\\
\end{split}
\end{equation}
where ${{b}_{i,n}}$ denotes the association status between MD $i$ and subchannel $n$; specifically, ${{b}_{i,n}}$ equals 1 when MD $i$ opts for subchannel $n$, and 0 otherwise; ${\bf{B}} = \left\{ {{b_{i,n}},\forall i \in {{\cal I}},\forall n \in {{\cal N}}} \right\}$; $\varsigma $ is the effective switched capacitance of chip architecture; four items on the right of the equal sign are the local computing EC, compressing EC, wireless transmitting EC, and encrypting EC, respectively.

\subsubsection{Offloading to SBS}
For any task $k$, if MD $i$ is associated with SBS $j$, it successively compresses, encrypts and transmits the part with size ${\bar d _{i,j,k}}$ to SBS $j$, and calculate the remaining part with size ${d_{i,k}} - {\bar d _{i,j,k}}$. After successively decrypting and decompressing the received data, SBS $j$ successively compresses, encrypts and transmits the part with size ${\hat d _{i,j,k}}$ to MBS, and calculates the remaining part with size $ {\bar d _{i,j,k}}-{\hat d _{i,j,k}}$. After successively decrypting and decompressing the received data, MBS calculates the part with size ${\hat d _{i,j,k}}$. Consequently, when MD $i$ is associated with SBS $j$, the time ${{\tau}_{i,j,k}^{\rm{SBS}}}$ used for processing the $\mathcal{T}_{i,k}$ can be given by
\begin{equation}\label{eq13}
\begin{split}
	&{{\tau}_{i,j,k}^{\rm{SBS}}} =
{\sum\limits_{n \in {{\cal N}}}\frac{{{{b}_{i,n}}}{{{\bar d }_{i,j,k}}}}{{{{\bar z }_{i,j,k}}{R_{i,j,n}}}} + \frac{{{{\hat d}_{i,j,k}}}}{{{{\hat z}_{i,j,k}}{\bar{R}}}} + \frac{{\hat{\mathscr{C}}_{i,j,k}^{\rm{SBS}}}}{{{{\bar f }_{i,j,k}}}} + \frac{{\bar{\mathscr{C}} _{i,j,k}^{\rm{SBS}}}}{{{{\bar f }_{i,j,k}}}}}\\
	&\  { + \frac{{\mathscr{C}_{i,j,k}^{\rm{MBS}}}}{{{{\hat f }_{i,j,k}}}} + } \sum\limits_{q \in {{\cal Q}}} {\frac{{{y_{i,k,q}}{{\hat \gamma }_q}{{\bar d }_{i,j,k}}}}{{{{\bar z }_{i,j,k}}{{\bar f }_{i,j,k}}}}}  + \sum\limits_{q \in {{\cal Q}}} \frac{{{y_{i,k,q}}{{\bar \gamma  }_q}{{\hat d}_{i,j,k}}}}{{{{\hat z}_{i,j,k}}{{\bar f }_{i,j,k}}}} +\\
	&\ \sum\limits_{q \in {{\cal Q}}} {\frac{{{y_{i,k,q}}{{\hat \gamma }_q}{{\hat d}_{i,j,k}}}}{{{{\hat z}_{i,j,k}}{{\hat f }_{i,j,k}}}}} + \frac{{( {{{\bar d }_{i,j,k}} - {{\hat d}_{i,j,k}}} ){c_{i,k}}}}{{{{\bar f }_{i,j,k}}}} + \frac{{{{\hat d}_{i,j,k}}{c_{i,k}}}}{{{{\hat f }_{i,j,k}}}},\\		
\end{split}
\end{equation}
where ${\bar{R}}$ represents the rate of data transmission through the wired backhaul connection from SBSs to MBSs; ${\bar f _{i,j,k}}$ signifies the CC assigned by SBS $j$ to $\mathcal{T}_{i,k}$, ${\hat f _{i,j,k}}$ is the one assigned to task ${k}$ of $\mathcal{E}_{i,j}$ by MBS. On the right of the equal sign in equation \eqref{eq13}, the first to second items are the transmission time from MD $i$ to SBS $j$ and from SBS $j$ to MBS, respectively; the third to fifth items are the decompressing time at SBS $j$, compressing time at SBS $j$, and decompressing time at MBS, respectively; the sixth to eighth items are the decrypting time at SBS $j$, encrypting time at SBS $j$, and decrypting time at MBS, respectively; the last two items are the computing time of tasks at SBS $j$ and MBS, respectively.
\par
Under the proportional resource assignment manner, the CC (CPU cycles) ${\bar f _{i,j,k}}$ assigned to $\mathcal{T}_{i,k}$ by SBS $j \in {\bar {\cal J}}$ can be given by
\begin{equation}\label{eq14}
{\bar f _{i,j,k}} ={ {\mathscr{A}_{i,j,k}^{\rm{SBS}}}f_j^{\rm{BS}}}/{\sum\nolimits_{u \in {{\cal I}}} {\sum\nolimits_{{n} \in {{\cal K}}}} {{x_{u,j}}{\mathscr{A}_{u,j,n}^{\rm{SBS}}}}},
\end{equation}
where $f_j^{\rm{BS}}$ is the total CC of SBS $j$;
\begin{equation}\label{eq15}
\mathscr{A}_{i,j,k}^{\rm{SBS}} = {\mathscr{B}_{i,j,k}^{\rm{SBS}}}+\hat{\mathscr{C}}_{i,j,k}^{\rm{SBS}} + \bar{\mathscr{C}} _{i,j,k}^{\rm{SBS}} + {\hat{\mathcal{C}}_{i,j,k}^{\rm{SBS}}} + {\bar{\mathcal{C}}_{i,j,k}^{\rm{SBS}}},
\end{equation}
${\mathscr{B}_{i,j,k}^{\rm{SBS}}}=( {{{\bar d }_{i,j,k}} - {{\hat d}_{i,j,k}}}){c_{i,k}}$ is the CPU cycles used for calculation at SBS $j$; $\hat{\mathcal{C}}_{i,j,k}^{\rm{SBS}} = \sum\nolimits_{q \in {{\cal Q}}} {{y_{i,k,q}}{{\hat \gamma }_q}{{\bar d }_{i,j,k}}/{{\bar z }_{i,j,k}}} $ is the CPU cycles used for decryption at SBS $j$; $\bar{\mathcal{C}}_{i,j,k}^{\rm{SBS}} = \sum\nolimits_{q \in {{\cal Q}}} {{y_{i,k,q}}{{\bar \gamma  }_q}{{\hat d}_{i,j,k}}/{{\hat z}_{i,j,k}}} $ is the CPU cycles used for encryption at SBS $j$.
\par
Since both MDs and SBSs can offload tasks to MBS for computing, MBS needs to tackle the data generated from the following two offloading models. In the one-step offloading, its CC needs to be assigned to the decryption, decompression and computation of tasks offloaded from MDs. In addition, its CC also needs to be assigned to the same operations of tasks offloaded from SBSs in the two-step offloading. Under the proportional resource assignment manner, the CC ${\hat f _{i,j,k}}$ assigned to the task ${k}$ of $\mathcal{E}_{i,j}$ by MBS can be given by
\begin{equation}\label{eq16}
	{\hat f _{i,j,k}} =\frac{{\mathscr{A}_{i,j,k}^{\rm{MBS}}}{f_0^{\rm{BS}}}}{\sum\limits_{u \in {{\cal I}}} {\sum\limits_{n \in {{\cal K}}} {({x_{u,0}}\mathscr{A}_{u,0,n}^{\rm{MBS}}+\sum\nolimits_{s \in {{\bar{\cal J}}}}{x_{u,s}}\mathscr{A}_{u,s,n}^{\rm{MBS}}) } }},
\end{equation}
where $f_0^{\rm{BS}}$ represents the total CC of MBS. The CPU cycles $\mathscr{A}_{i,0,k}^{\rm{MBS}}$ used for processing the part of task $k$ offloaded from MD $i$ at MBS can be given by
\begin{equation}\label{eq17}
\mathscr{A}_{i,0,k}^{\rm{MBS}} ={\mathscr{B}_{i,0,k}^{\rm{MBS}}} + {\hat{\mathcal{C}}_{i,0,k}^{\rm{MBS}}} +{\mathscr{C}_{i,0,k}^{\rm{MBS}}},
\end{equation}
where ${\mathscr{B}_{i,0,k}^{\rm{MBS}}}={{{\bar d }_{i,0,k}}{c_{i,k}}}$ is the CPU cycles used for calculating the part with size ${{\bar d }_{i,0,k}}$ offloaded from MD $i$ at MBS $0$, and  $\hat{\mathcal{C}}_{i,0,k}^{\rm{MBS}} = \sum\nolimits_{q \in {{\cal Q}}} {{y_{i,k,q}}{{\hat \gamma }_q}{{\bar d }_{i,0,k}}}/{{{\bar z }_{i,0,k}}} $ is the CPU cycles used for decrypting this part at MBS. The CPU cycles $\mathscr{A}_{i,j,k}^{\rm{MBS}}$ used for processing the part of task $k$ offloaded from SBS $j$ at MBS can be given by
\begin{equation}\label{eq18}
\mathscr{A}_{i,j,k}^{\rm{MBS}} = {\mathscr{B}_{i,j,k}^{\rm{MBS}}} + {\hat{\mathcal{C}}_{i,j,k}^{\rm{MBS}}} +\mathscr{C}_{i,j,k}^{\rm{MBS}},
\end{equation}
where ${\mathscr{B}_{i,j,k}^{\rm{MBS}}}={{{\hat d}_{i,j,k}}{c_{i,k}}}$ is the CPU cycles used for calculating the part with size ${{\hat d}_{i,j,k}}$ offloaded from SBS $j$ at MBS $0$, and  $\hat{\mathcal{C}}_{i,j,k}^{\rm{MBS}} = \sum\nolimits_{q \in {{\cal Q}}} {{y_{i,k,q}}{{\hat \gamma }_q}{{\hat d}_{i,j,k}}}/{{{\hat z}_{i,j,k}}}$ is the CPU cycles used for decrypting this part at MBS.
\par
Until now, when MD $i$ is associated with SBS $j$, the EC ${{\varepsilon}_{i,j,k}^{\rm{SBS}}}$ used for processing the $\mathcal{T}_{i,k}$ at SBS $j$ and its connected MBS can be given by
\begin{equation}\label{eq19}
\begin{split}
	&{\varepsilon}_{i,j,k}^{\rm{SBS}} =\frac{{\tilde \xi {{\hat d}_{i,j,k}}}}{{{{\hat z}_{i,j,k}}{\bar{R}}}}+{{\hat \xi }_j}\hat{\mathscr{C}}_{i,j,k}^{\rm{SBS}} + {{\hat \xi }_j}\bar{\mathscr{C}} _{i,j,k}^{\rm{SBS}} +{{\hat \xi }_0}\mathscr{C}_{i,j,k}^{\rm{MBS}}+  \\
	&{\sum\limits_{q \in {{\cal Q}}} \frac{{y_{i,k,q}}{{\tilde \gamma }_q}{{\bar d }_{i,j,k}}}{{{\bar z }_{i,j,k}}}  + \sum\limits_{q \in {{\cal Q}}} \frac{{y_{i,k,q}}{{\tilde \gamma }_q}{{\hat d}_{i,j,k}}}{{{\hat z}_{i,j,k}}}  + \sum\limits_{q \in {{\cal Q}}} \frac{{y_{i,k,q}}{{\tilde \gamma }_q}{{\hat d}_{i,j,k}}}{{{\hat z}_{i,j,k}}} }\\
	&{ + {{\hat \xi }_j}( {{{\bar d }_{i,j,k}} - {{\hat d}_{i,j,k}}} ){c_{i,k}} + {{\hat \xi }_0}{{\hat d}_{i,j,k}}{c_{i,k}},}\\
\end{split}
\end{equation}
where $\tilde \xi $ denotes the power consumption per second on wired line; ${\hat \xi _j}$ and ${\hat \xi _0}$ are the EC of each CPU cycle at SBS and MBS, respectively. On the right of the equal sign in the equation \eqref{eq19}, the first item is the uplink transmission EC from SBS $j$ to MBS; the second to fourth items are the decompressing EC at SBS $j$, compressing EC at SBS $j$ and decompressing EC at MBS, separately; the fifth to seventh items are decrypting EC at SBS $j$, encrypting EC at SBS $j$, and decrypting EC at MBS, respectively; the last two items are computing EC of $\mathcal{T}_{i,k}$ at SBS $j$ and MBS, respectively.
\subsubsection{Offloading to MBS}
For any task $k$, if MD $i$ is associated with MBS, it successively compresses, encrypts and transmits the part with size ${\bar d _{i,0,k}}$ to MBS, and calculates the remaining part with size ${d_{i,k}} - {\bar d _{i,0,k}}$. After successively decrypting and decompressing the received data, MBS calculates the part with size ${\bar d _{i,0,k}}$. Consequently, when MD $i$ is associated with MBS, the time ${{\tau}_{i,0,k}^{\rm{MBS}}}$ used for processing the $\mathcal{T}_{i,k}$ can be given by
\begin{equation}\label{eq20}
	\begin{split}
	&{\tau}_{i,0,k}^{\rm{MBS}} ={{{{\bar d }_{i,0,k}}}}/({{{{\bar z }_{i,0,k}}{R_{i,0}}}}) + {\mathscr{C}_{i,0,k}^{\rm{MBS}}}/{{{{\hat f}_{i,0,k}}}} +\\
&\ \ \ \ \ \ {{{{\bar d }_{i,0,k}}{c_{i,k}}}}/{{{{\hat f}_{i,0,k}}}} +\sum\nolimits_{q \in {{\cal Q}}} {{{{y_{i,k,q}}{{\hat \gamma }_q}{{\bar d }_{i,0,k}}}}/{{{{\bar z }_{i,j,k}}{{\hat f}_{i,0,k}}}} }.\\
	\end{split}
\end{equation}
where the four items on the right of the equal sign are the transmission time from MD $i$ to MBS, decompressing time, computing time and decrypting time, respectively. Under the proportional resource assignment manner, the CC ${\hat f _{i,0,k}}$ assigned to the task $k$ of $\mathcal{E}_{i,0}$ by this BS can be given by
\begin{equation}\label{eq21}
{\hat f _{i,0,k}} =\frac{{\mathscr{A}_{i,0,k}^{\rm{MBS}}}{f_0^{\rm{BS}}}}{\sum\limits_{u \in {{\cal I}}} {\sum\limits_{n \in {{\cal K}}} {({x_{u,0}}\mathscr{A}_{u,0,n}^{\rm{MBS}}+\sum\nolimits_{s \in {{\bar{\cal J}}}}{x_{u,s}}\mathscr{A}_{u,s,n}^{\rm{MBS}}) } }}.
\end{equation}
Then, when MD $i$ is associated with MBS, the EC ${\varepsilon}_{i,0,k}^{\rm{MBS}}$ used for processing the $\mathcal{T}_{i,k}$ at MBS can be given by
\begin{equation}\label{eq22}
{\varepsilon}_{i,0,k}^{\rm{MBS}} = {\hat \xi _0}\mathscr{C}_{i,0,k}^{\rm{MBS}} + {\hat \xi _0}{\bar d _{i,0,k}}{c_{i,k}} + \sum\limits_{q \in {{\cal Q}}} \frac{{y_{i,k,q}}{{\tilde \gamma }_q}{{\bar d }_{i,0,k}}}{{{\bar z }_{i,0,k}}},
\end{equation}
where the three items on the right of the equal sign are the EC caused by decompression, computation and decryption at MBS in the one-step offloading, respectively.
\par
In reality, the local execution and uplink transmission can be done simultaneously, and the local execution and BS processing can be also done simultaneously. In addition, we assume that the tasks can be processed one by one at MDs. Therefore, the overall time (delay) ${{\tau}_{i}}$ used for completing the tasks of MD $i$ can be given by
\begin{equation}\label{eq23}
  {{\tau }_{i}}=\sum\limits_{k\in \cal K}{\max }\Big(\sum\limits_{j\in \bar{\cal J}}{{{x}_{i,j}}\tau _{i,j,k}^{\rm{SBS}}}+{{x}_{i,0}}\tau _{i,0,k}^{\rm{MBS}},\sum\limits_{j\in \cal J}{{{x}_{i,j}}\tau _{i,j,k}^{\rm{LOC}}}\Big).
\end{equation}
Then, the overall EC ${{\varepsilon}_i}$ used for completing the tasks of MD $i$ can be given by
\begin{equation}\label{eq24}
\begin{split}
{{\varepsilon}_i} = &\sum\nolimits_{j \in {{\cal J}}} {\sum\nolimits_{k \in {{\cal K}}} {{{x}_{i,j}}{\varepsilon}_{i,j,k}^{\rm{LOC}}} }  + \sum\limits_{k \in {{\cal K}}} {{{x}_{i,0}}{\varepsilon}_{i,0,k}^{\rm{MBS}}}\\
&+\sum\nolimits_{j \in \bar {{\cal J}} } {\sum\nolimits_{k \in {{\cal K}}} {{{x}_{i,j}}{\varepsilon}_{i,j,k}^{\rm{SBS}}} }.
\end{split}
\end{equation}

\section{Problem Formulation}\label{sec3}
In pursuit of optimizing computational and communicational resources, reducing transmission workload, offloading security, and minimizing network-wide EC, the joint optimization of channel selection, the number of subchannels, secure service assignment, multi-step computation offloading, device association, DC control, power control, and frequency band partitioning is done under the proportional assignment of computational resources and the constraints related to processing delay and security breach cost. Specifically, the optimization problem is formulated as
\begin{equation}\label{eq25}
\begin{split}
&\underset{\boldsymbol{\Xi}}{\mathop{\min }}\, E\left( \boldsymbol{\Xi}\right) =\sum\nolimits_{i\in \cal I}{{{\varepsilon }_{i}}} \\
&{\text{s.t. }{{{C}}_1}{\rm{:}}}{{{\tau}_i} \le {\tau}_i^{\max },\forall i \in {\cal I},}\\
&{{{{C}}_2}:}{{\psi _i} \le \psi _i^{\max },\forall i \in {\cal I},}\\
&{{{{C}}_3}:}{\sum\nolimits_{j \in {{\cal J}}} {{x_{i,j}} = 1,\forall i \in {\cal I},} }\\
&{{{{C}}_4}:}{\sum\nolimits_{q \in {{\cal Q}}} {{y_{i,k,q}} = 1,\forall i \in {\cal I}, k \in {\cal K},} }\\
&{{{{C}}_5}:}{\sum\nolimits_{n \in {{\cal N}}} {{b_{i,n}} = 1,\forall i \in {\cal I},} }\\
&{{C_{6}}:}{\vartheta}_{1} \le \mu \le {\vartheta}_{2},\\
&{{C_{7}}:}1 \le N \le {N}^{\max},\\
&{{{{C}}_8}:}{{\vartheta}_{1}  \le {p_i} \le p_i^{\max },\forall i \in {\cal I},}\\
&{{{{C}}_9}:}{{{\bar z }^{\min }} \le {{\bar z }_{i,j,k}} \le {{\bar z }^{\max }},\forall i \in {\cal I}, j \in {{\cal J}}, k \in {\cal K},}\\
&{{{C}_{10}}:}{{{\hat z}^{\min }} \le {{\hat z}_{i,j,k}} \le {{\hat z}^{\max }},\forall i \in {\cal I}, j \in {{\cal J}}, k \in {\cal K},}\\
&{{C_{11}}:}{{x_{i,j}} \in \left\{ {0,1} \right\},\forall i \in {\cal I}, j \in {{\cal J}},}\\
&{{C_{12}}:}{{y_{i,k,q}} \in \left\{ {0,1} \right\},\forall i \in {\cal I}, k \in {\cal K}, q \in {{\cal Q}},}\\
&{{C_{13}}:}{{b_{i,n}} \in \left\{ {0,1} \right\},\forall i \in {\cal I}, n \in {{\cal N}},}\\
&{{C_{14}}:}{{\vartheta}_{1}  \le {{\hat d}_{i,j,k}} \le {{\bar d }_{i,j,k}} \le {d_{i,k}},\forall i \in {\cal I}, j \in {{\cal J}}, k \in {\cal K},}\\
\end{split}
\end{equation}
where $\boldsymbol{\Xi}=\{\mu ,N,\mathbf{p},\mathbf{X},\mathbf{Y},\bar{\mathbf{Z}},\hat{\mathbf{Z}},\mathbf{B},\bar{\mathbf{D}},\hat{\mathbf{D}}\}$; to avoid zero division, ${\vartheta}_{1}$ and ${\vartheta}_{2}$ take two constants that are infinitely close to 0 and 1, respectively; $C_1$ indicates that the task processing time of MD $i$ can't exceed its deadline ${\tau}_i^{\max }$; $C_2$ indicates that the cost incurred from a security breach on MD $i$ should not exceed its maximum acceptable cost threshold $\psi _i^{\max }$; $C_3$ and $C_{11}$ stipulate that each MD can be associated with only one BS; $C_4$ and $C_{12}$ dictate that $\mathcal{T}_{i,k}$ can utilize single cryptographic algorithm; $C_5$ and $C_{13}$ ensure that each MD connects to just one subchannel; $C_{6}$ sets constraints on the range of values for $\mu$, with lower and upper bounds denoted by ${\vartheta}_{1} $ and ${\vartheta}_{2} $; ${C_{7}}$ gives a lower bound (i.e., 1) and an upper bound (i.e., $N^{\max }$) of $N$; $C_{8}$ gives a lower bound (${\vartheta}_{1} $) and an upper bound ($p_i^{\max }$) of ${p}_{i}$; $C_{9}$ gives a lower bound (${\bar z ^{\min }}$) and an upper bound (${\bar z ^{\max }}$) of ${{{\bar z }_{i,j,k}}}$; $C_{10}$ gives a lower bound (${\hat z^{\min }}$) and an upper bound (${\hat z^{\max }}$) of ${{{\hat z}_{i,j,k}}}$; $C_{14}$ indicates that the offloaded parts with sizes ${\hat d_{i,j,k}}$ and ${\bar d _{i,j,k}}$ are greater than or equal to ${\vartheta}_{1} $, but less than or equal to ${d _{i,k}}$. Meanwhile, ${\hat d_{i,j,k}}$ must be less than or equal to ${\bar d _{i,j,k}}$.

\section{Algorithm Design}\label{sec4}
As revealed in \cite{Y.Zheng2015Mar}, the WWO algorithm has the advantages of a simple algorithm framework, easy implementation, small population size, few control parameters and low execution time, compared to traditional heuristics algorithms. In addition, WWO has an adaptive mechanism to balance its exploration and exploitation behaviors, which increases the probability of avoiding falling into local optimum. Consequently, it has been regarded as an effective meta-heuristic algorithm used for solving large-scale complex problems.
\par
The WWO algorithm consists of propagation, refraction, and breaking operations. The propagation performs a coarse-grained search, and the refraction and breaking operations perform a fine-grained search. Furthermore, the refraction also can escape from the local optimum to some extent. However, when the scale of the optimization problem is extremely large, the global search ability of the WWO algorithm is relatively weak and its convergence speed is relatively low. Given this, the propagation behavior of the WWO algorithm is replaced with genetic operations including selection, diversity-guided mutation, adaptive crossover and mutation, and thus AGWWO algorithm is established in this paper. The selection is used for generating a new population from the old population and retaining the historical best individual, the diversity-guided mutation is used for enhancing the global search ability and thus avoiding premature convergence, adaptive behaviors are used for improving the convergence speed, and crossover and mutation are used for exploitation and exploration respectively.
\par
To utilize the AGWWO algorithm for solving \eqref{eq25}, we must first encode individuals as waves and establish a fitness function. Following that, we initialize their values. The propagation, refraction, and breaking procedures can then be executed, each of which will be described in detail.

\subsection{Encode wave}
The population consisting of $M$ waves (individuals) is denoted as ${{\cal M}} = \left\{ {1,2, \cdots ,M} \right\}$, where the height of wave $m$ is ${h_m}$ and used for controlling the algorithm process, and its initial value takes the maximal constant ${h^{\max }}$. Then, the optimization parameters $\mu$, $N$, ${\bf{p}}$, ${\bf{X}}$, ${\bf{Y}}$, $\bar {\bf{Z}} $, $\hat {\bf{Z}}$, $\bf{B}$, $\bar {\bf{D}}$ and $\hat {\bf{D}}$ of \eqref{eq25} are encoded as $\dot{\mu}_{m}$, $\dot{N}_{m}$, ${\dot{{\bf{P}}}_m}$, ${\dot{{\bf{X}}}_m}$, ${\dot{{\bf{Y}}}_m}$, ${{\dot{\bf{Z}}}_m}$, ${{\ddot{\bf{Z}}}_m}$, $\dot{\bf{B}}_{m}$, ${\dot{\bf{D}}_m}$ and ${{\ddot{\bf{D}}}_m}$, respectively, where the latter can be regarded as the wavelets of wave $m$ or the chromosomes of individual $m$; ${\dot{\mu}_{m}}$ is the frequency band partitioning factor in the wave $m$; ${\dot{N}_{m}}$ is the number of subchannels in the wave $m$; ${\dot{{\bf{P}}}_m} = \left\{ {{\dot{p}_{m,i}},i \in {\cal I}} \right\}$, ${\dot{p}_{m,i}}$ is the transmission power of MD $i$ in the wave $m$; ${\dot{{\bf{X}}}_m} = \left\{ {{\dot{x}_{m,i}},i \in {\cal I} } \right\}$, ${\dot{x}_{m,i}}$ is the index of the BS selected by MD $i$ in the wave $m$; ${\dot{{\bf{Y}}}_m} = \left\{ {{\dot{y}_{m,i}},i \in \bar {\cal I} } \right\}$, ${\dot{y}_{m,i}}$ is the index of cryptographic algorithm selected by virtual MD $i$ in the wave $m$, and $\bar {\cal I}  = \left\{ {1,2, \cdots ,K,K + 1, \cdots ,2K, \cdots ,UK} \right\}$ represents the index-set of virtual MDs; ${{\dot{\bf{Z}}}_m} = \left\{ {{\dot{z}_{m,i}},i \in \bar {\cal I} } \right\}$, ${\dot{z}_{m,i}}$ is the DC ratio used by virtual MD $i$ in the wave $m$; ${{\ddot{\bf{Z}}}_m} = \left\{ {{\ddot{z}_{m,i}},i \in \bar {\cal I} } \right\}$, ${\ddot{z}_{m,i}}$ is the DC ratio used by the SBS associated with virtual MD $i$ in the wave $m$; ${\dot{{\bf{B}}}_m} = \{ {{\dot{b}_{m,i}},i \in {\cal I} } \}$, ${\dot{b}_{m,i}}$ is the index of the subchannel selected by MD $i$ in the wave $m$; ${\dot{\bf{D}}_m} = \{ {{\dot{d}_{m,i}},i \in \bar {\cal I} } \}$, ${\dot{d}_{m,i}}$ is the amount of data offloaded from virtual MD $i$ to its associated BS in the wave $m$; ${{\ddot{\bf{D}}}_m} = \{ {{\ddot{d}_{m,i}},i \in \bar {\cal I} } \}$, ${\ddot{d}_{m,i}}$ is the amount of data offloaded from the SBS associated with virtual MD $i$ to MBS.

\subsection{Fitness function}
To accurately evaluate the fitness value of the waves, a reasonable fitness function needs to be designed. As observed in \eqref{eq25}, it is evident that the constraints ${C_1}$ and ${C_2}$ take the form of nonlinear, mixed-integer coupling, presenting challenges for adherence in the behavior of waves. Consequently, they are incorporated into the fitness function as penalty terms, serving to deter the waves from converging to infeasible regions. This approach ensures that the resulting population consistently identifies a feasible optimal solution.
\par
To minimize the network-wide EC under the constraints ${C_1}$ and ${C_2}$, the fitness function of wave $m$ is defined as
\begin{equation}\label{eq26}
\begin{split}
&{F}( \dot{\boldsymbol{\Xi}}_{m})= - E( \dot{\boldsymbol{\Xi}}_{m} )- \sum\nolimits_{i \in {\cal I}} {{\alpha _i}\max \left( {0,{{\tau}_i} - {\tau}_i^{\max }} \right)}\\
 &\ \ \ \ \ \ \ \ \ \ \ \ - \sum\nolimits_{i \in {\cal I}} {{\beta _i}\max \left( {0,{\psi _i} - \psi _i^{\max }} \right)},\\
\end{split}
\end{equation}
where $\max \left( 0, \theta  \right)$ represents a function that returns the larger value between 0 and $\theta$; ${\alpha _i}$ and ${\beta _i}$ are the penalty factors of MD $i$; $\dot{\boldsymbol{\Xi}}_{m}=\{\dot{\mu}_{m},\dot{N}_{m},\dot{\mathbf{p}}_{m},\dot{\mathbf{X}}_{m},\dot{\mathbf{Y}}_{m},\dot{\mathbf{Z}}_{m},\ddot{\mathbf{Z}}_{m},\dot{\mathbf{B}}_{m},\dot{\mathbf{D}}_{m},\ddot{\mathbf{D}}_{m}\}$.

\subsection{Population initialization}
To satisfy the conditions from ${C_3}$ to ${C_{14}}$, the initial population can be created based on the following guidelines. Specifically, any wave $m$ can be initialized into
\begin{equation}\label{eq27}
\left\{
\begin{split}
&\dot{\mu}_{m}^0 = {\rm{rand}}\left( 1 \right),\\
&\dot{N}_{m}^0 = {\rm{randi}}\left( {N}^{\max} \right), \\
&\dot{x}_{m,i}^0 = {\rm{randi}}\left( {{\cal J}} \right),\forall i \in {{\cal I}} ,\\
&\dot{y}_{m,i}^0 = {\rm{randi}}\left( {{\cal Q}} \right),\forall i \in \bar {{\cal I}} ,\\
&\dot{p}_{m,i}^0 = {\rm{rand}}(p_i^{\max }),\forall i \in {\cal I},\\
&\dot{z}_{m,i}^0 = {\bar z ^{\min }} + {\rm{rand}}\left( {{{\bar z }^{\max }} - {{\bar z }^{\min }}} \right),\forall i \in \bar {{\cal I}} ,\\
&\ddot{z}_{m,i}^0 = {\hat z^{\min }} + {\rm{rand}}\left( {{{\hat z}^{\max }} - {{\hat z}^{\min }}} \right),\forall i \in \bar {{\cal I}} ,\\
&\dot{b}_{m,i}^0 = {\rm{randi}}\left( {{\cal N}} \right),\forall i \in {{\cal I}} ,\\
&\dot{d}_{m,i}^0 = {\rm{rand}}\left( {{d_{u,k}}} \right),\forall i \in \bar {{\cal I}} ,\\
&\ddot{d}_{m,i}^0 = {\rm{rand}}(\dot{d}_{m,i}^0),\forall i \in \bar {{\cal I}} ,\\
&[u,k] = {\rm{ind2sub}}([U,K],i),\forall i \in \bar {{\cal I}} ,\\
\end{split}
\right.
\end{equation}
where $[u,k] = {\rm{ind2sub}}(\left[ {U,K} \right],i)$ returns the row index $u$ and column index $k$ of a matrix with dimensions $U \times K$, corresponding to the linear index $i$; ${\rm{randi}}\left( {{\cal I}} \right)$ randomly selects an element from the set ${{\cal I}}$; ${\rm{rand}}\left( \theta \right)$ generates a random number uniformly distributed between 0 and $\theta$.

\subsection{Propagation}
As revealed in the previous section, to enhance global search ability, improve convergence speed and avoid premature convergence, the propagation behavior of the WWO algorithm is replaced with genetic operations consisting of selection, diversity-guided mutation, adaptive crossover and mutation.

\subsubsection{Selection}
The selection operation plays a crucial role in filtering individuals (waves) within the population. Rather than choosing the absolute best individuals, it aims to select excellent parents for the next generation. In this study, a tournament-based selection strategy is employed, wherein two random waves are compared, and the one with the highest fitness value is chosen to move to the next generation. During the refraction operation, the wave learns from the best wave and eventually determines the best location. Consequently, the historical best wave, which possesses the highest fitness value among waves from both previous and current generations, is always retained during the selection process. Specifically, if the historical best wave is not initially selected for the next generation, it replaces the worst wave in the current population, which has the lowest fitness value.

\subsubsection{Crossover}
In GA, the crossover operation involves swapping gene segments between a pair of chromosomes to create two new individuals, based on the specified crossover probability. This process maintains population diversity, aids convergence, and prevents the algorithm from getting stuck in local optima. During each crossover, two adjacent individuals, denoted as $m$ and $\bar m  = m + 1$, are chosen to exchange their corresponding gene segments. The crossover probability \cite{M.Y.Li2004Sep} for propagating the superior individuals' building blocks can be given by
\begin{equation}\label{eq28}
{\bar{\mathscr{P}}_{m,\bar m }} =
\left\{
\begin{split}
	&{a_1}\frac{{{{\bar {\mathscr{F}} }_{m,\bar m }} - {\mathscr{F}^{\min }}}}{{{\mathscr{F}^{{\rm{ave}}}} - {\mathscr{F}^{\min }}}},&{\bar{\mathscr{F}} _{m,\bar m }} < {\mathscr{F}^{{\rm{ave}}}},\\
	&{a_2},&{\bar{\mathscr{F}} _{m,\bar m }} \ge {\mathscr{F}^{{\rm{ave}}}},\\
\end{split}
\right.
\end{equation}
where ${a}_{1}$ and ${a}_{2}$ are constants satisfying $0 \le {a}_{1}\le{a}_{2} \le 1$; ${\bar {\mathscr{F}}_{m,\bar m }}$ represents the minimum of fitness values of individuals $m$ and $\bar m$; ${\mathscr{F}^{\min }}$ and ${\mathscr{F}^{{\rm{ave}}}}$ denote the minimum and average of fitness values of all individuals in the population, respectively.

\subsubsection{Mutation}
In GA, the mutation operation is that some chromosomes of individuals are changed according to the mutation rules under a certain probability, finally generating new individuals. It endows GA with the local search ability. In each mutation, some chromosomes of a randomly selected individual $m$ are mutated with a certain probability ${\hat{\mathscr{P}}_m}$ under the constraints of \eqref{eq25}. Specifically, it can be given by
\begin{equation}\label{eq29}
{\hat{\mathscr{P}}_m} = \left\{
\begin{split}
&{a_3}\frac{{{\mathscr{F}^{\max }} - {{ {\mathscr{F}}}_m}}}{{{\mathscr{F}^{\max }} - {\mathscr{F}^{\rm{ave}}}}},&{{\mathscr{F}}_m} \ge {\mathscr{F}^{\rm{ave}}},\\
&{a_4},&{{\mathscr{F}}_m} < {\mathscr{F}^{\rm{ave}}},\\
\end{split}
\right.
\end{equation}
where ${a}_{3}$ and ${a}_{4}$ are constants satisfying $0 \le {a}_{3}\le{a}_{4} \le 1$; ${\mathscr{F}_{m}}$ is the fitness values of individual $m$; ${\mathscr{F}^{\max }}$ is the maximum of fitness values of all individuals in the population.
\par
Under the probability ${\hat{\mathscr{P}}_m}$, the chromosomes of individual $m$ can be updated using
\begin{equation}\label{eq30}
{\dot{\mu}_{m}} = \left\{
\begin{split}
&{r_1}{\vartheta}_{2} + \left( {1 - {r_1}} \right){\dot{\mu}_{m}},\ \ {r_2} > 0.5,\\
&{r_1}{\vartheta}_{1}+\left( {1 - {r_1}} \right){\dot{\mu}_{m}},\ \ {r_2} \le 0.5,\\
\end{split}
\right.
\end{equation}
\begin{equation}\label{eq31}
{\dot{N}_{m}} = \left\{
\begin{split}
&{\cal{R}}\left( {{r_1}{N}^{\max} + \left( {1 - {r_1}} \right){\dot{N}_{m}}} \right),\ \ {r_2} > 0.5,\\
&{\cal{R}}\left( {{r_1}+\left( {1 - {r_1}} \right){\dot{N}_{m}}} \right),\ \ {r_2} \le 0.5,\\
\end{split}
\right.
\end{equation}
\begin{equation}\label{eq32}
{\dot{x}_{m,i}} = \left\{
\begin{split}
&{\cal{R}}\left( {{r_1}S + \left( {1 - {r_1}} \right){\dot{x}_{m,i}}} \right),\ \ {r_2} > 0.5,\forall i \in  {\cal I} ,\\
&{\cal{R}}\left( {{r_1} + \left( {1 - {r_1}} \right){\dot{x}_{m,i}}} \right),\ \ {r_2} \le 0.5,\forall i \in  {\cal I} ,\\
\end{split}
\right.
\end{equation}
\begin{equation}\label{eq33}
{\dot{y}_{m,i}} = \left\{
\begin{split}
&{\cal{R}}\left( {{r_1}L + \left( {1 - {r_1}} \right){\dot{y}_{m,i}}} \right),\ \ {r_2} > 0.5,\forall i \in \bar {{\cal I}} ,\\
&{\cal{R}}\left( {{r_1} + \left( {1 - {r_1}} \right){\dot{y}_{m,i}}} \right),\ \ {r_2} \le 0.5,\forall i \in \bar {{\cal I}} ,\\
\end{split}
\right.
\end{equation}
\begin{equation}\label{eq34}
{\dot{p}_{m,i}} = \left\{
\begin{split}
&{r_1}{{p}_{i}^{\max}} + \left( {1 - {r_1}} \right){\dot{p}_{m,i}},\ \ {r_2} > 0.5,\forall i \in {{\cal I}},\\
&{r_1}{\vartheta}_{1} +\left( {1 - {r_1}} \right){\dot{p}_{m,i}},\ \ {r_2} \le 0.5,\forall i \in {{\cal I}},\\
\end{split}
\right.
\end{equation}
\begin{equation}\label{eq35}
{\dot{z}_{m,i}} = \left\{
\begin{split}
&{r_1}{\bar z ^{\max }} + \left( {1 - {r_1}} \right){\dot{z}_{m,i}},\ \ {r_2} > 0.5,\forall i \in \bar {{\cal I}} ,\\
&{r_1}{\bar z ^{\min }} + \left( {1 - {r_1}} \right){\dot{z}_{m,i}},\ \ {r_2} \le 0.5,\forall i \in \bar {{\cal I}} ,\\
\end{split}
\right.
\end{equation}
\begin{equation}\label{eq36}
{\ddot{z}_{m,i}} = \left\{
\begin{split}
&{r_1}{\hat z^{\max }} + \left( {1 - {r_1}} \right){\ddot{z}_{m,i}},\ \ {r_2} > 0.5,\forall i \in \bar {{\cal I}} ,\\
&{r_1}{\hat z^{\min }} + \left( {1 - {r_1}} \right){\ddot{z}_{m,i}},\ \ {r_2} \le 0.5,\forall i \in \bar {{\cal I}} ,\\
\end{split}	
\right.
\end{equation}
\begin{equation}\label{eq37}
{\dot{b}_{m,i}} = \left\{
\begin{split}
&{\cal{R}}\left( {{r_1}{\dot{N}_{m}} + \left( {1 - {r_1}} \right){\dot{b}_{m,i}}} \right),{r_2} > 0.5,\forall i \in {\cal I} ,\\
&{\cal{R}}\left( {{r_1} + \left( {1 - {r_1}} \right){\dot{b}_{m,i}}} \right),{r_2} \le 0.5,\forall i \in {\cal I} ,\\
\end{split}
\right.
\end{equation}
\begin{equation}\label{eq38}
{\dot{d}_{m,i}} = \left\{
\begin{split}
&{r_1}{d_{u,k}} + \left( {1 - {r_1}} \right){\dot{d}_{m,i}},\ \ {r_2} > 0.5,\forall u \in \bar{{\cal I}},\\
&\left( {1 - {r_1}} \right){\dot{d}_{m,i}},\ \ {r_2} \le 0.5,\forall i \in \bar {{\cal I}} ,\\
\end{split}
\right.
\end{equation}
\begin{equation}\label{eq39}
{\ddot{d}_{m,i}} = \left\{
\begin{split}
&{r_1}{\dot{d}_{m,i}} + \left( {1 - {r_1}} \right){\ddot{d}_{m,i}},\ \ {r_2} > 0.5,\forall i \in \bar {{\cal I}} ,\\
&\left( {1 - {r_1}} \right){\ddot{d}_{m,i}},\ \ {r_2} \le 0.5,\forall i \in \bar {{\cal I}} ,\\
\end{split}
\right.
\end{equation}
where $[u,k] = {\rm{ind2sub}}\left( {[U,K],i} \right),\forall i \in \bar {{\cal I}} $; $\cal{R}\left(\theta\right)$ denotes a rounding operation on $\theta$; ${r_1}$ and ${r_2}$ are random numbers obeying 0$-$1 uniform distribution, and they are used for the control of the mutation magnitude and searching direction respectively; ${r_2} > 0.5$ means a mutation toward the maximum, otherwise the minimum.
\par
Before adaptive crossover and mutation, the diversity-guided mutation is used for enhancing the global search ability and thus avoiding premature convergence. As defined in \cite{R.K.U2022Sep}, the population diversity for an N-dimensional problem is measured by
\begin{equation}\label{eq40}
\begin{split}
&\mathscr{D} =\frac{1}{10M}\left[ {\sum\limits_{m \in {{\cal M}}} {\left({\frac{1}{J_1}}\sqrt { {{{\left( {{\dot{\mu}_{m}} - \dot{\mu}^{\rm{ave}}} \right)}^2}} }+ {{\frac{1}{J_2}}\sqrt { {{{\left( {{\dot{N}_{m}} - \dot{N}^{\rm{ave}}} \right)}^2}} } }\right.} } \right.\\
&+{{\frac{1}{J_3}}\sqrt {\sum\nolimits_{i \in  {{\cal I}} } {{{\left( {{\dot{x}_{m,i}} - \dot{x}_i^{\rm{ave}}} \right)}^2}} } }+{{\frac{1}{J_4}}\sqrt {\sum\nolimits_{i \in \bar {{\cal I}} } {{{\left( {{\dot{y}_{m,i}} - \dot{y}_i^{\rm{ave}}} \right)}^2}} } } \\
&+{\frac{1}{J_5}}{\sqrt {\sum\nolimits_{i \in {{\cal I}}} {{{\left( {{\dot{p}_{m,i}} - \dot{p}_i^{\rm{ave}}} \right)}^2}} } }+{\frac{1}{J_6}}{\sqrt {\sum\nolimits_{i \in \bar {{\cal I}} } {{{\left( {{\dot{z}_{m,i}} - \dot{z}_i^{\rm{ave}}} \right)}^2}} } } \\
&+{\frac{1}{J_7}}{\sqrt {\sum\nolimits_{i \in \bar {{\cal I}} } {{{\left( {{\ddot{z}_{m,i}} - \ddot{z}_i^{\rm{ave}}} \right)}^2}} } }+{{\frac{1}{J_8}}{\sqrt {\sum\nolimits_{i \in  {{\cal I}} } {{{\left( {{\dot{b}_{m,i}} - \dot{b}_i^{\rm{ave}}} \right)}^2}} } } }\\
&\left. \left. {+{\frac{1}{J_9}}{\sqrt {\sum\limits_{i \in \bar {{\cal I}} } {{{\left( {{\dot{d}_{m,i}} - \dot{d}_i^{\rm{ave}}} \right)}^2}} } }+{\frac{1}{J_{10}}}{\sqrt {\sum\limits_{i \in \bar {{\cal I}} } {{{\left( {{\ddot{d}_{m,i}} - \ddot{d}_i^{\rm{ave}}} \right)}^2}} } } } \right)\right],\\
\end{split}
\end{equation}
where ${J_1}$, ${J_2}$, ${J_3}$, ${J_4}$, ${J_5}$, ${J_6}$, ${J_7}$, ${J_8}$, ${J_9}$ and ${J_{10}}$ represent the lengths of diagonals of feasible domains of ${\dot{\mu}_m}$, ${\dot{N}_m}$, ${\dot{{\bf{X}}}_m}$, ${\dot{{\bf{Y}}}_m}$, ${\dot{{\bf{P}}}_m}$, ${{\dot{\bf{Z}}}_m}$, ${{\ddot{\bf{Z}}}_m}$, ${\dot{\bf{B}}_m}$, ${\dot{\bf{D}}_m}$ and ${{\ddot{\bf{D}}}_m}$ respectively. In addition,
\begin{equation}\label{eq41}
\left\{
\begin{split}
&\dot{\mu}^{\rm{ave}}= \sum\nolimits_{m \in {{\cal M}}} {{\dot{\mu}_{m}}/M},\ \dot{N}^{\rm{ave}} = \sum\nolimits_{m \in {{\cal M}}} {{\dot{N}_{m}}/M};\\
&\dot{x}_i^{\rm{ave}} = \sum\nolimits_{m \in {{\cal M}}} {{\dot{x}_{m,i}}/M},\ \dot{p}_i^{\rm{ave}} = \sum\nolimits_{m \in {{\cal M}}} {{\dot{p}_{m,i}}/M},\\
&\ \ \dot{b}_i^{\rm{ave}} = \sum\nolimits_{m \in {{\cal M}}} {{\dot{b}_{m,i}}/M},\ \forall i \in  {{\cal I}};\\
&\dot{y}_i^{\rm{ave}}= \sum\nolimits_{m \in {{\cal M}}} {{\dot{y}_{m,i}}/M},\ \dot{z}_i^{\rm{ave}} = \sum\nolimits_{m \in {{\cal M}}} {{{\dot{z}_{m,i}}}/M},\\
&\ \ \ddot{z}_i^{\rm{ave}} = \sum\nolimits_{m \in {{\cal M}}} {{\ddot{z}_{m,i}}/M},\ \dot{d}_i^{\rm{ave}} = \sum\nolimits_{m \in {{\cal M}}} {{\dot{d}_{m,i}}/M},\\
&\ \ \ddot{d}_i^{\rm{ave}}= \sum\nolimits_{m \in {{\cal M}}} {{\ddot{d}_{m,i}}/M},\ \forall i \in \bar {{\cal I}}.\\
\end{split}
\right.
\end{equation}
Then, the diversity-guided mutation takes place under the probability
\begin{equation}\label{eq42}
\tilde{\mathscr{P}} = \left\{
\begin{split}
{a_5}&,\quad{\rm{  if \;}}{\mathscr{D}} < {\mathscr{D}_{1}},\\
{a_6}&,\quad{\rm{  if \;}}{{{\mathscr{D}_1}}} \le {{{\mathscr{D}}}} < {\mathscr{D}_{2}},\\
{a_7}&,\quad{\rm{  otherwise}},\\
\end{split}
\right.
\end{equation}
where ${a}_{5}$, ${a}_{6}$ and ${a}_{7}$ are constants satisfying $0 < {a}_{6}<{a}_{5}< 1$, and ${a}_{7}$ satisfying $0 < {a}_{7} < 1$ is almost equal to 0; ${{\mathscr{D}}_{1}}$ and ${{\mathscr{D}}_{2}}$ satisfying $0<{\mathscr{D}_{1}}<{\mathscr{D}_{2}}<1$ are the diversity thresholds.

\subsection{Refraction}
In the propagation of wave $m$, if its fitness value in the newly generated location is lower than in the old location, we reduce the height of this wave, i.e., ${h_m} = {h_m} - 1$, rather than update the location of this wave. When the height $h_m$ is zero, it means that the wave $m$ cannot be further updated. To avoid the stagnation of waves' search, the refraction operation is performed on the wave $m$ when ${h_m} = 0$. Specifically, the wave $m$ is updated using
\begin{equation}\label{eq43}
\dot{\mu}_{m} = {\mathbb{N}\left( {( \dot{\mu}^{_{\text{*}}} + {\dot{\mu}_{m}})/2,\left| { \dot{\mu}^{_{\text{*}}} - {\dot{\mu}_{m}}} \right|/2} \right)},
\end{equation}
\begin{equation}\label{eq44}
\dot{N}_{m} = {\cal{R}}\left( {\mathbb{N}\left( {( \dot{N}^{_{\text{*}}} + {\dot{N}_{m}})/2,\left| { \dot{N}^{_{\text{*}}} - {\dot{N}_{m}}} \right|/2} \right)}\right),
\end{equation}
\begin{equation}\label{eq45}
\dot{x}_{m,i} = {\cal{R}}\left( {\mathbb{N}\left( {( \dot{x}_i^{_{\text{*}}} + {\dot{x}_{m,i}})/2,\left| { \dot{x}_i^{_{\text{*}}} - {\dot{x}_{m,i}}} \right|/2} \right)} \right),\forall i \in {{\cal I}},
\end{equation}
\begin{equation}\label{eq46}
\dot{y}_{m,i} = {\cal{R}}\left( {\mathbb{N}\left( {(\dot{y}_i^{_{\text{*}}} + {\dot{y}_{m,i}})/2,\left| {\dot{y}_i^{_{\text{*}}} - {\dot{y}_{m,i}}} \right|/2} \right)} \right),\forall i \in \bar {{\cal I}},
\end{equation}
\begin{equation}\label{eq47}
\dot{p}_{m,i} = \mathbb{N}\left( {\left( {\dot{p}_i^{_{\text{*}}} + {\dot{p}_{m,i}}} \right)/2,\left| {\dot{p}_i^{_{\text{*}}} - {\dot{p}_{m,i}}} \right|/2} \right),\forall i \in {{\cal I}},
\end{equation}
\begin{equation}\label{eq48}
\dot{z}_{m,i} = \mathbb{N}\left( {\left( {\dot{z}_i^{_{\text{*}}} + {\dot{z}_{m,i}}} \right)/2,\left| {\dot{z}_i^{_{\text{*}}} - {\dot{z}_{m,i}}} \right|/2} \right),\forall i \in \bar {{\cal I}},
\end{equation}
\begin{equation}\label{eq49}
 \ddot{z}_{m,i}= \mathbb{N}\left( {\left( {\ddot{z}_i^{_{\text{*}}} + {\ddot{z}_{m,i}}} \right)/2,\left| {\ddot{z}_i^{_{\text{*}}} - {\ddot{z}_{m,i}}} \right|/2} \right),\forall i \in \bar {{\cal I}},
\end{equation}
\begin{equation}\label{eq50}
\dot{b}_{m,i} = {\cal{R}}\big( {\mathbb{N}( {( \dot{b}_i^{_{\text{*}}} + {\dot{b}_{m,i}})/2,| { \dot{b}_i^{_{\text{*}}} - {\dot{b}_{m,i}}} |/2} )} \big),\forall i \in {{\cal I}},
\end{equation}
\begin{equation}\label{eq51}
\dot{d}_{m,i} = \mathbb{N}\big( {( {\dot{d}_i^{_{\text{*}}} + {\dot{d}_{m,i}}} )/2,| {\dot{d}_i^{_{\text{*}}} - {\dot{d}_{m,i}}} |/2} \big),\forall i \in \bar {{\cal I}},
\end{equation}
\begin{equation}\label{eq52}
\ddot{d}_{m,i}= \mathbb{N}\big( {( {\ddot{d}_i^{_{\text{*}}} + {\ddot{d}_{m,i}}} )/2,| {\ddot{d}_i^{_{\text{*}}} - {\ddot{d}_{m,i}}} |/2} \big),\forall i \in \bar {{\cal I}},
\end{equation}
where $\dot{\mu}^{_{\text{*}}}$, $\dot{N}^{_{\text{*}}}$, $\dot{x}_i^{_{\text{*}}}$, $\dot{y}_i^{_{\text{*}}}$, $\dot{p}_i^{_{\text{*}}}$, $\dot{z}_i^{_{\text{*}}}$, $\ddot{z}_i^{_{\text{*}}}$, $\dot{b}_i^{_{\text{*}}}$ $\dot{d}_i^{_{\text{*}}}$ and $\ddot{d}_i^{_{\text{*}}}$ are the locations of the point $i$ of wavelets of the current best wave ${m_{}^{_{^{\text{*}}}}}$, respectively; $\mathbb{N}\left( {{\theta}_1 ,{{\theta}_2}} \right)$ denotes a function that generates a random number obeying normal distribution with mean ${\theta}_1$ and variance ${\theta}_2$. Seen from \eqref{eq43}-\eqref{eq52}, the refraction is a local search process. Due to the nature of the normal distribution, there exists a small probability of escaping the local optimum. Significantly, in order to start a new rounding search, the height of wave $m$ needs to be changed into ${h_m} = {h^{\max }}$ after a refraction operation.

\subsection{Breaking}
When the propagation generates next-generation waves with higher fitness values than the current best wave, the braking operation is performed for the former. Such an operation is used for fine-grained search, i.e., further searching around the potential optimal solutions area of these next-generation waves. In the breaking operation, the newly generated waves are regarded as solitary waves. The rule of traditional breaking is to randomly generate $V$ solitary waves, each of which updates partial points. Unlike this, an improved rule is used for generating $V$ solitary waves, each of which updates all points in this paper. The solitary wave obtained by performing breaking for wave $m$ can be given by
\begin{equation}\label{eq53}
	\dot{\mu}_{m}^{_{'}}= {\dot{\mu}^{_{\text{*}}} + \zeta u\left( {\vartheta}_{2} - {\vartheta}_{1} \right)},
\end{equation}
\begin{equation}\label{eq54}
	\dot{N}_{m}^{_{'}}= {\cal{R}}\left( {\dot{N}^{_{\text{*}}} + \zeta u\left( {{N}^{\max} - 1} \right)} \right),
\end{equation}
\begin{equation}\label{eq55}
	\dot{x}_{m,i}^{_{'}}= {\cal{R}}\left( {\dot{x}_i^{_{\text{*}}} + \zeta u \left( {S - 1} \right)} \right),\forall i \in  {{\cal I}} ,
\end{equation}
\begin{equation}\label{eq56}
	 \dot{y}_{m,i}^{_{'}} = {\cal{R}}\left( {\dot{y}_i^{_{\text{*}}} + \zeta u \left( {Q - 1} \right)} \right),\forall i \in \bar {{\cal I}} ,
\end{equation}
\begin{equation}\label{eq57}
	\dot{p}_{m,i}^{_{'}} = \dot{p}_i^{_{\text{*}}} + \zeta u \left( p_i^{\max } - {\vartheta}_{1} \right),\forall i \in {{\cal I}},
\end{equation}
\begin{equation}\label{eq58}
	\dot{z}_{m,i}^{_{'}} =\dot{z}_i^{_{\text{*}}} + \zeta u \left( {{{\bar z }^{\max }} - {{\bar z }^{\min }}} \right),\forall i \in \bar {{\cal I}} ,
\end{equation}
\begin{equation}\label{eq59}
	\ddot{z}_{m,i}^{_{'}}  = \ddot{z}_i^{_{\text{*}}} + \zeta u \left( {{{\hat z}^{\max }} - {{\hat z}^{\min }}} \right),\forall i \in \bar {{\cal I}} ,
\end{equation}
\begin{equation}\label{eq60}
	\dot{b}_{m,i}^{_{'}}= {\cal{R}}( {\dot{b}_i^{_{\text{*}}} + \zeta u \left( {\dot{N}_{m}^{_{'}} - 1} \right)}),\forall i \in  {{\cal I}} ,
\end{equation}
\begin{equation}\label{eq61}
	 \dot{d}_{m,i}^{_{'}} = \dot{d}_i^{_{\text{*}}} + \zeta u {d_{u,k}},\forall i \in \bar {{\cal I}} ,
\end{equation}
\begin{equation}\label{eq62}
 \ddot{d}_{m,i}^{_{'}}  =\ddot{d}_i^{_{\text{*}}}+ \zeta u \dot{d}_{m,i}^{_{'}},\forall i \in \bar {{\cal I}} ,
\end{equation}
where $[u,k] = {\rm{ind2sub}}\left( {[U,K],i} \right),\forall i \in \bar {{\cal I}} $; $\zeta$ is the random number obeying a normal distribution with mean 0 and variance 1; $u $ is the breaking coefficient, increasing linearly from the minimum ${u ^{\min }}$ to the maximum ${u ^{\max }}$ with the number of iterations. Significantly, $\dot{\mu} ^{_{\text{*}}}$,  $\dot{N}^{_{\text{*}}}$, $\dot{x}_i^{_{\text{*}}}$, $\dot{y}_i^{_{\text{*}}}$, $\dot{p}_i^{_{\text{*}}}$, $\dot{z}_i^{_{\text{*}}}$, $\ddot{z}_i^{_{\text{*}}}$, $\dot{b}_i^{_{\text{*}}}$, $\dot{d}_i^{_{\text{*}}}$ and $\ddot{d}_i^{_{\text{*}}}$ are the locations of point $i$ of wavelets of the current best wave ${m_{}^{_{^{\text{*}}}}}$ of the newly generated waves after propagation. If none of the solitary waves has a higher fitness value than the current best wave ${m_{}^{_{^{\text{*}}}}}$, the wave ${m_{}^{_{^{\text{*}}}}}$ is retained; otherwise, the wave ${m_{}^{_{^{\text{*}}}}}$ is replaced with a solitary wave that has the highest fitness value among all solitary waves.
\par
To summarize, the entire procedure for the AGWWO algorithm can be represented as Algorithm 1, with $T$ denoting the number of iterations.
\begin{table}[!t]
	\centering
	\begin{tabular}{ll}
		\toprule[1pt]
		\textbf{Algorithm 1: AGWWO} \\ \midrule[0.5pt]
		1: \textbf{Input:} $M$, $N$, $J$, $I$, $K$, $T$, $Q$, $\varpi $, ${\sigma ^2}$, ${{\tau}_i^{\max }}$, $\bar{R}$, ${\eta _{i,k}}$, ${\psi _i^{\max }}$, ${\hat{v} _{i,k}}$,\\
           \ \ \ \ \ ${\bar{v} _{i,k}}$, ${u ^{\min }}$, ${u ^{\max }}$, ${\bar z ^{\min}}$, ${\bar z ^{\max}}$, ${\hat z^{\min}}$, ${\hat z^{\max}}$, ${{f_i}}$, $f_j^{\rm{BS}}$, ${\xi}$, $\tilde \xi $, ${{d_{i,k}}}$,\\
           \ \ \ \ \ ${{c_{i,k}}}$, $\alpha_i$, $\beta_i$, $h^{\max}$, $a_1$, $a_2$, $a_3$, $a_4$, $a_5$, $a_6$, $a_7$, $\mathscr{D}_1$, $\mathscr{D}_2$, $\varsigma $, ${\hat \xi _j}$, $\boldsymbol{\bar \gamma}$,\\
           \ \ \ \ \ $\boldsymbol{\hat \gamma}$, $\boldsymbol{\tilde \gamma}$, $\lambda _1^{\rm{LOC}}$, $\lambda _2^{\rm{LOC}}$, $\lambda _3^{\rm{LOC}}$, $\bar \lambda _1^{\rm{BS}}$, $\bar \lambda _2^{\rm{BS}}$, $\bar \lambda _3^{\rm{BS}}$, $\hat \lambda _1^{\rm{BS}}$, $\hat \lambda _2^{\rm{BS}}$, $\hat \lambda _3^{\rm{BS}}$\\
                      \ \ \ \ \ and $\varpi $.\\
		2: \textbf{Output:} $\dot{\boldsymbol{\Xi}}_{m}$ for all $m \in \cal{M}$ at $t$-th iteration.\\
		3: \textbf{Initialization:}\\
		4: \ \ Initialize the iteration index: $t = 1$.\\
		5: \ \ Initialize the population ${{\cal M}}$ consisting of $M$ waves using \eqref{eq27}, and \\
		   \ \ \ \ \ \ \ let heights of these waves be ${h^{\max }}$.\\
		6: \ \ Calculate the fitness values of all waves using \eqref{eq26}.\\
		7: \ \ Find the current best wave in the population ${{\cal M}}$.\\
		8: \ \ Replace the historical best wave with the current best wave ${{m_{}^{_{^{\text{*}}}}}}$ if \\
		\ \ \ \ \ \ \ the former has a smaller fitness value than the latter.\\
		9: \textbf{While} $t \le {T}$ \textbf{do}\\
		10: \ Select $M$ waves from the population ${{\cal M}}$ to generate a new \\
		\ \ \ \ \ \ \ \ population $\bar {{\cal M}}$ using the tournament method.\\
		11: \ Replace the historical best wave with the current best wave ${{m_{}^{_{^{\text{*}}}}}}$ if \\
		\ \ \ \ \ \ \ \ the former has a smaller fitness value than the latter in $\bar {{\cal M}}$.\\
		12: \ Execute the diversity-guided mutation using \eqref{eq30}-\eqref{eq39} under \eqref{eq42}.\\
		13: \ Calculate the fitness values of all waves using \eqref{eq26}.\\
		14: \ Adaptively crossover any two neighboring waves under \eqref{eq28}.\\
		15: \ Adaptively mutate using \eqref{eq30}-\eqref{eq39} under the probability \eqref{eq29}.\\
		16: \ Calculate fitness value $\tilde{\mathscr{F}}_{m}={F}( \dot{\boldsymbol{\Xi}}_{m} )$ of any $m  \in \bar {{\cal M}} $ using \eqref{eq26}.\\
		17: \ \textbf{For} each $m \in {{\cal M}}$ \textbf{do} \\
		18: \ \ \ \ \textbf{If} ${\tilde{\mathscr{F}}_{m }} > {{\mathscr{F}}_m}$ holds, \textbf{then}\\
		19: \ \ \ \ \ \ \ \textbf{If} ${{\tilde{\mathscr{F}}_{m }} > {\mathscr{F}}_{{m_{}^{_{^{\text{*}}}}}}}$ holds, \textbf{then}\\
		20:  \ \ \ \ \ \ \ \ \ \ \ Break the wave $m$ of the population ${{\cal M}}$ to generate $V$ \\
		\ \ \ \ \ \ \ \ \ \ \ \ \ \ \ \ \ solitary waves using \eqref{eq53}-\eqref{eq62}.\\
		21:  \ \ \ \ \ \ \ \ \ \ \ The wave ${m_{}^{_{^{\text{*}}}}}$ is replaced with the best solitary wave if \\
		\ \ \ \ \ \ \ \ \ \ \ \ \ \ \ \ \ the latter has a larger fitness value than the former.\\
		22:  \ \ \ \ \ \ \ \ \textbf{EndIf}\\
		23:  \  \ \ \ \ \ \ \ The wave ${m}$ in the population ${{\cal M}}$ is replaced with the one in \\
		\ \ \ \ \ \ \ \ \ \ \ \ \ \ \ \ the population $\bar {{\cal M}}$.\\
		24:  \ \ \ \ \textbf{Else}\\
		25:  \ \ \ \ \ \ \ \ Update the height of wave $m$ in the population ${{\cal M}}$: \\
		\ \ \ \ \ \ \ \ \ \ \ \ \ \ \ \ \ ${h_m} = {h_m} - 1$.\\
		26:  \ \ \ \ \ \ \ \ For $m  \in {{\cal M}}$, \textbf{if} ${h_m} = 0$ holds, \textbf{then}\\
		27:  \ \ \ \ \ \ \ \ \ \ \ A new wave is generated using \eqref{eq43}-\eqref{eq52} for wave $m$ \\
		\ \ \ \ \ \ \ \ \ \ \ \ \ \ \ \ \  and then used for replacing the latter.\\
		28:  \ \ \ \ \ \ \ \ \ \ \ Let the height of wave $m$ be ${h_m} = {h^{\max }}$.\\
		29:  \ \ \ \ \ \ \ \ \textbf{EndIf}\\
		30:  \ \ \ \ \textbf{EndIf}\\
		31:  \ \ \textbf{EndFor}\\
		32:  \ \ Calculate fitness value ${{\mathscr{F}}_m} = {F}( \dot{\boldsymbol{\Xi}}_{m} )$ of any $m \in {{\cal M}}$ using \eqref{eq26}.\\
		33:  \ \ Find the current best wave in the population ${{\cal M}}$.\\
		34:  \ \ Replace the historical best wave with the current best wave ${{m_{}^{_{^{\text{*}}}}}}$ if \\
		\ \ \ \ \ \ \ \ the former has a smaller fitness value than the latter in ${{\cal M}}$.\\
		35:  \ \ Update the iteration index: $t = t + 1$.\\
		36:\textbf{EndWhile}\\
		\bottomrule[0.5pt]
	\end{tabular}
	\label{alg1}
\end{table}

\section{Algorithm Analysis}\label{sec5}
In the following section, we will sequentially analyze the convergence, computational complexity, and parallel implementation of AGWWO.
\subsection{Convergence Analysis}
\par
As revealed in \cite{B.Zhang2016Apr}, to prove the convergence of AGWWO, two special cases of the changes in waves' (individual) fitness values during population evolution need to be analyzed. In the first case, the fitness values always increase with the number of iterations. In this case, the propagation operation is always performed, where each newly generated wave will replace the old wave. In the second case, the fitness values cannot be updated all the time. In this case, the refraction operation is always done, making the newly generated waves continuously approach the best wave. Next, we will analyze the two cases separately.
\par
\noindent
\textit{\textbf{Theorem 1: }} After numerous iterations, the propagation operation reaches the best solution globally.
\par
\textit{Proof}: In the first case, the algorithm just does a propagation operation. At this time, the AGWWO algorithm is equivalent to the adaptive genetic algorithm (GA) with diversity-guided mutation. As revealed in \cite{M.Y.Li2004Sep}, if the optimal individual can always be maintained after or before the selection operation, GA will converge to the global optimum. In the propagation operation, the current worst individual is replaced with the historical best individual after selecting individuals using the tournament method. It is evident that the propagation operation reaches the best solution globally after numerous iterations. \ding{113}
\par
When the diversity-guided mutation probabilities have ${a_5} = 0$, ${a_6} = 0$ and ${a_7} = 0$ in \eqref{eq42}, the propagation operation is equivalent to the traditional GA. Then, we have the following corollary.
\par
\noindent
\textit{\textbf{Corollary 1:}} When ${a_5} = 0$, ${a_6} = 0$ and ${a_7} = 0$, the propagation operation always converges to the global optimum if the best individual is maintained after/before the selection operation.
\par
\textit{Proof}: By following the procedure in \cite{M.Y.Li2004Sep}, Corollary 1 can be easily proved. \ding{113}
\par
Next, we will analyze the convergence of the refraction operation.
\par
\noindent
\textit{\textbf{Theorem 2: }} After numerous iterations, the refraction operation reaches the best solution globally.
\par
\textit{Proof}: By following the procedure in \cite{B.Zhang2016Apr}, the convergence of the refraction operation can be established as follows.
\par
As discussed in the previous section, the fitness values cannot be updated in the second case. At this time, the refraction operation is continuously performed. In addition, we assume that the current best wave ${m_{}^{_{^{\text{*}}}}}$ is always not updated at each iteration. That is to say, its $\dot{\mu} ^{_{\text{*}}}$,  $\dot{N}^{_{\text{*}}}$, $\dot{x}_i^{_{\text{*}}}$, $\dot{y}_i^{_{\text{*}}}$, $\dot{p}_i^{_{\text{*}}}$, $\dot{z}_i^{_{\text{*}}}$, $\ddot{z}_i^{_{\text{*}}}$, $\dot{b}_i^{_{\text{*}}}$, $\dot{d}_i^{_{\text{*}}}$ and $\ddot{d}_i^{_{\text{*}}}$ are not updated at each iteration. To simplify the proof process, the equation \eqref{eq47} is used as an example. Consequently, the equation \eqref{eq47} at the $t$-th iteration can be rewriten as
\begin{equation}\label{eq63}
	{ \dot{p}_{m,i}}^{t} =  0.5\left({\dot{p}_i^{_{\text{*}}} + {\dot{p}_{m,i}^{t-1}}}\right) + 0.5\left({\bar{r}\left( {\dot{p}_i^{_{\text{*}}} - {\dot{p}_{m,i}}^{t-1}} \right)}\right),
\end{equation}
where $\bar{r}$ represents a random number obeying a normal distribution. The further derivation of \eqref{eq63} can be given by
\begin{equation}\label{eq64}
\begin{split}
&{ \dot{p}_{m,i}}^{t} =0.5\left( {1 + \bar{r}} \right) \dot{p}_i^{_{\text{*}}} + 0.5\left( {1 - \bar{r}} \right){ \dot{p}_{m,i}}^{t-1}\\
&=0.5\left( {1 + \bar{r}} \right) \dot{p}_i^{_{\text{*}}} + { 0.5^2}\left( {1 - \bar{r}} \right)\left( {1 + \bar{r}} \right) \dot{p}_i^{_{\text{*}}}+{ {0.5}^2}{\left( {1 - \bar{r}} \right)^2}{ \dot{p}_{m,i}}^{t-2}\\
&=\cdots \\
&=0.5\left( {1 + \bar{r}} \right) \dot{p}_i^{_{\text{*}}}\sum\nolimits_{j = 0}^{{t} - 1} {{{0.5}^j}{{\left( {1 - \bar{r}} \right)}^j}}+ {0.5^{{t}}}{\left( {1 - \bar{r}} \right)^{{t}}}{ \dot{p}_{m,i}}^{0} \\
&=[ {1 - {{0.5 }^{{t}}}{{\left( {1 - \bar{r}} \right)}^{{t}}}} ] \dot{p}_i^{_{\text{*}}}+ { 0.5 ^{{t}}}{\left( {1 - \bar{r}} \right)^{{t}}}{ \dot{p}_{m,i}}^{0}\\
\end{split}
\end{equation}
where the mathematical expectation of $\bar{r}$ is 0; when ${t} \to \infty $, ${\left( {1/2} \right)^{{t}}}{\left( {1 - \bar{r}} \right)^{{t}}} \to 0$, $\mathop {\lim }\limits_{{t} \to \infty } {\dot{p}_{m,i}}^{t} = \dot{p}_i^{_{\text{*}}}$. In other words, ${\dot{p}_{m,i}}$ reaches the optimum $\dot{p}_i^{_{\text{*}}}$ after a large number of iterations.
\par
Other refraction rule equations can be similarly processed for the proof of convergence. Since the current best wave is not updated, the historical best wave is changeless. Therefore, the refraction operation always reaches the best solution globally after numerous iterations. \ding{113}
\par
\noindent
\textit{\textbf{Theorem 3:}} After numerous iterations, the AGWWO algorithm reaches the best solution globally.
\par
\textit{Proof}: As revealed in previous theorems, the propagation and refraction operations always converge to the global optimum after a large number of iterations. In AGWWO, the breaking operation is used for updating the current best wave using the best solitary wave with a higher fitness value. Such an operation is beneficial to finding a better and better historical best wave. In Steps 32-34 of AGWWO, the historical best wave is always replaced with the current best wave if the latter wave has a higher fitness value than the former wave. Evidently, such operations will help to find a better and better solution. In general, the propagation, refraction and breaking operations are used for global or local searching, the solution is always updated toward the maximum fitness value at each iteration. After a large number of iterations, the global optimum can be found. At this time, the AGWWO algorithm reaches the best solution globally. \ding{113}

\subsection{Complexity Analysis}
The computational complexity of AGWWO can be given by the following proposition.
\par
\noindent
\textit{\textbf{Proposition 1:}} In the worst-case scenario that all MDs share each channel, the computational complexity of AGWWO is $\mathcal{O}(\max\{2TM, TIJK, 3TIK, TMVIK, TMNVI^2\} )$ after $T$ iterations.
\par
\textit{Proof}: To reduce the computation complexities of calculating fitness values, some skillful operations need to be considered. It is easy to find that these fitness values are heavily dependent on the delay, EC and security breach costs. Furthermore, it is evident that the complexities of calculating delay and EC are mainly caused by calculating computation capacities assigned to MDs and data rates of associated MDs. To reduce the computational complexity of calculating these computation capacities and data rates, $\dot{\mathbf{X}}_{m}$ and $\dot{\mathbf{B}}_{m}$ first need to be converted into $\mathbf{o}=\{{o}_{i},\forall i\in \mathcal{I}\}$ and $\bar{\mathbf{o}}=\{{o}_{i},\forall i\in \mathcal{I}\}$ for any wave $m$, respectively. Moreover, $\dot{\mathbf{Y}}_{m}$ needs to be converted into the indices of cryptographic algorithms selected by MDs in any individual $m$. To further reduce the computational complexity of calculating computation capacities and data rates, we just need to take account of the utilized subchannels, BSs and cryptographic algorithms for each MD.
\par
To reduce the computational complexity, ${{R}_{i,j,n}}$ can be calculated after calculating $\sum\nolimits_{u\in {{\tilde{\mathcal{I}}}_{i,j,n}}}{{{p}_{u}}{{\hbar}_{u,s}}}$ for SBS $j$. Similarly, ${{R}_{i,0,n}}$ can be calculated after calculating $\sum\nolimits_{u\in \mathcal{I}}{{{x}_{u,0}}}$ for MBS. Consequently, under the given ${o}_{i}$ and ${\bar{o}}_{i}$,  the computational complexity of ${{R}_{i,j,n}}$ is $\mathcal{O}(NI^2)$ for all MDs, BSs and subchannels in the worst-case scenario that all MDs share each channel, and the computational complexity of ${{R}_{i,0,n}}$ is $\mathcal{O}(NI)$ for all MDs and subchannels, and MBS. In \eqref{eq14}, the CC ${\bar f _{i,j,k}}$ assigned to the $\mathcal{T}_{i,k}$ by SBS $j$ can be calculated after calculating ${ {\mathscr{A}_{i,j,k}^{\rm{SBS}}}f_j^{\rm{BS}}}$ and ${\sum\nolimits_{u \in {{\cal I}}} {\sum\nolimits_{{n} \in {{\cal K}}}} {{x_{u,j}}{\mathscr{A}_{u,j,n}^{\rm{SBS}}}}}$; in \eqref{eq16},  the CC ${\hat f _{i,j,k}}$ assigned to the task ${k}$ of $\mathcal{E}_{i,j}$ by MBS can be calculated after calculating ${{\mathscr{A}_{i,j,k}^{\rm{MBS}}}{f_0^{\rm{BS}}}}$ and $\sum\nolimits_{u \in {{\cal I}}} \sum\nolimits_{n \in {{\cal K}}} ({x_{u,0}}\mathscr{A}_{u,0,n}^{\rm{MBS}}+\sum\nolimits_{s \in {{\bar{\cal J}}}}{x_{u,s}}\mathscr{A}_{u,s,n}^{\rm{MBS}})$; in \eqref{eq21},  the CC ${\hat f _{i,0,k}} $ assigned to the task $k$ of $\mathcal{E}_{i,0}$ by this BS can be calculated after calculating ${{\mathscr{A}_{i,0,k}^{\rm{MBS}}}{f_0^{\rm{BS}}}}$ and $\sum\nolimits_{u \in {{\cal I}}} \sum\nolimits_{n \in {{\cal K}}} ({x_{u,0}}\mathscr{A}_{u,0,n}^{\rm{MBS}}+\sum\nolimits_{s \in {{\bar{\cal J}}}}{x_{u,s}}\mathscr{A}_{u,s,n}^{\rm{MBS}})$. Under the given indices of cryptographic algorithms selected by MDs, the computational complexity of ${\mathscr{A}_{u,j,n}^{\rm{SBS}}}$ and ${\mathscr{A}_{u,j,n}^{\rm{SBS}}}$ is $\mathcal{O}(1)$; under the given indices of BSs and subchannels selected by MDs, the computational complexity of ${\sum\nolimits_{u \in {{\cal I}}} {\sum\nolimits_{{n} \in {{\cal K}}}} {{x_{u,j}}{\mathscr{A}_{u,j,n}^{\rm{SBS}}}}}$ and $\sum\nolimits_{u \in {{\cal I}}} \sum\nolimits_{n \in {{\cal K}}} ({x_{u,0}}\mathscr{A}_{u,0,n}^{\rm{MBS}}+\sum\nolimits_{s \in {{\bar{\cal J}}}}{x_{u,s}}\mathscr{A}_{u,s,n}^{\rm{MBS}})$ is $\mathcal{O}(IK )$. Consequently, under the given index of the selected cryptographic algorithm, the computational complexity of ${{\bar{f}}_{i,j,k}}$ and ${{\hat{f}}_{i,j,k}}$ is $\mathcal{O}(IK )$ for all users, BSs and tasks in any wave.
\par
Considering the analyses provided earlier, given the indices of cryptographic algorithms, $\mathbf{o}$ and $\bar{\mathbf{o}}$, the computational complexity of calculating delay using \eqref{eq23} is $\mathcal{O}(\max\{IK, NI^2\} )$ for all MDs in the worst-case scenario, and the computational complexity of calculating EC using \eqref{eq24} is still $\mathcal{O}(\max\{IK, NI^2\} )$ for all MDs in the worst-case scenario. In addition, the computational complexity of evaluating the security breach cost using \eqref{eq10} is $\mathcal{O}(IK)$ for all MDs. Therefore, the computational complexity of calculating fitness values of all waves using \eqref{eq26} is $\mathcal{O}(\max\{MIK, MNI^2\} )$ for all waves (individuals) in the worst-case scenario.
\par
Next, we give the complexity analyses for AGWWO step by step. In such an algorithm, the computational complexity of Steps 4, 25 and 35 is ${{\cal O}}( 1)$, the one of Step 5 is ${{\cal O}}( {MIK} )$, the one of Steps 6, 13, 16 and 32  is  $\mathcal{O}(\max\{MIK, MNI^2\} )$ in the worst-case scenario, the one of Steps 7 and 33 is $\mathcal{O}(M )$, and the one of Steps 8, 11, 23, 27 and 34 is ${{\cal O}}( {MIK} )$. During the propagation process, the new population is created using the tournament method in each round, where two waves (individuals) are selected. As demonstrated in \cite{F.Guo2018Dec}, the computational complexity of this operation is ${{\cal O}}( {2M} )$. It means that the computational complexity of Step 10 is ${{\cal O}}( {2M} )$. In addition, for the chromosomes whose lengths are $I$ and $IK$, the computation complexities of the crossover operation are ${{\cal O}}( {3I} )$ and ${{\cal O}}( {3IK} )$ respectively, and the ones of the mutation operation are ${{\cal O}}( {IS} )$ and ${{\cal O}}( {IJK} )$ respectively. That is to say, the computational complexity of Steps 12 and 15 is ${{\cal O}}( {IJK} )$, and the one of Step 14 is ${{\cal O}}( {3IK} )$. Since the wave $m$ needs to be broken into $V$ wavelets, the computational complexity of Step 20 is ${{\cal O}}( {VIK} )$. Since Step 21 implies the calculation of fitness values of $V$ solitary waves, its computational complexity is $\mathcal{O}(\max\{VIK, NVI^2\} )$ in the worst-case scenario. Evidently, in Steps 17-31, the computational complexity mainly comes from Steps 20 and 21. The computational complexity of them is $\mathcal{O}(\max\{MVIK, MNVI^2\} )$ for all waves in the worst-case scenario.
\par
In general, in the worst-case scenario that all MDs share each channel, the computational complexity of AGWWO is $\mathcal{O}(\max\{2TM, TIJK, 3TIK, TMVIK, TMNVI^2\} )$ after $T$ iterations.\ding{113}

\subsection{Parallel Implement}
It is evident that the high computational complexity of AGWWO is mainly caused by calculating the fitness values of all waves. In addition, such a computational complexity will increase with the population size dramatically. To simplify the process and enhance the operational efficiency of AGWWO, we can allow all waves to calculate their fitness concurrently. In fact, such a consideration has been widely advocated in reality.

\section{Numerical Results}\label{sec6}
In the simulation, we consider that 20 SBSs are randomly deployed at one macrocell. In addition, $\varpi =20$ MHz, ${\sigma ^2}=10^{-11}$ mW, ${{\tau}_i^{\max }}=5$$\sim$10 s, $M=20$, $\bar{J}=30$, $K=3$, ${N}^{\max}=5$, $\bar{R}=1$ Gbps, ${\xi} =50$, $\alpha_i = \beta_i = 10^{20}$, and $h^{\max}=5$, $a_1 = a_2=0.8$, $a_3 =a_4=0.3$, $a_5=0.6$, $a_6 = 0.03$, $a_7=10^{-5}$, $\mathscr{D}_1 =0.01$, $\mathscr{D}_2=0.25$ \cite{M.Y.Li2004Sep}; $\varsigma = 10^{-25}$, ${\hat \xi _j}=1$ W/GHz, $\lambda _1^{\rm{LOC}}=1.027$$\times$$10^{-15}$, $\lambda _2^{\rm{LOC}}=32.28$, $\lambda _3^{\rm{LOC}}=0.3$; $\bar \lambda _1^{\rm{BS}}=0.076$, $\bar \lambda _2^{\rm{BS}}=0.7116$, $\bar \lambda _3^{\rm{BS}}=0.5794$; $\hat \lambda _1^{\rm{BS}}=0.115$, $\hat \lambda _2^{\rm{BS}}=-0.9179$, $\hat \lambda _3^{\rm{BS}}=0.046$ \cite{T.N.2020Jan}, ${\eta _{i,k}}=1000$$\sim$5000 USDs (United States dollars), ${\psi _i^{\max }}=5000$$\sim$10000 USDs, $L=6$, ${\hat{v} _{i,k}}=1$$\sim$3, ${\bar{v} _{i,k}}=\left\{5,6\right\}$, ${u ^{\min }}=0.001$, ${u ^{\max }}=0.25$, ${\bar z ^{\min}}=2.3$, ${\bar z ^{\max}}=2.9$, ${\hat z ^{\min}}=3.4$, ${\hat z ^{\max}}=11.2$, ${{f_i}}=1$ GHz, $f_j^{\rm{BS}}=20$ GHz, $\tilde \xi =1$ mW, ${{d_{i,k}}}=200$$\sim$500 KB, ${{c_{i,k}}}=50$$\sim$100 cycles/bit, $\boldsymbol{\bar \gamma}=\left[100,200,250,300,350,1050\right]$ cycles/bit, $\boldsymbol{\hat \gamma}=\left[90,280,350,300,400,1700\right]$ cycles/bit, $\boldsymbol{\tilde \gamma}=\left[2.5296, 5.0425, 6.837, 7.8528, 8.7073, 26.3643\right]$$\times$$10^{-7}$ J/bit\cite{Y.Zhang2020Nov}. In addition, the pathloss between MBS $j$ and MD $i$ is 128.1 + 37.6 $\text{log}_{10}\left(\ell_{i,j}\right)$, the one between SBS $j$ and MD $i$ is 140.7 + 36.7 $\text{log}_{10}\left(\ell_{i,j}\right)$, where $\ell_{i,j}$ is the distance between MD $i$ and BS $j$. Moreover, the log-normal shadowing fading with standard deviation of 8 dB is considered. Without loss of generality, it is assumed that ${p}^\text{MD}$ and ${f}^\text{MD}$ are the maximal transmission power and CC of any MD, respectively.
\par
In order to emphasize the effectiveness of AGWWO, we introduce the following algorithms for comparison in the simulation.
\par
\noindent
\textbf{\textit{Computing at Mobile Terminals (CMT):}} All tasks of MDs are executed locally in the maximal CC.
\par
\noindent
\textbf{\textit{Adaptive genetic algorithm (AGA):}} To solve \eqref{eq25}, AGA with diversity-guided mutation is introduced.
\par
\noindent
\textbf{\textit{Water Wave Optimization (WWO):}} To solve \eqref{eq25}, WWO in \cite{Y.Zheng2015Mar} is introduced.
\par
The following results can be found in the simulation. In CMT, all tasks of MDs are executed in the maximal allowable CC. However, partial tasks of MDs in other algorithms are offloaded to BSs for computing under the proportional resource assignment manner. Evidently, CMT may achieve higher total local EC and network-wide EC than other algorithms due to the utilization of more local CC in the former, where total local EC and network-wide EC refer to the energy consumed by all MDs and the one consumed by all MDs and SBSs respectively. Due to the lack of global search capability, WWO may find a worse solution than AGA and AGWWO. Consequently, WWO may achieve higher total local EC and network-wide EC than AGA and AGWWO. By replacing the propagation operation in WWO with genetic operations including selection, diversity-guided mutation, adaptive crossover, and mutation, AGWWO has a global search capability. In addition, the refraction and breaking operations let AGWWO have stronger local search capability than AGA. Therefore, AGWWO may find a better solution than AGA. That is to say, AGWWO may achieve lower total local EC and network-wide EC than AGA.
\begin{figure}[!t]
	\centering
	\includegraphics[width=3.2in]{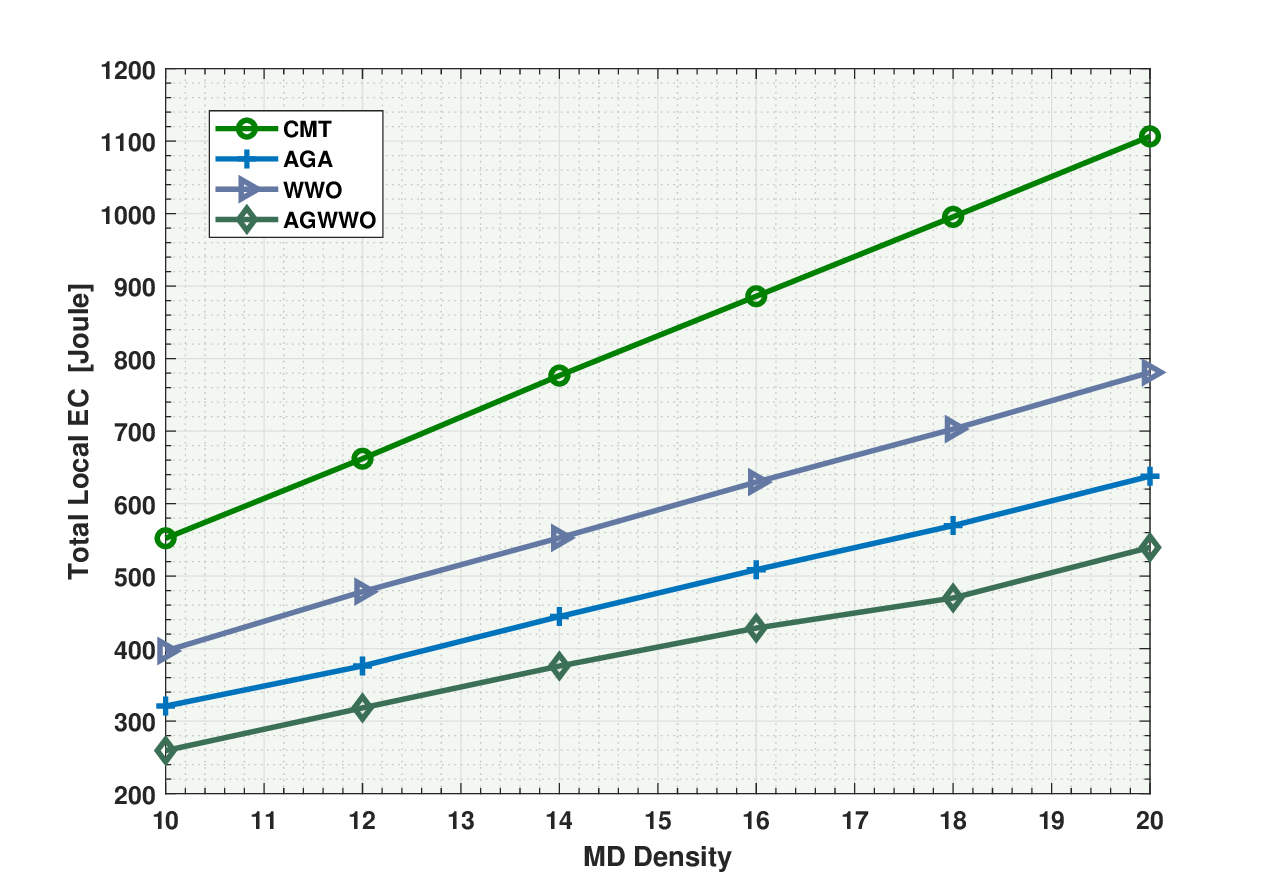}
	\caption{Impacts of ${\rho}^{\text{MD}}$ on total local EC.}
	\label{fig3}
\end{figure}
\par
In the simulation, besides WWO, the delay and cost constraints of other algorithms can almost always be satisfied. In other words, besides WWO, the time support ratio and cost support ratio of other algorithms are almost always one, where the time/cost support ratio is a measurement that quantifies the proportion of MDs whose time/cost requirements are within or equal to the maximum allowable time/cost, relative to the total number of MDs. Since CMT locally executes all tasks in the maximal allowable CC, its processing time is very short and has no processing cost. The global search capability of AGA and AGWWO lets the delay and cost constraints of MDs be almost guaranteed strictly. However, WWO may achieve a low time support ratio and a low cost support ratio because it is easy to fall into local optimum.
\begin{figure}[!t]
	\centering
	\includegraphics[width=3.2in]{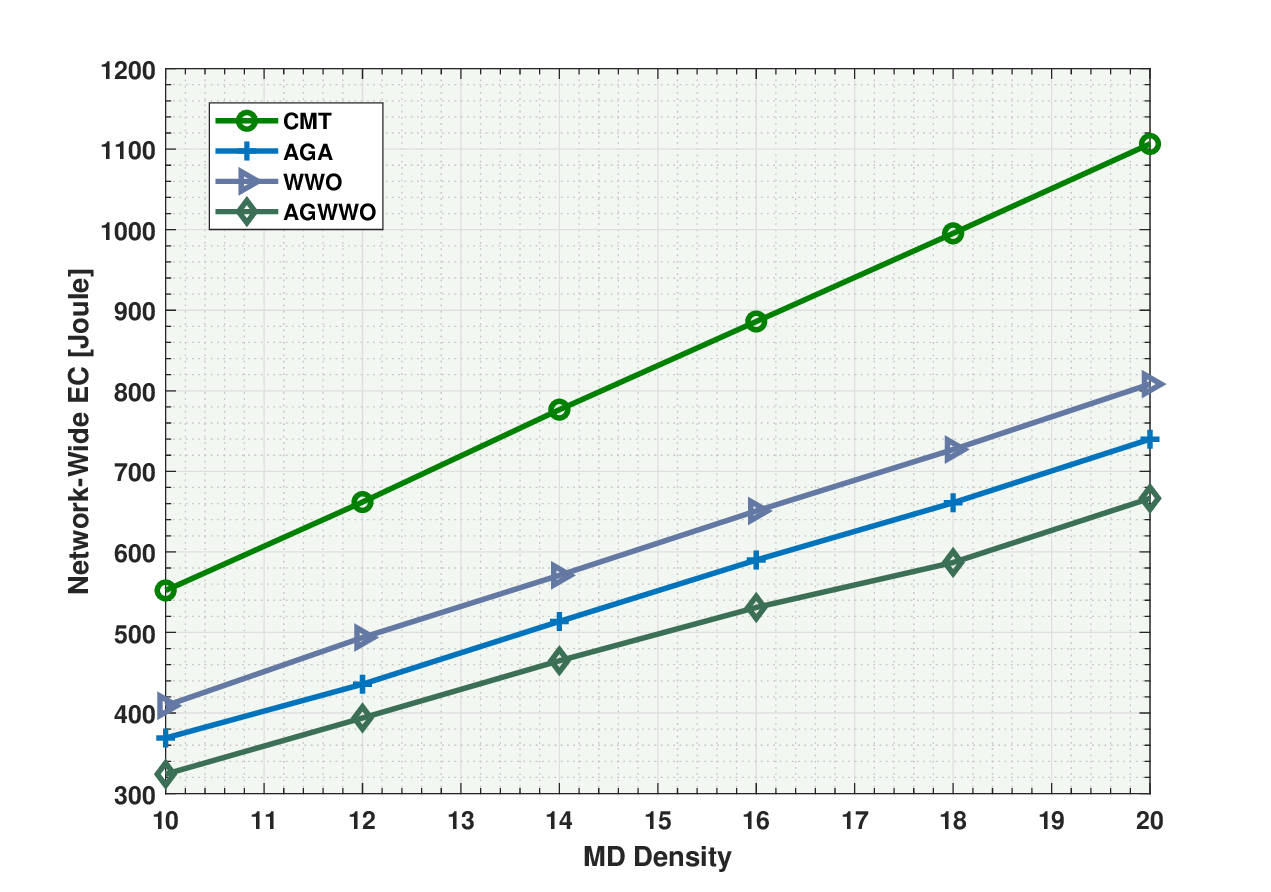}
	\caption{Impacts of ${\rho}^{\text{MD}}$ on network-wide EC.}
	\label{fig4}
\end{figure}
\par
Under ${p}^\text{MD}=23$ dBm and ${f}^\text{MD}= 1$ GHz, Figs. \ref{fig3} and \ref{fig4} show the impacts of MD density ${\rho}^{\text{MD}}$ on total local EC and network-wide EC respectively, where ${\rho}^{\text{MD}}$ denotes the MD density referring to the number of MDs at each macrocell. As shown in Figs. \ref{fig3} and \ref{fig4}, total local EC and network-wide EC should increase with MD density since more MDs mean more EC.
\begin{figure}[!t]
    \centering
    \includegraphics[width=3.2in]{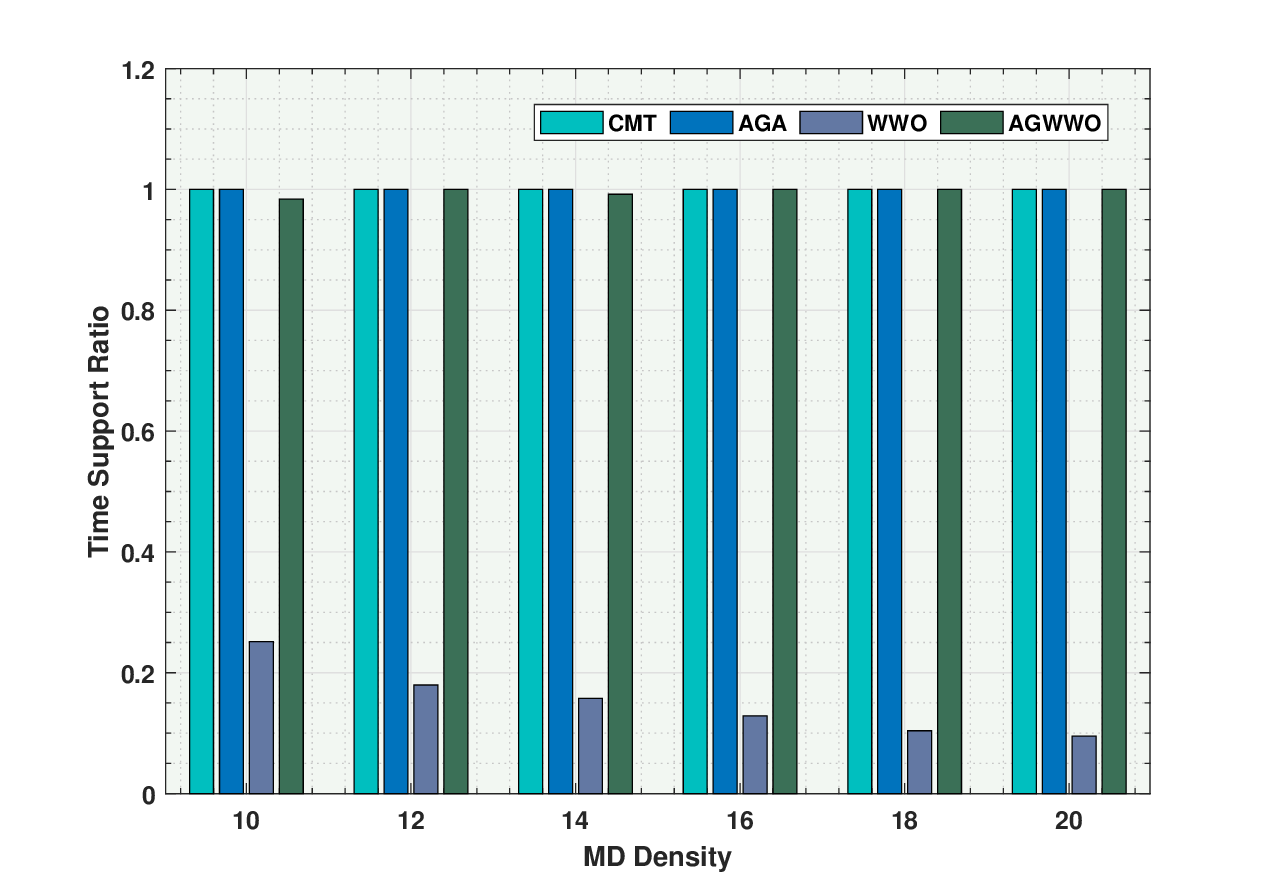}
    \caption{Impacts of ${\rho}^{\text{MD}}$ on time support ratio.}
    \label{fig5}
\end{figure}
\begin{figure}[!t]
	\centering
	\includegraphics[width=3.2in]{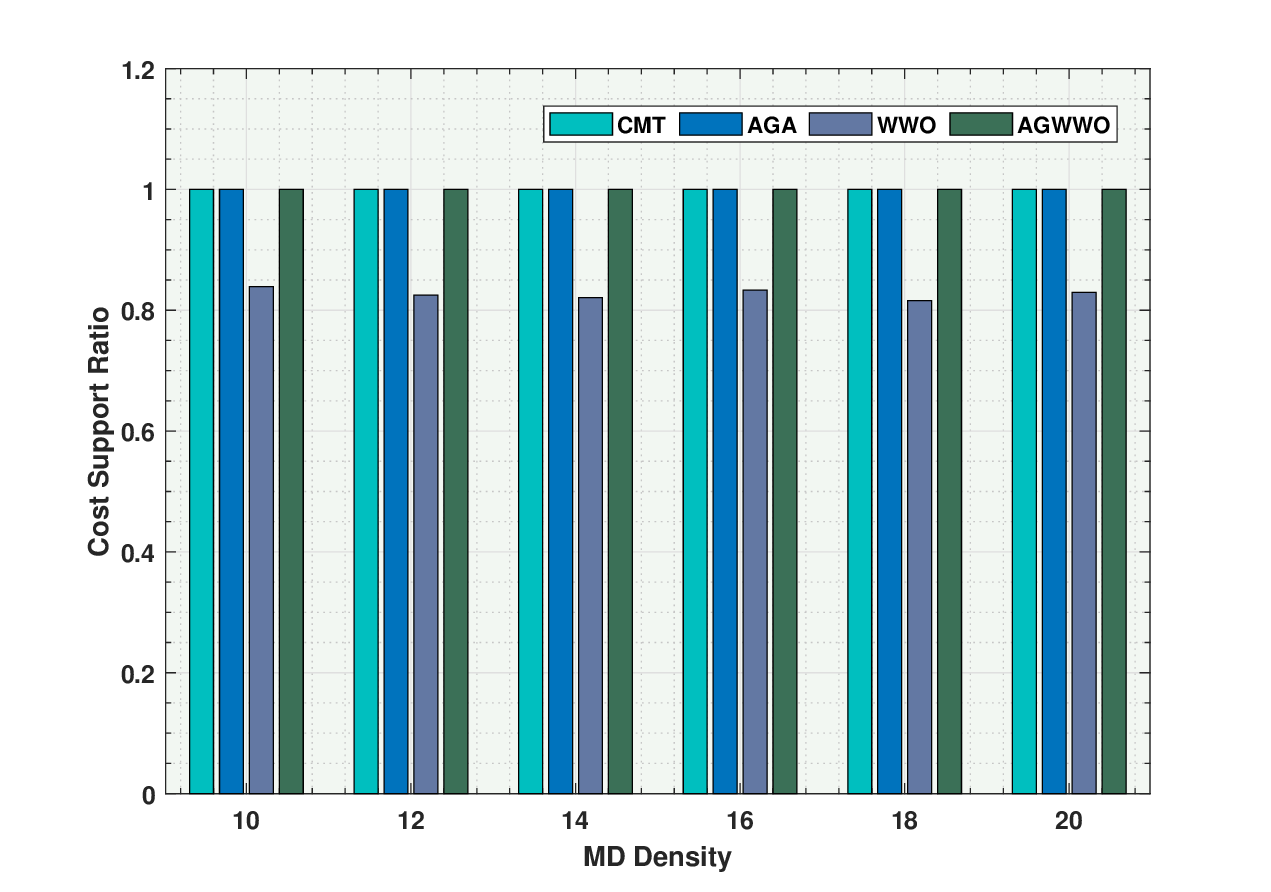}
	\caption{Impacts of ${\rho}^{\text{MD}}$ on cost support ratio.}
	\label{fig6}
\end{figure}
\par
Under ${p}^\text{MD}=23$ dBm and ${f}^\text{MD}= 1$ GHz, Figs. \ref{fig5} and \ref{fig6} show the impacts of MD density ${\rho}^{\text{MD}}$ on the time support ratio and cost support ratio respectively. As revealed earlier, the delay and cost constraints of WWO cannot be guaranteed, but other algorithms can almost always be satisfied. In the simulation, we find that WWO may achieve extremely high processing time, which is far higher than the processing cost. That is to say, the time constraints mainly decide the fitness function value of WWO. When ${\rho}^{\text{MD}}$ increases, the number of MDs whose time constraints cannot be satisfied may increase, but the number of MDs whose cost constraints cannot be satisfied may have no significant change. As shown in Fig. \ref{fig5}, the time support ratio of WWO may decrease with ${\rho}^{\text{MD}}$ since more MDs mean more MDs whose time constraints cannot be satisfied in WWO, but the cost support ratio of WWO is not the case.
\begin{figure}[!t]
	\centering
	\includegraphics[width=3.2in]{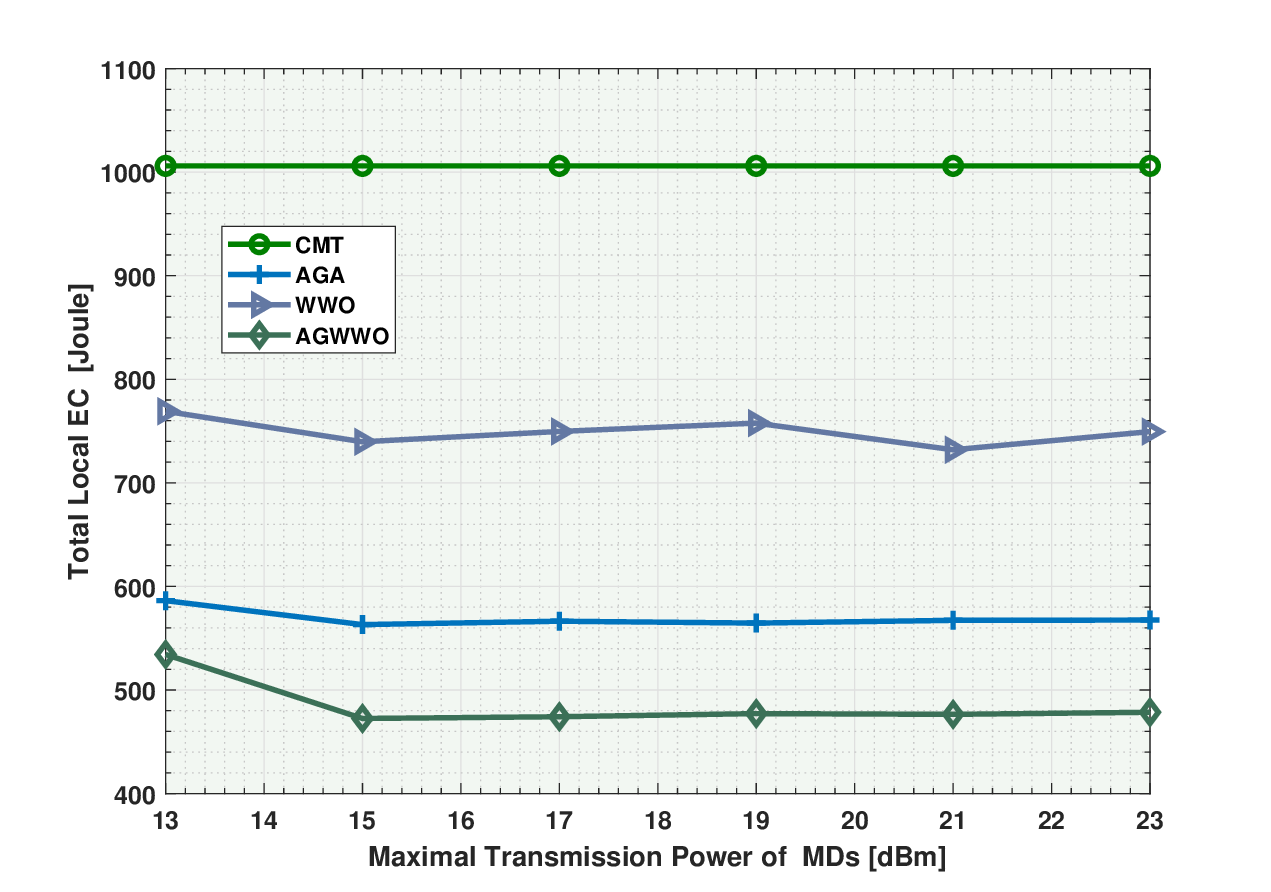}
	\caption{Impacts of ${p}^{\text{MD}}$ on total local EC.}
	\label{fig7}
\end{figure}
\begin{figure}[!t]
	\centering
	\includegraphics[width=3.2in]{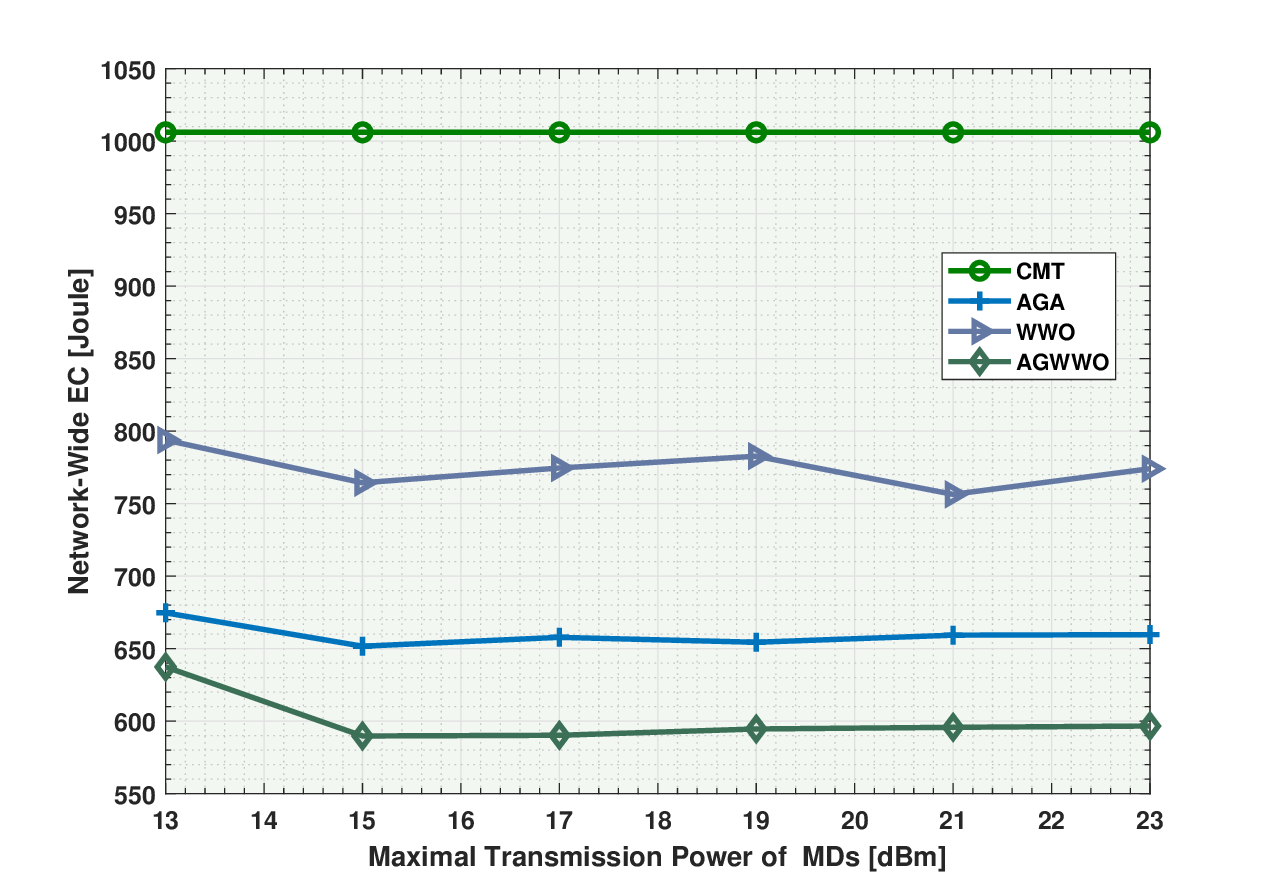}
	\caption{Impacts of ${p}^{\text{MD}}$ on network-wide EC.}
	\label{fig8}
\end{figure}
\par
Under ${\rho}^{\text{MD}} = 20$ and ${f}^\text{MD}= 1$ GHz, Figs. \ref{fig7} and \ref{fig8} show the impacts of maximal transmission power ${p}^\text{MD}$ on total local EC and network-wide EC respectively. As shown in Figs. \ref{fig7} and \ref{fig8}, total local EC and network-wide EC of CMT should not change with ${p}^{\text{MD}}$ since such a parameter has no relation to them. However, the total local EC and network-wide EC of other algorithms may decrease with ${p}^{\text{MD}}$ in general. The reason for this may be that more and more MDs may offload their tasks for computing when ${p}^{\text{MD}}$ increases and local processing EC is often higher than remote processing EC .
\begin{figure}[!t]
    \centering
    \includegraphics[width=3.2in]{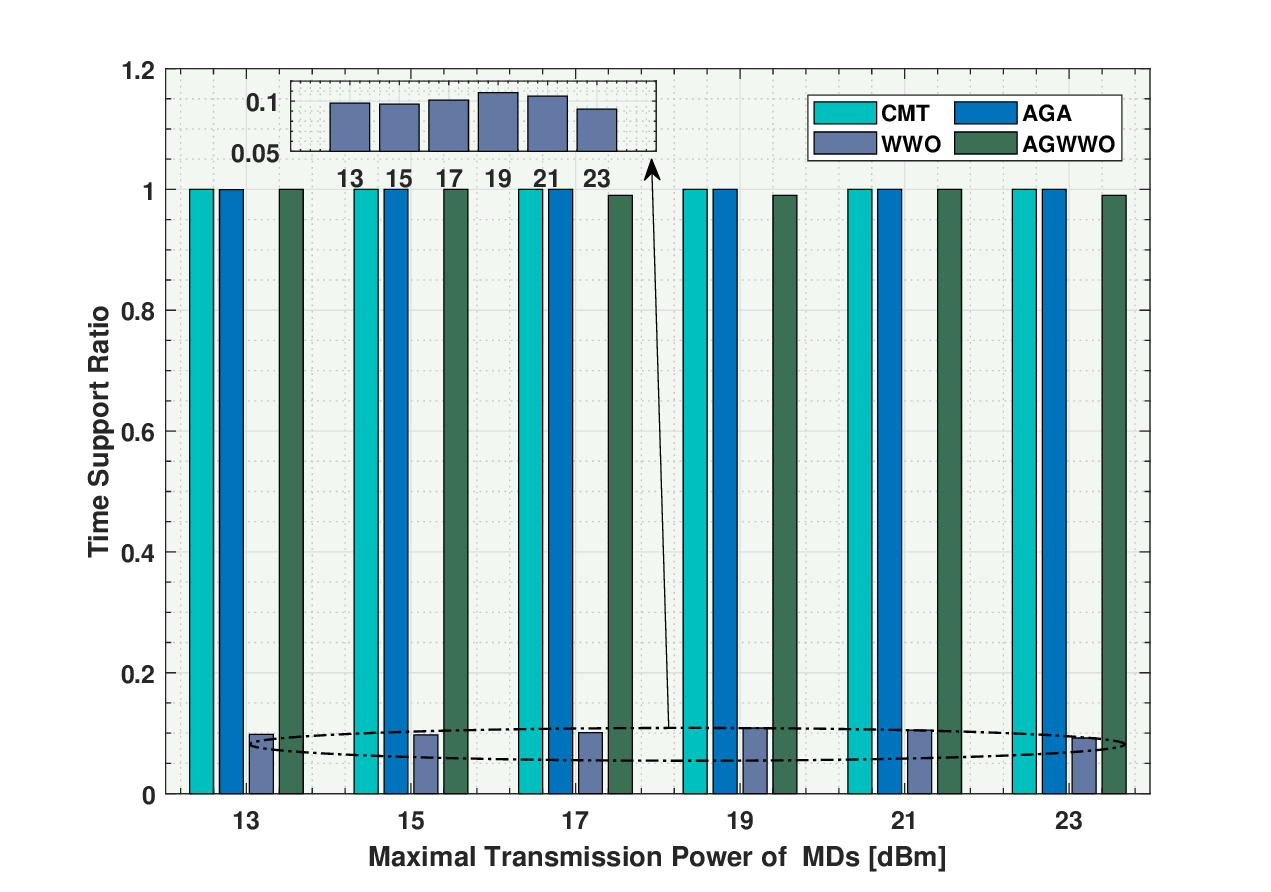}
    \caption{Impacts of ${p}^{\text{MD}}$ on time support ratio.}
    \label{fig9}
\end{figure}
\begin{figure}[!t]
	\centering
	\includegraphics[width=3.2in]{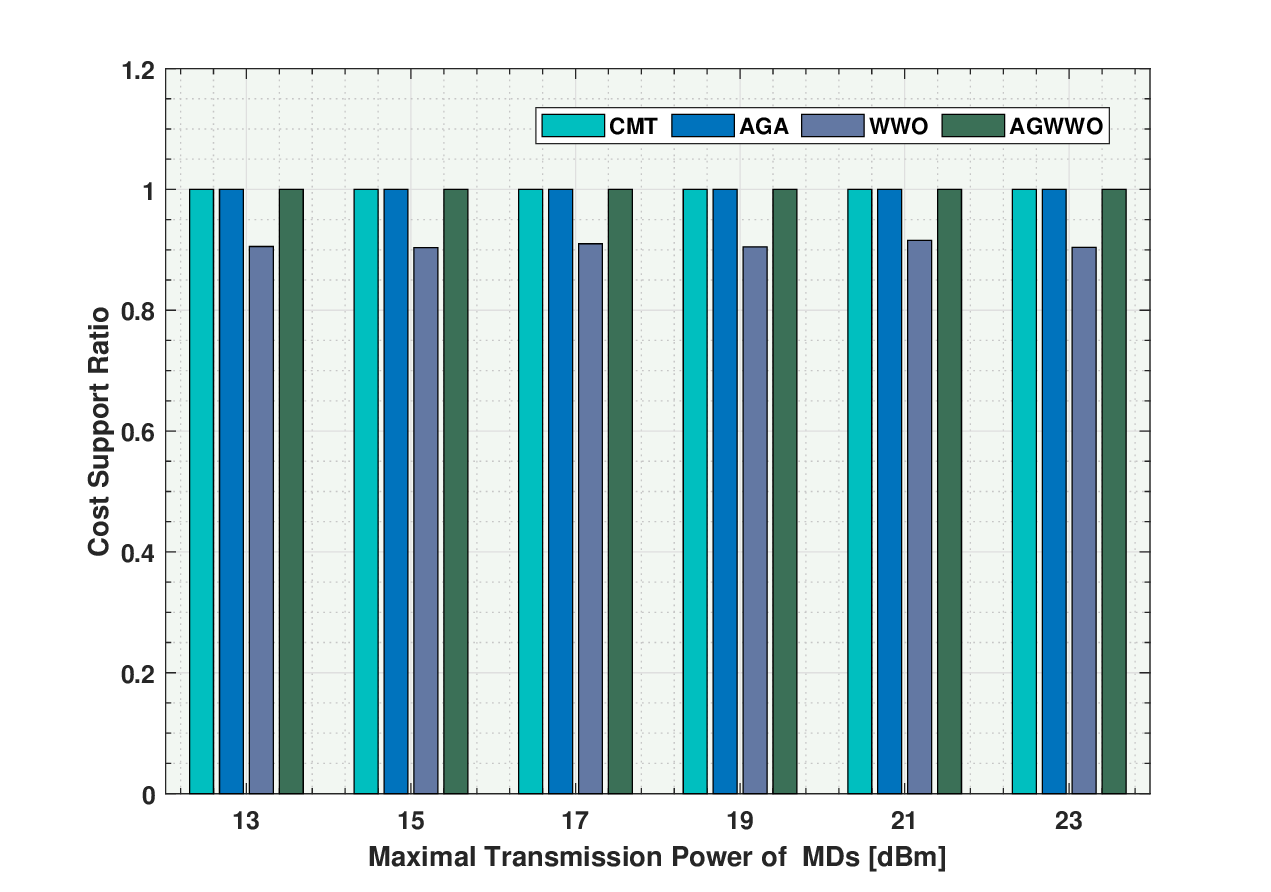}
	\caption{Impacts of ${p}^{\text{MD}}$ on cost support ratio.}
	\label{fig10}
\end{figure}
\par
Under ${\rho}^{\text{MD}} = 20$ and ${f}^\text{MD}= 1$ GHz, Figs. \ref{fig9} and \ref{fig10} show the impacts of maximal transmission power ${p}^\text{MD}$ on the time support ratio and cost support ratio respectively. Similar to Figs. \ref{fig5} and \ref{fig6}, the delay and cost constraints of WWO cannot be guaranteed, but other algorithms can almost always be satisfied. The time support ratio of WWO may initially increase with ${p}^\text{MD}$ but then decrease with it. As we know, the initially increased ${p}^\text{MD}$ may result in an increased uplink data rate when it cannot cause severe interference. However, when ${p}^\text{MD}$ increases, the network interference becomes severer and severer, resulting in a lower and lower uplink data rate. Similar to Fig. \ref{fig6}, ${p}^\text{MD}$ may have no significant impact on the cost support ratio.
\begin{figure}[!t]
	\centering
	\includegraphics[width=3.2in]{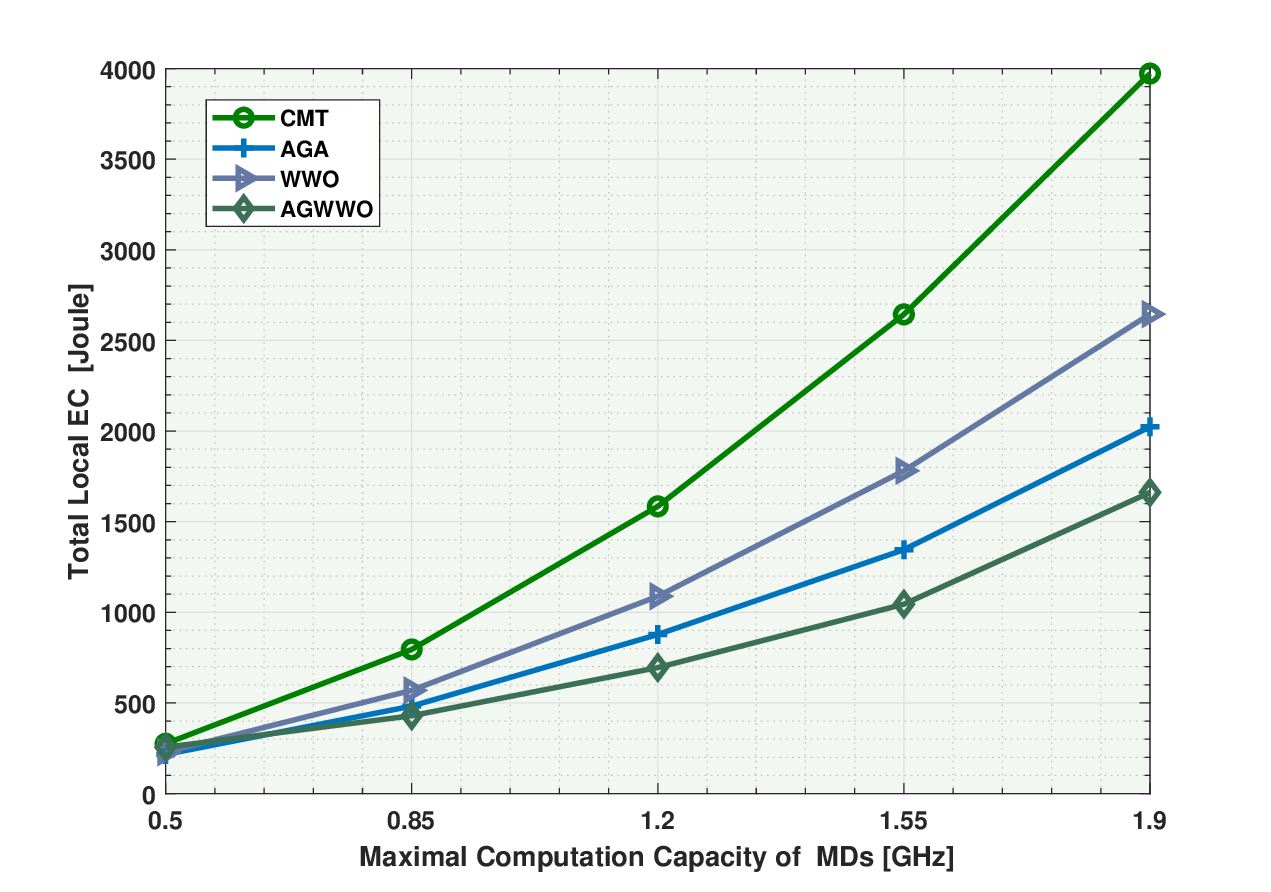}
	\caption{Impacts of $f^{\text{MD}}$ on total local EC.}
	\label{fig11}
\end{figure}
\begin{figure}[!t]
	\centering
	\includegraphics[width=3.2in]{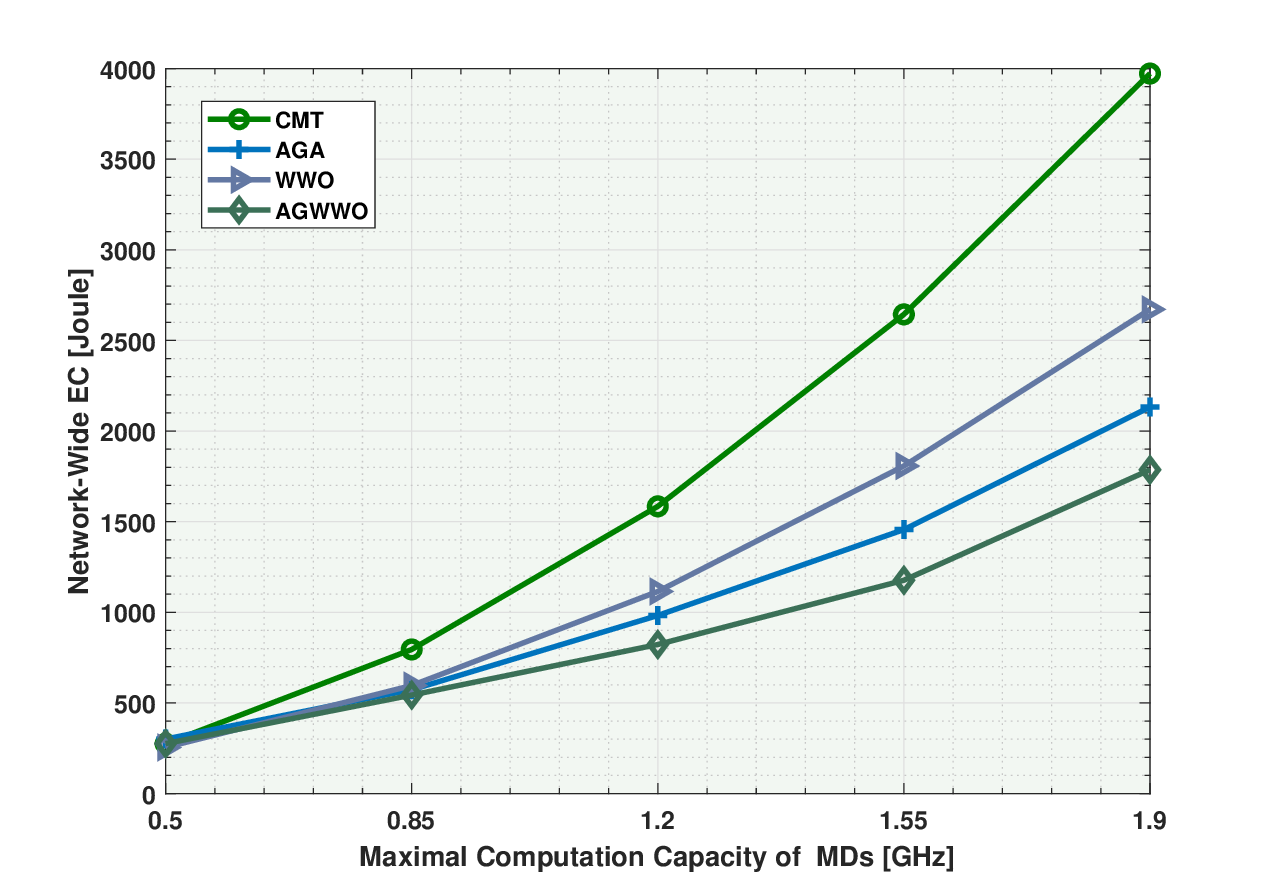}
	\caption{Impacts of $f^{\text{MD}}$ on network-wide EC.}
	\label{fig12}
\end{figure}
\par
Under ${p}^\text{MD} = 23$ dBm and ${\rho}^{\text{MD}} = 20$, Figs. \ref{fig11} and \ref{fig12} show the impacts of maximal CC $f^{\text{MD}}$ on total local EC and network-wide EC respectively. As shown in Figs. \ref{fig11} and \ref{fig12}, total local EC and network-wide EC of all algorithms may increase with $f^{\text{MD}}$ in general. The reason for this may be that a higher $f^{\text{MD}}$ means a higher local EC.
\begin{figure}[!t]
    \centering
    \includegraphics[width=3.2in]{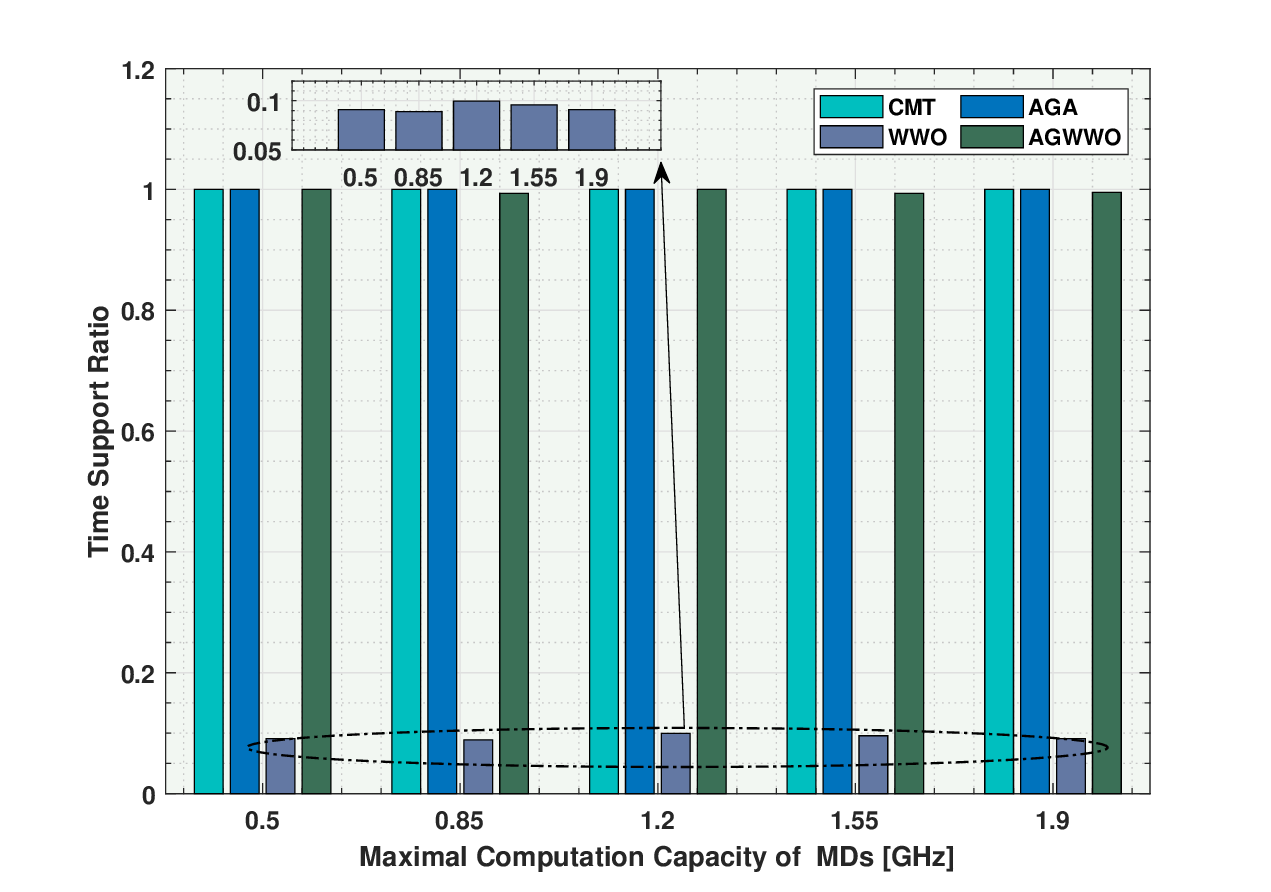}
    \caption{Impacts of $f^{\text{MD}}$ on time support ratio.}
    \label{fig13}
\end{figure}
\begin{figure}[!t]
	\centering
	\includegraphics[width=3.2in]{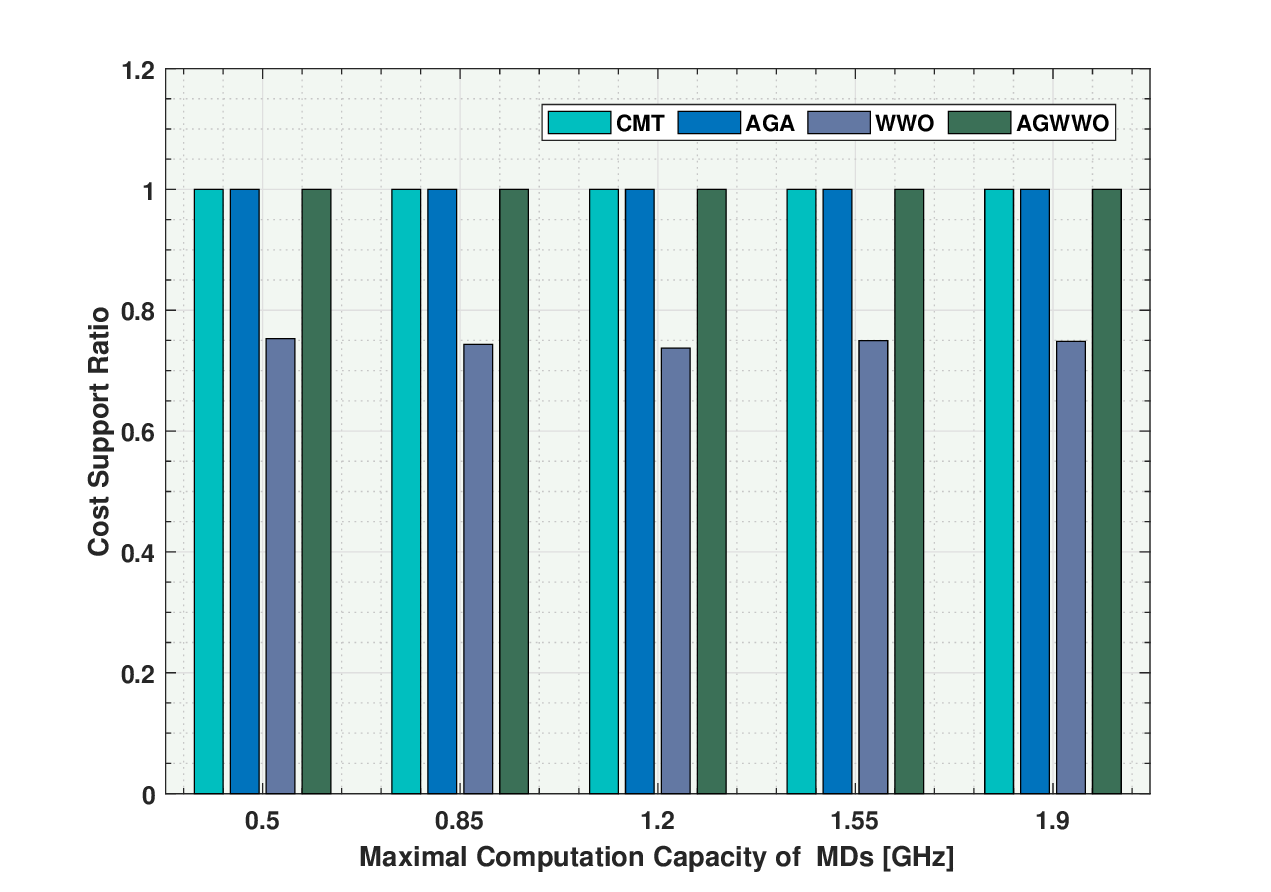}
	\caption{Impacts of $f^{\text{MD}}$ on cost support ratio.}
	\label{fig14}
\end{figure}
Under ${p}^\text{MD} = 23$ dBm and ${\rho}^{\text{MD}} = 20$, Figs. \ref{fig13} and \ref{fig14} show the impacts of maximal CC $f^{\text{MD}}$ on the time support ratio and cost support ratio respectively. Similar to Figs. \ref{fig5} and \ref{fig6}, the delay and cost constraints of WWO cannot be guaranteed, but other algorithms can almost always be satisfied. The time support ratio of WWO may initially increase with $f^{\text{MD}}$ but then decrease with it. In the simulation, it is easy to find that more MDs locally execute tasks in WWO than in AGA and AGWO. An initially increased $f^{\text{MD}}$ may result in a shorter local executing time, but more EC. To further reduce EC, when $f^{\text{MD}}$ increases, some MDs will be forced to offload their tasks for computing, resulting in a lower and lower time support ratio. Similar to Fig. \ref{fig6}, $f^{\text{MD}}$ may have no significant impact on the cost support ratio.
\begin{figure}[!t]
	\centering
	\includegraphics[width=3.2in]{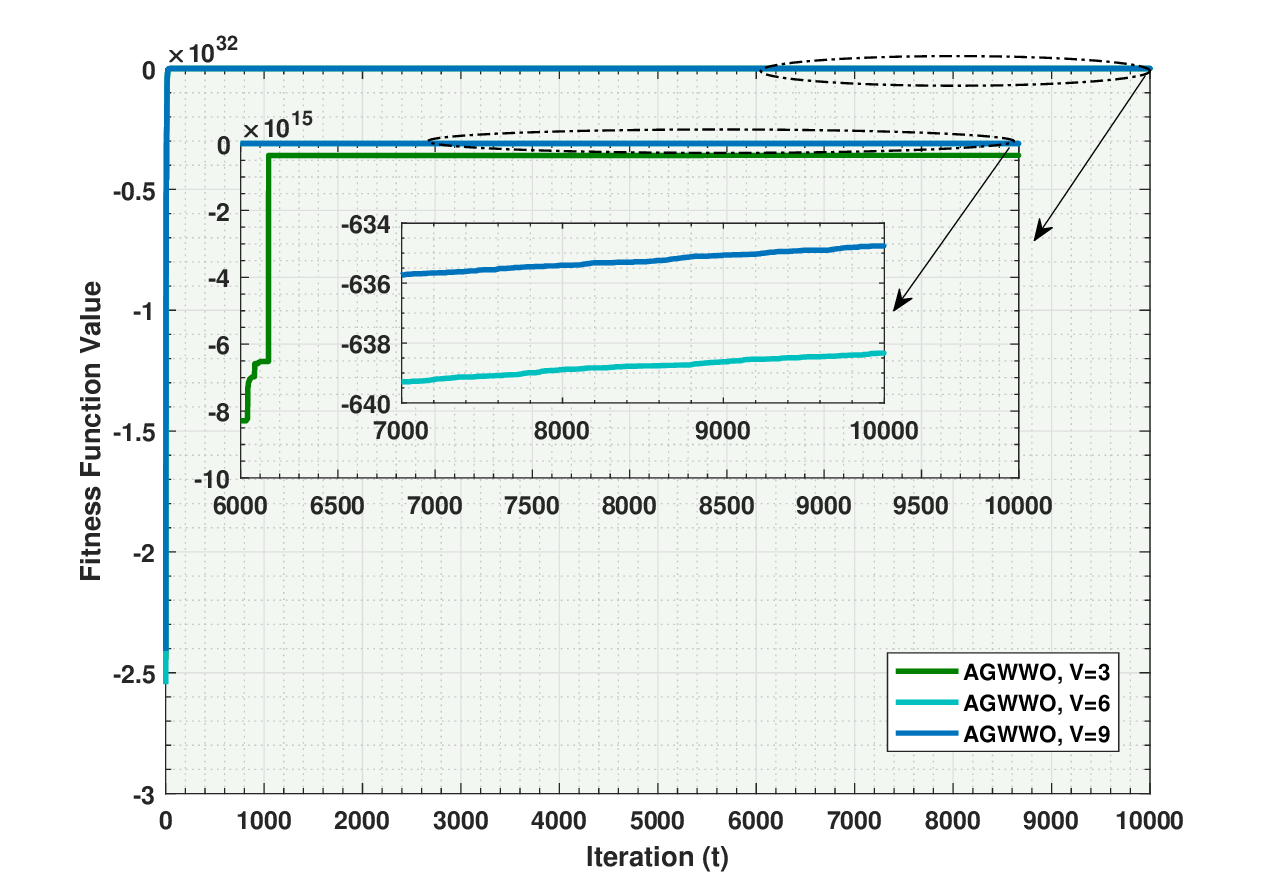}
	\caption{Impacts of $V$ on fitness function value.}
	\label{fig15}
\end{figure}
\begin{figure}[!t]
	\centering
	\includegraphics[width=3.2in]{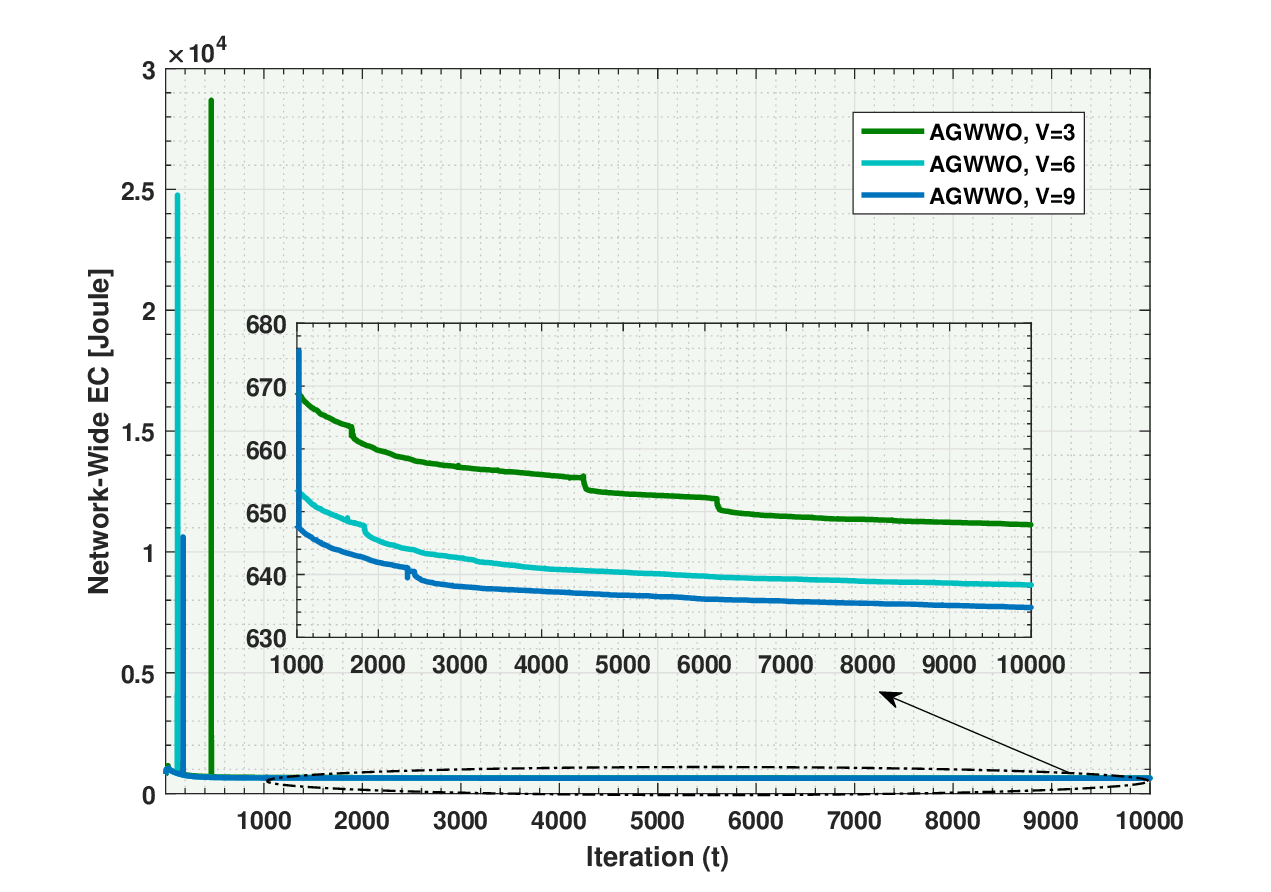}
	\caption{Impacts of $V$ on network-wide EC.}
	\label{fig16}
\end{figure}
\par
Under ${p}^\text{MD} = 23$ dBm, ${\rho}^{\text{MD}} = 20$ and ${f}^\text{MD}= 1$ GHz, Figs. \ref{fig15} and \ref{fig16} show the impacts of $V$ (number of solitary waves) on the fitness function value and network-wide EC respectively. As illustrated in Figs. \ref{fig15} and \ref{fig16}, the fitness function value and network-wide EC are relatively stable after a few iterations, which means that AGWWO is convergent. In addition, the fitness function value may increase with $V$, but the network-wide EC may decrease with it. That's because a larger $V$ means more opportunities for searching for a better solution with a higher fitness function value and a lower network-wide EC.
\begin{figure}[!t]
	\centering
	\includegraphics[width=3.2in]{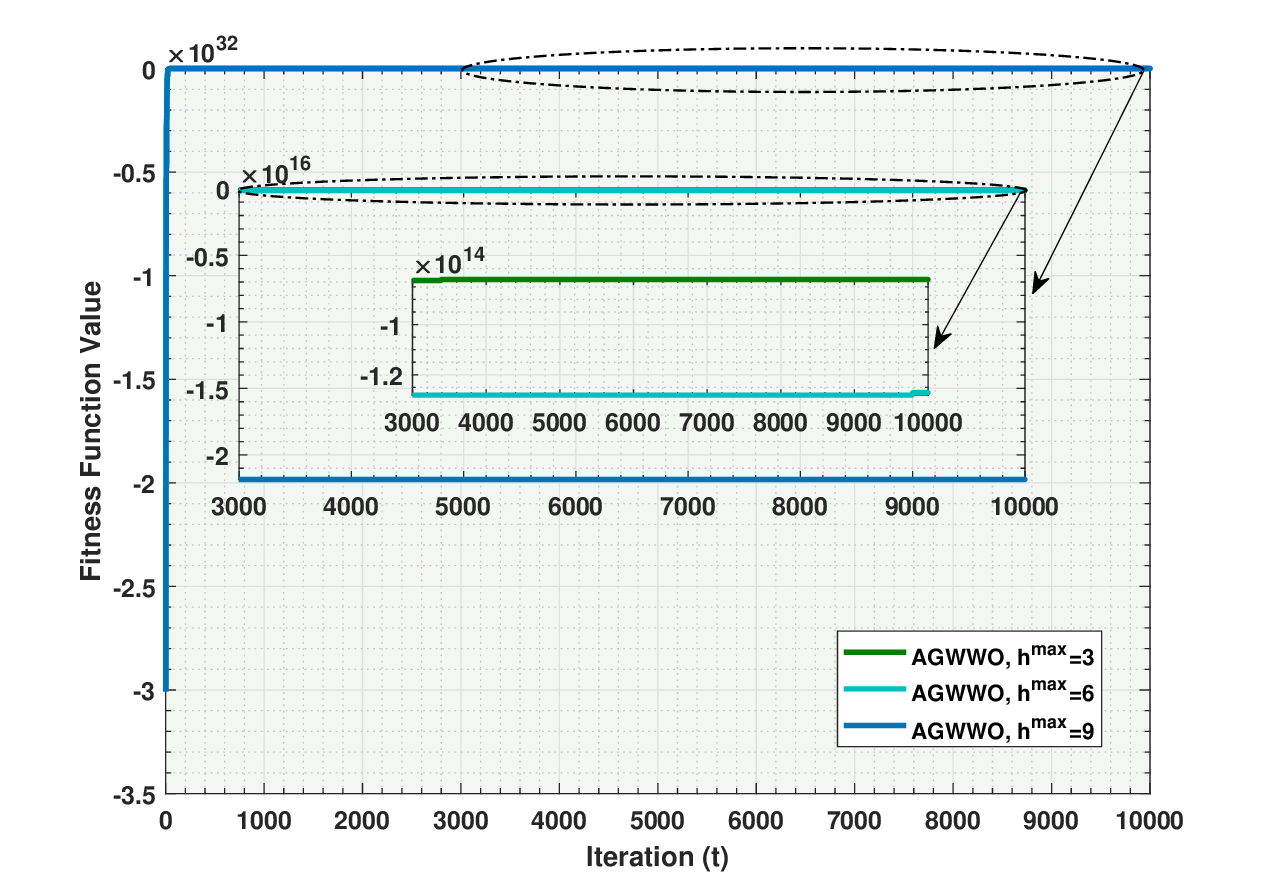}
	\caption{Impacts of ${h}^{\max}$ on fitness function value.}
	\label{fig17}
\end{figure}
\begin{figure}[!t]
	\centering
	\includegraphics[width=3.2in]{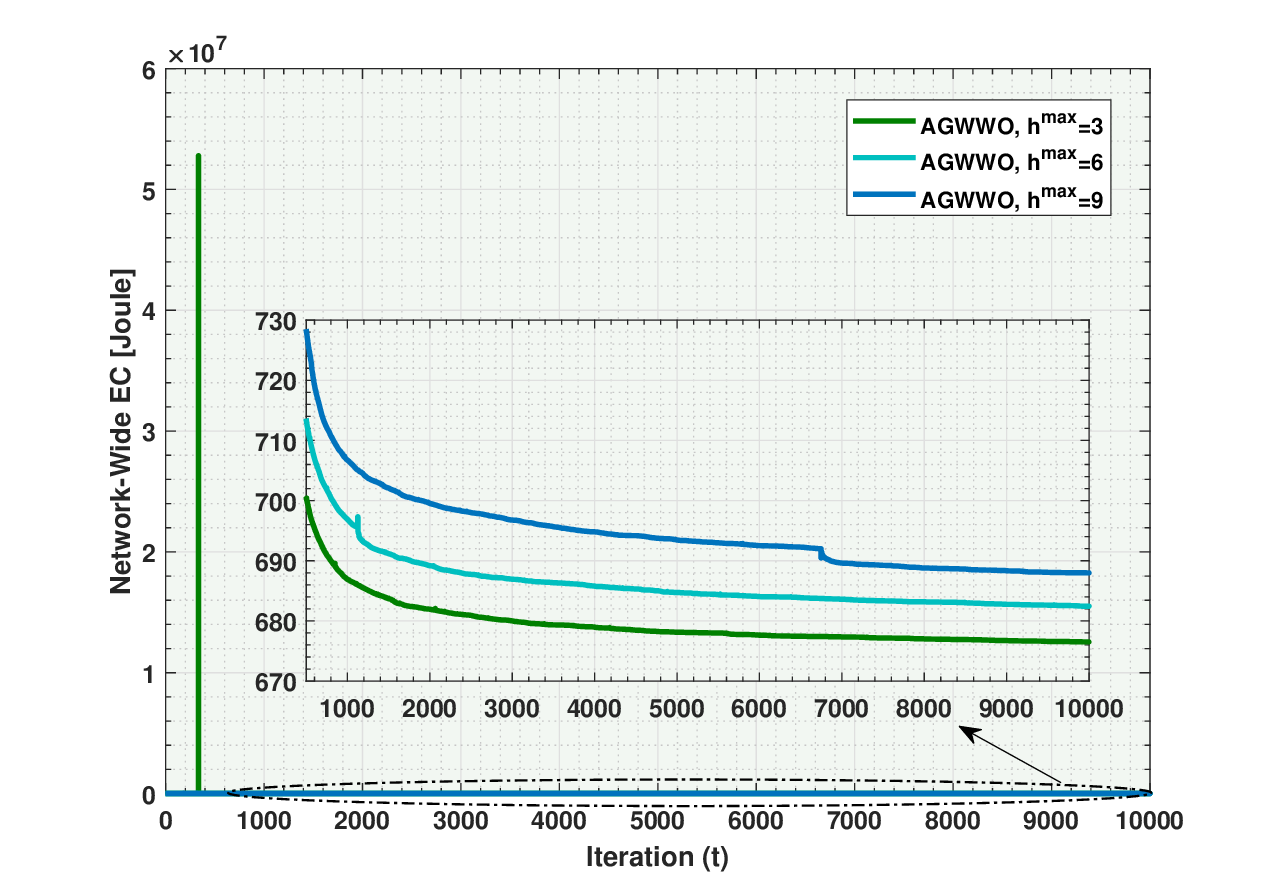}
	\caption{Impacts of ${h}^{\max}$ on network-wide EC.}
	\label{fig18}
\end{figure}
\par
Under ${p}^\text{MD} = 23$ dBm, ${\rho}^{\text{MD}} = 20$ and ${f}^\text{MD}= 1$ GHz, Figs. \ref{fig17} and \ref{fig18} show the impacts of ${h}^{\max}$ (maximal height of waves) on the fitness function value and network-wide EC respectively. As illustrated in Figs. \ref{fig17} and \ref{fig18}, the fitness function value and network-wide EC are relatively stable after a few iterations, which means that AGWWO is convergent. In addition, the fitness function value may decrease with ${h}^{\max}$, but the network-wide EC may increase with it. That's because a larger ${h}^{\max}$ also means more opportunities for searching for a better solution with a higher fitness function value and a lower network-wide EC.
\begin{figure}[!t]
	\centering
	\includegraphics[width=3.2in]{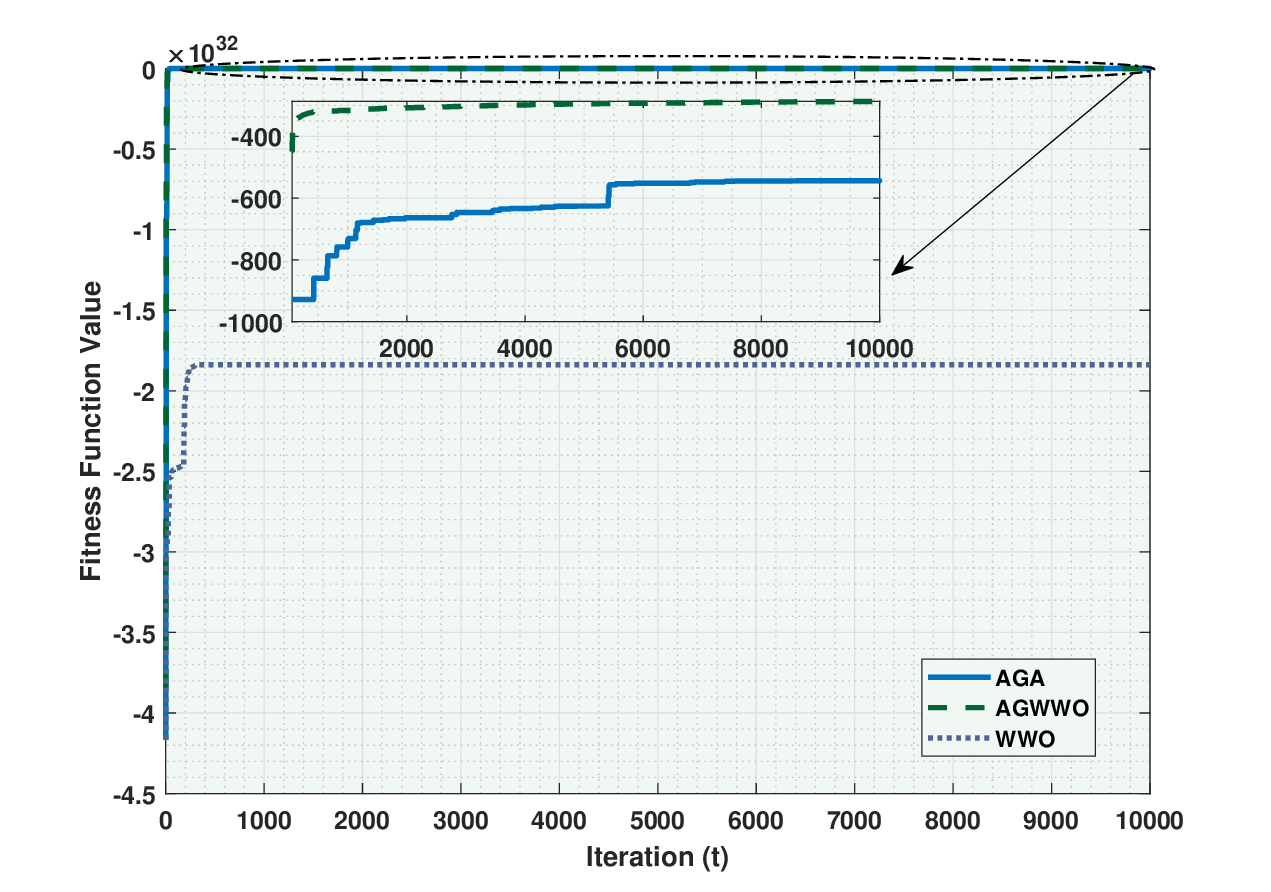}
	\caption{Convergence of algorithms.}
	\label{fig19}
\end{figure}
\par
Under ${p}^\text{MD} = 23$ dBm, ${\rho}^{\text{MD}} = 20$ and ${f}^\text{MD}= 2$ GHz, Fig. \ref{fig19} shows the convergence of AGA, AGWWO, and WWO. As illustrated in Fig. \ref{fig19}, AGWWO may have a higher convergence speed and find a better solution than other AGA and WWO because of global search capability with enhanced local search capability. In addition, WWO may be premature and find a worse solution than other algorithms since it is easy to fall into the local optimum.
\section{Conclusion}\label{sec7}
In this paper, a secure data-compressed multi-step computation offloading model is first established for multi-task ultra-dense networks with both OFDMA and NOMA. Then, we optimize joint DC, secure multi-step computation offloading, and resource assignment to minimize network-wide EC. To this end, we develop the AGWWO algorithm, which incorporates genetic operations to replace the propagation behavior of the traditional WWO algorithm. As for the AGWWO algorithm, we make detailed analyses concentrating on convergence, computational complexity, and potential for parallel implementation. Simulation results demonstrate that, in comparison to existing algorithms, AGWWO shows promise in achieving lower total local EC and network-wide EC while maintaining time and cost constraints. Future work may focus on developing compression and encryption models with greater flexibility and reduced resource requirements.

\vspace{-0.6in}
\begin{IEEEbiography}[{\includegraphics[width=1in,height=1.1in,clip,keepaspectratio]{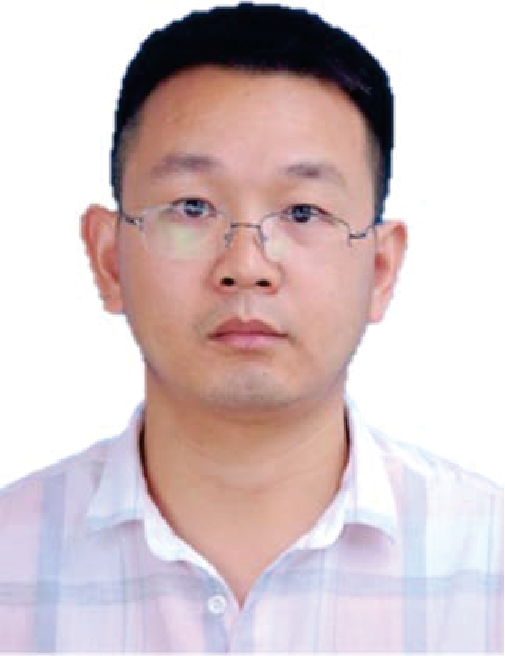}}]{Tianqing Zhou}
 received the Ph.D. degree in Information and Communication Engineering from Southeast University, Nanjing, China, in 2016. He is currently an associate Professor with the School of Information and Software Engineering, East China Jiaotong University. He is the author or coauthor of more than 40 journal papers indexed by SCI, and holds more than 10 patents. His current research interests include mobile edge computing and caching, artificial intelligence algorithms, wireless resource management and ultra-dense networks.
\end{IEEEbiography}
\vspace{-0.65in}
\begin{IEEEbiography}[{\includegraphics[width=1in,height=1.1in,clip,keepaspectratio]{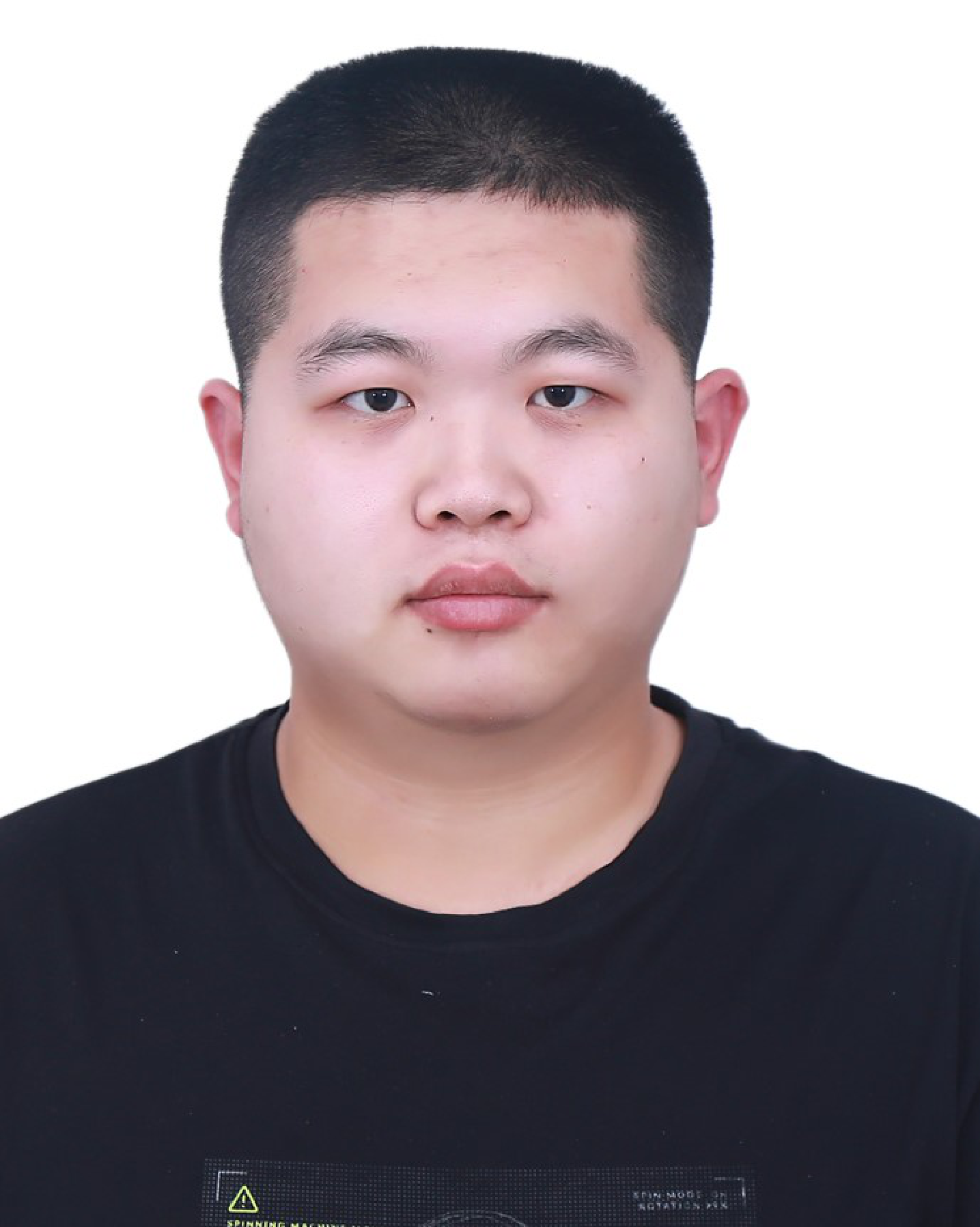}}]{Kangle Liu}
received the B.E. degree in Electronic and Information Engineering from China University of Geosciences, Beijing, China, in 2021. He is currently pursing the M.E. degree in Information and Communication Engineering at East China Jiaotong University, Nanchang, China. His current research interests include ultra dense network.
\end{IEEEbiography}
\vspace{-0.6in}
\begin{IEEEbiography}[{\includegraphics[width=1in,height=1.1in,clip,keepaspectratio]{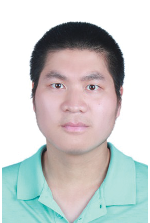}}]{Dong Qin} received the Ph.D. degree in Information and Communication Engineering from Southeast University, Nanjing, China, in 2016. He is currently an associate Professor with the School of Information and Software Engineering, East China Jiaotong University. His current research interests lie in the area of cooperative communication and OFDM techniques.
\end{IEEEbiography}
\vspace{-0.65in}
\begin{IEEEbiography}[{\includegraphics[width=1in,height=1.1in,clip,keepaspectratio]{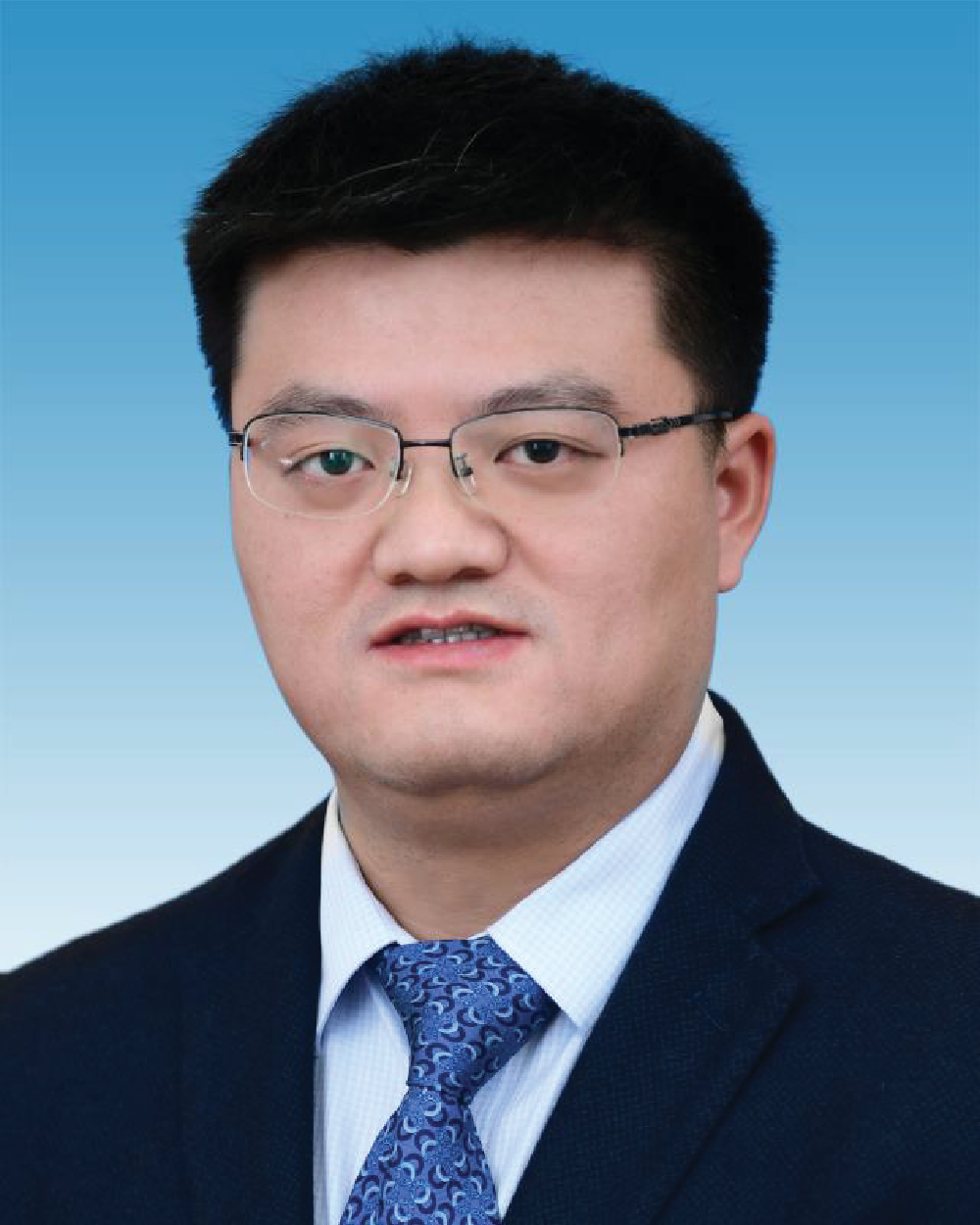}}]{Nan Jiang}
received the PhD degree in Nanjing University of Aeronautics and Astronautics, Nanjing, China, in 2008. Currently he is a professor in School of Information and Software Engineering, East China Jiaotong University. He was the research scholar in Complex Networks and Security Research Lab at Virginia Tech between 2013 and 2014. His current research interests lie in wireless sensor networks, the scalable sensor networks, wireless protocol and architecture, distributed computing.
\end{IEEEbiography}
\vspace{-0.6in}
\begin{IEEEbiography}[{\includegraphics[width=1in,height=1.1in,clip,keepaspectratio]{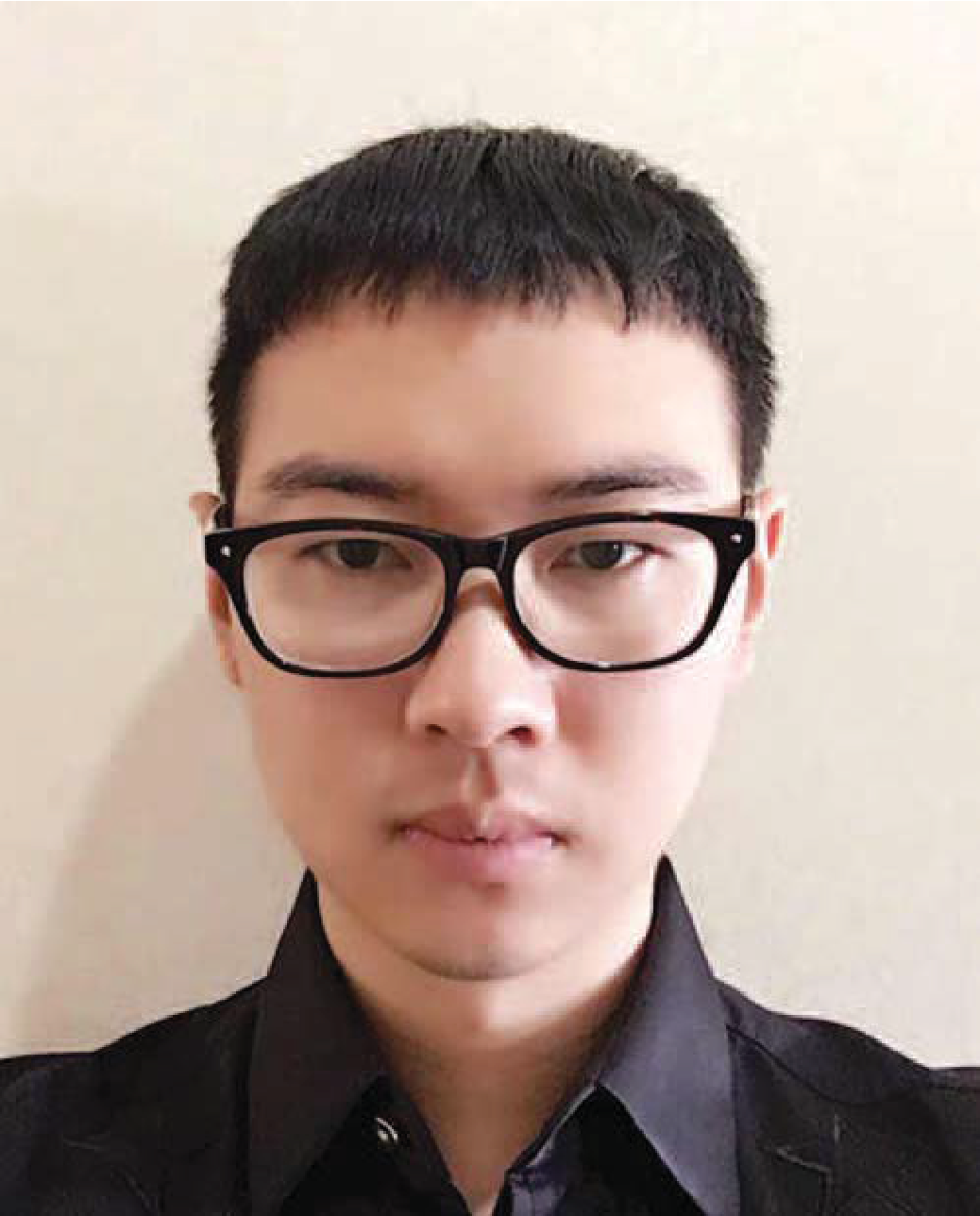}}]{Xuan Li} received Ph.D. degree in Communication and Information System from Xidian University, Xian, China, in 2017. He is currently an associate Professor with the School of Information and Software Engineering, East China Jiaotong University. His research interests include interference cancellation, topology control, spectrum sharing of graph theory, and cognitive radio networks.
\end{IEEEbiography}
\vspace{-0.6in}
\begin{IEEEbiography}[{\includegraphics[width=1in,height=1.15in,clip,keepaspectratio]{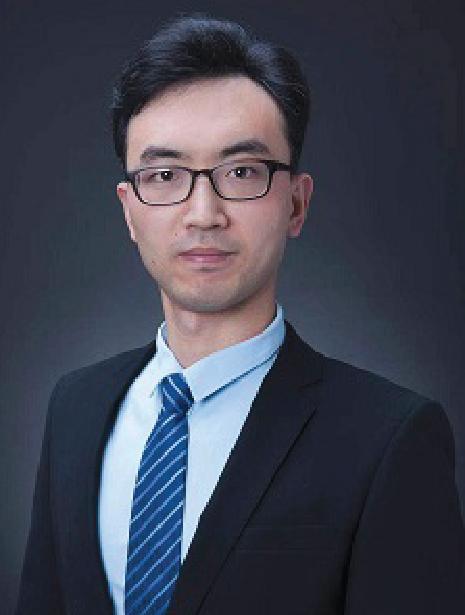}}]{Chunguo Li}
 (SM'16) received the bachelor's degree in Wireless Communications from Shandong University in 2005, and the Ph.D. degree in Wireless Communications from Southeast University in 2010. In July 2010, he joined the School of Information Science and Engineering, Southeast University, Nanjing China, where he is currently an Advisor of Ph.D candidates and Full Professor. From June 2012 to June 2013, he was the Post-Doctoral researcher with Concordia University, Montreal, Canada. From July 2013 to August 2014, he was with the DSL laboratory of Stanford University as visiting associate professor. From August 2017 to July 2019, he was the adjunct professor of Xizang Minzu University under the supporting Tibet program organized by China National Human Resources Ministry. He is the Fellow of IET, the Fellow of China Institute of Communications (CIC), and IEEE CIS Nanjing Chapter Chair. His research interests are in cell-free distributed MIMO wireless communications for 6G, and machine learning based image/video signal processing algorithm design.
\end{IEEEbiography}





\end{document}